\documentclass[11pt,a4paper,oneside,openany]{book}
\usepackage{epsfig}
\usepackage{amssymb,amsmath}
\usepackage{eucal}
\newsavebox{\uuunit}
\sbox{\uuunit}
    {\setlength{\unitlength}{0.825em}
     \begin{picture}(0.6,0.7)
        \thinlines
        \put(0,0){\line(1,0){0.5}}
        \put(0.15,0){\line(0,1){0.7}}
        \put(0.35,0){\line(0,1){0.8}}
       \multiput(0.3,0.8)(-0.04,-0.02){12}{\rule{0.5pt}{0.5pt}}
     \end {picture}}
\newcommand {\unity}{\mathord{\!\usebox{\uuunit}}}

\newcommand{\eq}{\begin{equation}}
\newcommand{\feq}{\end{equation}}
\newcommand{\eqn}{\begin{eqnarray}}
\newcommand{\feqn}{\end{eqnarray}}
\newcommand{\arr}{\begin{eqnarray*}}
\newcommand{\farr}{\end{eqnarray*}}

\newcommand{\M}{{\cal M}}

\newcommand{\BB}{{\cal B}}

\newcommand{\DD}{{\cal D}}

\newcommand{\RR}{{\cal R}}
\newcommand{\VV}{{\cal V}}

\newcommand{\nn}{\nonumber}
\def\half{{\frac 1 2}}

\newtheorem{note}{Remark}[section]
\newtheorem{lemma}{Lemma}[section]
\newtheorem{corollary}{Corollary}[section]
\newtheorem{proposition}{Proposition}[section]

\newenvironment{prop}{\begin{proposition}}{\end{proposition}}

\csname @addtoreset\endcsname{equation}{section}

\def\rmd{{\rm d}}
\newcommand{\ft}[2]{{\textstyle\frac{#1}{#2}}}

\def\bfone{\relax{\rm 1\kern-.35em 1}}
\newtheorem{definizione}{Definition}[section]

\usepackage[latin1]{inputenc}
\begin{document}
\clearpage
\thispagestyle{empty}
\def\cnt#1{\begin{center}{#1}\end{center}}
{\Large
\vskip-24mm
\cnt{
Universit\`a degli Studi di Milano\\
Facolt\`a di Scienze Matematiche, Fisiche e Naturali\\
Corso di dottorato in Fisica, XVI ciclo\\
Anno 2003}
\vskip1cm
\cnt{Tesi di Dottorato}
\vskip.7cm
\cnt{\Huge\bf Toward the classification\\ \vspace{0.5cm}
 of BPS solutions of \\
\vspace{0.5cm}
${\cal N}=2$, $d=5$  gauged supergravity \\ \vspace{0.5cm}
 with matter couplings}
\vskip6.2cm
\newlength{\meta}
\setlength{\meta}{8.3cm}
\def\due#1#2{\par\noindent
\makebox[\linewidth][l]{\makebox[\meta][l]{#1\hfill}\makebox{#2\hfill}}\par}
\due{CANDIDATO:}{RELATORE:}
\due{Alessio Celi}{Prof. Daniela Zanon}
}

 \clearpage \ \clearpage
\tableofcontents

\addcontentsline{toc}{chapter}{Motivation}
\chapter*{Motivation}

The quest for a quantum theory of gravity is one of the major challenges of theoretical physics of the last four decades. 

Supergravity in different dimensions has a special role in this story: though it is not a fundamental theory it maintains a great importance because it describes the low-energy effective limit of Superstring theory, which represents the best candidate to be the right framework for  unifying all fundamental interactions.

In particular the solutions to the supergravity equations of motion, especially those that preserve some supersymmetry, are of interest for many reasons. One reason  is that they are useful in studying compactifications from $d=10$ or $11$ down to a lower dimensional space. By compactifying down to four spacetime  dimensions, for example, one might hope to make contact with particle physics phenomenology.      

Another reason resides in the existence of black holes: from the point of view of General Relativity these are simple spacetime configurations that have a singularity. Classically this is not a problem if the singularities are hidden behind event horizons, since this means that nothing can come out from the region containing the singularity. However, Hawking showed \cite{hawking} that, under very general assumptions, quantum mechanics implies that black holes emit particles. Because the radiation is thermal, this leads to the problem of ``information loss'' and besides that it implies that black holes are thermodynamic objects. So it is a strong test for string theory to explain which are the microscopic states that justify the thermodynamic properties of the macroscopic state (the black hole), such as the area law \cite{arealaw}. This question is very difficult to solve in general because it is intrinsically a strong coupling problem. The  BPS (extreme) black holes allow, by way of the non-renormalization theorems given by supersymmetry,  to address it.  
Moreover, the second string revolution, with the discovery of D-branes and the understanding of duality transformations, has led to a greater comprehension of the non-perturbative structure of string  theory and has thrown new light on black holes  and on many others questions. 

In this context supergravity plays an important role, again.
The string duality transformations show that the five different superstring theories in ten dimensions represent indeed  the patches at different regime of the same theory living in eleven dimensions, called M-theory. M-theory is much less understood than string theory, but one of the most important things that is known about it, is that its low-energy effective action is given  by   $d=11$ supergravity. 


$Dp$-branes are solitonic $p+1$-dimensional extended objects, which couple to Neveu Schwarz or to Ramond $p+1$-forms.
As a consequence of open/closed string duality they admit a double interpretation: microscopically the $p$--brane
  degrees of freedom are described by a suitable {\it gauge theory}
  $\mathcal{GT}_{p+1}$
  living on the  $p+1$ dimensional \emph{world volume} $\mathcal{WV}_{p+1}$
  that can be either conformal or not;
  macroscopically each $p$-brane is a generalized soliton in the
  following sense. It  is a classical solution
  of  $d=10$ or $d=11$ supergravity interpolating between two
  asymptotic geometries that with some abuse of language can respectively be 
  named   \emph{the geometry at infinity} $geo^\infty$ and \emph{the near 
  horizon geometry} $geo^H$.
  The latter, which only occasionally corresponds to a true event horizon,
  is  instead  universally characterized by the following property: it can be interpreted as a
  solution of some suitable $p+2$ dimensional supergravity $\mathcal{SG}_{p+2}$ times
  an appropriate \emph{internal space} $\Omega_{d-p-2}$.
In this spirit the quantum properties of black holes can be effectively described in terms of intersecting brane solutions \cite{Gauntlett:2001qs}.
  
Because of the statement above, all space--time dimensions
  $11 \geq d \geq 3 $ are relevant and
  supergravities in these diverse dimensions  describe
  various perturbative and non--perturbative aspects of superstring
  theory. In particular we have a most intriguing $gauge/gravity$
  correspondence implying that classical supergravity $\mathcal{SG}_{p+2}$ expanded around
  the vacuum solution $geo^H$ is dual to the quantum gauge theory $\mathcal{GT}_{p+1}$
  in one lower dimension.
The first remarkable example of this is offered by AdS/CFT correspondence of Maldacena \cite{maldapasto}.

There are two ways to study a supergravity theory:
by compactification from higher dimensions (in this case the
  scalar manifold is identified as the moduli space of the internal compact space), or  by direct construction of each supergravity theory in the
  chosen space--time dimension. In this case one uses all the \emph{a
  priori} constraints provided by supersymmetry,
  namely the field content of the various multiplets and the global and
  local symmetries that the action must have.
The first method makes direct contact with important aspects of
superstring theory but provides answers that are specific to the
chosen compact internal space $\Omega_{10-d}$ (and not fully general).
The second method gives instead fully general answers. Obviously the
specific answers obtained by compactification must fit into the
general scheme provided by the second method.

In this thesis I want to discuss the derivation of the BPS solutions of the most general ${\cal N}=2$, $d=5$ gauged supergravity with matter, first derived in \cite{ceresole:2000}. In particular I will concentrate on the dependence of BPS equations on the choice of gauging, in presence of non trivial hypermultiplets (not constant prepotential).

In the last few years five dimensional gauged supergravities have received a lot of attention mainly because of two  developments. On one hand we have the $AdS_5/CFT_4$
correspondence between
\begin{enumerate}
  \item [a] superconformal gauge theories in $d=4$, viewed as
the world volume description of a stack of $\mathrm{D3}$--branes
  \item [b] type IIB supergravity compactified on  $AdS_5$ times a five--dimensional
internal manifold $X^5$ which yields a gauged supergravity
model in $d=5$
\end{enumerate}
On the other hand we have the quest for supersymmetric realizations
of the Randall-Sundrum scenarios \cite{RS2},\cite{RS1} which also correspond to domain
wall solutions of appropriate $d=5$ gauged supergravities\footnote{Basic elements of these subjects are recalled in the appendix \ref{domain wall}.}. The ${\cal N}=2$ version furnishes the suitable set-up to address this last question and gravity/gauge theory correspondence in general.


Indeed, theories with 8 supercharges are specially interesting because this is the maximal supersymmetry that allows matter couplings with coupling functions that can continuously vary. The theory with 32 supercharges are completely fixed. There is only the supergravity multiplet that contains all other fields. The theory with 16 supercharges allow matter couplings. Here, there are discrete choices, essentially the number of multiplets that are included and possibly some signs related to compact or non-compact gauging. For 8 (and all lower numbers) of supercharges there are coupling functions that determine the manifold. These manifolds are not necessary symmetric spaces, and to determine the model, one has to specify certain functions. But the 8 supersymmetries impose a lot of structure on the couplings, and the remaining geometry leads to the concept of ``special geometry''.

An other aspect regarding the ${\cal N}=2$, $d=5$ gauged supergravity due to the presence of ``less supersymmetry'' is the following: if we consider the whole set of supergravity theories in diverse
dimensions we discover an important general property. With the
caveat of three noteworthy exceptions in all the other cases the
constraints imposed by supersymmetry imply that the scalar manifold
$\mathcal{M}_{scalar}$ is necessarily a homogeneous coset manifold
$ \mathcal{G}/\mathcal{H}$
of the  non--compact type, namely a suitable non compact Lie group
$ \mathcal{G}$ modded by the action of a maximal compact subgroup
$ \mathcal{H}\subset \mathcal{G}$.  By $\mathcal{M}_{scalar}$ we mean the
manifold parametrized by the scalar fields $\phi^I$ present in the theory. The
metric $g_{IJ}(\phi)$ defining the Riemannian structure of the scalar
manifold appears in the supergravity lagrangian through
the scalar kinetic term which is of the $\sigma$--model type:
\begin{equation*}
  \mathcal{L}_{scalar}^{kin} =\frac{1}{2} \,
g_{IJ}(\phi) \, \partial_\mu \phi^I \, \partial^\mu \phi^J
\end{equation*}
The three noteworthy exceptions where the scalar manifold is allowed to be
something more general than a coset $ \mathcal{G}/\mathcal{H}$ are
the following
\begin{enumerate}
  \item $\mathcal{N}=1$ supergravity in $d=4$ where
  $\mathcal{M}_{scalar}$ is simply required to be a \emph{Hodge K\"ahler
  manifold}.
  \item $\mathcal{N}=2$ supergravity in $d=4$ where
  $\mathcal{M}_{scalar}$ is simply required to be the product of a \emph{special K\"ahler
  manifold} $ \mathcal{SK}_n$
   containing the $n_V$ complex scalars of
  the $n_V$ vector multiplets with a \emph{quaternionic manifold}
  $\mathcal{QM}_{n_H}$ containing the $4n_H$ real scalars of the $n_H$
  hypermultiplets. \footnote{The notion of quaternionic geometry, as it enters
  the formulation of hypermultiplet coupling was introduced by
  Bagger and Witten in \cite{witten:1983} and formalized by Galicki in \cite{gal} who also explored the relation
  with the notion of HyperK\"ahler quotient, whose use in the construction of supersymmetric ${\cal N}=2$ lagrangians
  had already been emphasized in \cite{hklr}. The general problem of classifying quaternionic homogeneous spaces
  had been addressed in the mathematical literature by Alekseevski
  \cite{alex}.}
  \item $\mathcal{N}=2$ supergravity in $d=5$ where
  $\mathcal{M}_{scalar}$ is simply required to be the product of a \emph{very
  special manifold} $ \mathcal{VS}_n$ \footnote{The notion of very special geometry
  is essentially due to the work of G\"unaydin Sierra and Townsend who discovered it
  their work on coupling $d=5$ supergravity to vector multiplets
  \cite{GST1,GST2}. The notion was subsequently refined and properly related to special
  K\"ahler geometry in four dimensions through the work by de Wit and Van Proeyen
  \cite{deWit:1992nm,deWit:1993wf,deWit:1995tf}.} containing the $n_V$ real scalars of
  the $n_V$ vector multiplets with a \emph{quaternionic manifold}
  $\mathcal{QM}_{n_H}$ containing the $4n_H$ real scalars of the $n_H$
  hypermultiplets.
\end{enumerate}
However some families of homogeneous spaces exist and they are the ones usually considered in practical applications.\footnote{Recently non homogeneous manifold was used in the construction of the Randall-Sundrum scenario \cite{cardoso:2002}.}



In spite of the great number of works devoted to the study of BPS solutions
(especially concerning models with vector multiplets only \cite{renandrei},\cite{Behrndt:2000tr}), interesting classes of solutions like black holes still remain to be analyzed in full generality. In particular, the addition of the hypermultiplets enriches the structure of the scalar manifold and increases the number of viable gauging. As a consequence of promoting the global R--symmetry group $SU(2)$ to a local symmetry, new terms enter in the potential enlarging the variety of admissible solutions. This happens, for example, for domain wall configurations as clearly explained in \cite{ceresole:2001}. Although at least one hypermultiplet always appears in Calabi--Yau compactifications of M--theory \cite{9806051}, \cite{9506144} solutions of this type are the only ones already studied in presence of hypermultiplets \cite{Behrndt:2000kz},\cite{flat-domainwall},\cite{cardoso:2002},\cite{berhndt:2002},\cite{curved-domainwall} apart from \cite{gutperle:2001}, \cite{noialtri}.

My aim is to partially cover this lack, trying to extract from the analysis of  special interesting cases (see chapter \ref{black} and \ref{vet})
some feature that hold also in general with the hope to obtain 
 a ``dictionary'' of all possible  solutions.

 Very recently a great improvement to this program was given by Gauntlett et al.  in \cite{Gauntlett:2002nw} where they introduced a powerful method based on the group structure that characterizes supergravity models and that is applicable in different dimensions. Until now, it has been applied to the ungauged five and six dimensional ${\cal N}=2$ theory in \cite{Gauntlett:2002nw} and \cite{0306235}, to the eleven dimensional supergravity \cite{0212008} and to the minimal gauged ones with eight supercharges in $d=4$ and $5$ respectively in \cite{0307022} and \cite{0304064}.

My purpose, explicitly realized in chapter \ref{gen}, is to extend the results of \cite{0304064} to the theory with a generic number of hypermultiplets using together the informations deduced on the group structure of the base space and the analysis of hyperini variation\footnote{In the following I refer to it simple as ``hyperini equation''.} under supersymmetry.

Someone could express some doubts on the utility of classifying BPS solutions of gauged supergravity with matter in lower dimensions (than 11) arguing that all configurations of interest for string/M-theory are coded by the eleven dimensional supergravity hence it would seem a problem already solved in \cite{0212008}. This impression is not true, in fact the same authors in \cite{0212008} write ``{\itshape the classification we are advocating would still leave the very challenging task of determining all of the explicit supersymmetric solutions that can arise within the classes we will discuss. For example, in special case when the flux is zero, this corresponds to explicitly classifying all special holonomy manifolds, which seems to require fundamentally new mathematical ideas in order  to make progress}''. Moreover no explicit knowledge of all possible compactifications down to a lower dimension exists.

So I think that it is not a mere exercise to address the problem for case ${\cal N}=2$ in $d=5$ in presence of matter multiplets.
As I have already argued this case has a lot of possible applications and
the comparison with eleven dimensional results could be fruitful to elucidate better the mathematical structure and put in evidence new features of low-energy limit of string/M theory.
Furthermore this study could favor a supersymmetric realization of phenomenological models like the Randall-Sundrum scenario.

The plan of the thesis is the following: in the next chapter the basic elements of the theory are reviewed,  putting the accent on its geometric properties that are impressed on the scalar manifold. Starting from the ungauged theory it is reported how the gauging procedure works in  this case  and which effects it has on the susy transformations and on  the scalar potential. The role of the \emph{moment map} is emphasized.  The chapter ends with the presentation of the complete lagrangian and the supersymmetry variation of the fermions that are necessary to study BPS solutions. This concludes the introductory part of the thesis. 

The chapter \ref{black} is devoted to the study of electrostatic spherical symmetric solutions with an arbitrary number of hypermultiplets and it is mainly based on my article \cite{noialtri} with S. Cacciatori and my tutor D. Zanon. This particular class of configurations is relevant because it contains the Reissern--Nordstr\"om black hole. It is shown how, starting from a generic ansatz of such type, it is possible to derive (first order) BPS equations using the properties of the quaternionic geometry without specifying the scalar manifold\footnote{This  is important because our knowledge of the metric of the quaternionic manifolds is limited to few cases also for coset space.} and explicitly choosing the gauge isometry, although the presence of a non zero graviphoton complicates  the calculations a lot, with respect to the uncharged case. The form of the BPS equation for scalars is discussed and compared with the one obtained for a flat domain wall in \cite{ceresole:2001}: the similarity suggests the existence of a general structure  characterizing the hyperini equation which will be completely explained in the chapter \ref{gen}.  
The above study is extended in chapter \ref{vet} to include vector multiplets, too \cite{vet:to be published}. The introduction of these matter fields is important not only because it allows us to consider gauge groups greater that $U(1)$,  but also it is necessary to have solutions with two fixed points (like AdS black holes). The abelian case i.e. with gauge group $U(1)^{n_V+1}$ where $n_V$ is the number of vector multiplet is analyzed.
With the chapter \ref{gen} we enter in the most ambitious part of thesis. Here the problem of classifying BPS solutions in presence of a generic number of hypermultiplets is addressed. Two ingredients are fundamental in this effort: rewriting the BPS constraints on the killing spinor in term of bosonic quantities as first introduced in \cite{Gauntlett:2002nw} and the analysis of hyperini equation. It is remarkable that this last result holds also in four dimensions  the hyperini equation being identical; it should be possible without great trouble to extend the classification in $d=4$ too.
The configurations with a time--like killing spinor \footnote{This is a short way to say that the vector $V_\mu=1/2 {\bar \epsilon}^i\gamma_\mu\epsilon_i$ constructed with the killing spinor is time--like.} are explicitly treated \cite{gen:to be published}. 

I want to stress that the contributions presented in this work do not constitute the ultimate word on the subject, but are in fact a starting point:
for example it remains to consider light--like killing spinors and to include vector multiplets. These and other remarks are pointed out in the conclusion. For the notations, which are in large part the same as \cite{noialtri}, I refer the reader to the appendix. 


\chapter{General elements of the theory}\label{teo} 

In this chapter I review the main  features of gauged supergravity paying attention to the geometric structure of the theory that is a consequence of extended supersymmetry. After a general discussion on the subject, I will specialize to the case ${\cal N}=2$ in five dimensions giving the explicit Lagrangian and the quantities  of interest for the second part of the work. The aim here is not to give the exact procedure to construct a supergravity theory, a subject which would need more than one book for each one of the different methods existing.We only want  to emphasize which are the basic considerations an explicit construction starts from, skipping technical details. 

\section{Introduction}

Roughly speaking Supergravity is the gauged theory of supersymmetry, that is it describes systems which are left invariant by the action  of local supersymmetric transformations on space--time.
As a starting point we consider the (rigid) supersymmetry algebra: taking as example the ${\cal N}=1$ case in four dimensions it is of the form
\begin{eqnarray}
[ J_{\mu\nu}, J_{\rho\sigma} ] &=& 2\eta_{\rho[\nu}\delta^{\tau\lambda}_{\mu]\sigma}J_{\tau\lambda}
\nn\\
\left[ J_{\mu\nu}, p_\rho \right] &=& \eta_{\rho[\nu}p_{\mu]\nn}\\
\left[ J_{\mu\nu},  Q \right] &=& \frac 12 \gamma_{\mu\nu} Q\nn\\
\left[ p_\mu,p_\nu \right]&=& 0 \nn\\
\left[ p_\mu,  Q \right] &=&0\nn\\
\left\{ Q , \bar Q \right\}&=&  {\rm i}\mathcal{C}\gamma^\mu\,p_\mu
\label{susy}
\end{eqnarray}
where $J_{\mu\nu}, p_\mu$ are the generators of the Poincar\'e algebra ($\mu=0,1,2,3$);
 $\gamma_\mu$ are a set of Dirac matrices satisfying the Clifford algebra
$$\{ \gamma_\mu , \gamma_\nu \} = 2 \eta_{\mu\nu};$$ $\gamma_{\mu\nu} \equiv \half \bigl[\gamma_\mu , \gamma_\nu \bigr]$; $\mathcal{C}$ is the charge conjugation matrix satisfying
$$ \mathcal{C}\gamma_\mu \mathcal{C}^{-1} = - \gamma_\mu^\dagger;$$
finally, the supercharge $Q$  is a Majorana spinor, satisfying:
$$\bar Q \equiv Q^\dagger \gamma_0 = Q^t \mathcal{C}.$$
The supersymmetry generators $Q$ act by turning bosonic degrees
of freedom to fermionic ones, and viceversa.
Note that, due to the presence of anticommutators, what (\ref{susy}) defines is not an algebra (as the Coleman--Mandula theorem states \cite{coma}) but an its generalization called superalgebra.
As appears from the above relations considering supersymmetry, carried by the supercharge $Q$, as a local
symmetry, implies considering also the translations $p_\mu$ as generators of local transformations, that is it involves {\em general covariance}.
This is the reason for the name {\em supergravity}; it conciliates supersymmetry with general relativity.

From this consideration it is also clear that the construction of supergravity as a gauged theory of the SuperPoincar\`e group is not as straightforward as for general relativity. Indeed not all the ``gauge fields'' are independent, e.g. the spin connection $w_\mu^{ab}$ is a function of $e^a_\mu$.
 Moreover general coordinate transformations should appear rather than local transformations. This means for examples that diffeomorphisms should replace local translations.
In general relativity this is realized by the introduction of the constraint $De^a=0$, implying that the metric $g_{\mu\nu}= e^a_\mu e_{\nu a} $ is torsionless. 
In the supergravity context this acquires the name of ``conventional'' constraint or supertorsion condition. Its exact form depends explicitly on the particular supergravity theory considered and the way to achieve it changes for the different methods.

However there are some consequences of constraints that are general and characterize each supergravity theory. First of all, together with the  Bianchi identities, they impose the introduction  of new terms in the curvatures that are necessary to the closure of the algebra. This can be summarized saying that \emph{ the algebra becomes soft} in the sense that its structure constants become structure functions. Furthermore the gravitini variation under supersymmetry takes the form of a covariant derivative operator acting on the local parameter of the transformation $\epsilon$
\eqn
\delta_\epsilon \psi = {\cal D} \epsilon
\feqn 
For the simple supergravity theory we have chosen as example ${\cal D}$ reduces to the ordinary covariant derivative of general relativity. If we consider extended supergravities and/or we introduce also matter, the situation changes. ${\cal D}$ will take additional contributions from the other fields of the theory.  

As already mentioned before, the concrete construction of a supergravity theory is a hard task that requires the use of sophisticated techniques but always the starting point is given by the application of these general ideas on the field of the theory, which are organized in multiplets of rigid supersymmetry.
The choice of these multiplets (weyl+matter) determines the degrees of freedom  present in the theory.

In this chapter I will not discuss how to construct the ${\cal N}=2$, $d=5$ supergravity\footnote{I refer the reader interested to these aspects to the book \cite{castdauriafre} and to great number of good reviews and lectures like \cite{review_sugra}. 
For the construction of theory with eight supercharges in four, five and six dimensions in the superconformal tensor calculus approach we suggest \cite{vp_lessons}. 
}
but I will try to motivate (without an explicit demonstration) why this theory is how it is. In particular I will focus my attention on the surprisingly sophisticated geometric structure that characterizes the theory, first to stand on its feet at the ungauged level and, secondly, to be gauged producing non abelian symmetries and the scalar potential.
Obviously all such structures are imposed on the theory by supersymmetry and the presence of the
fermions. Yet since the fermions are ugly objects to deal with
while their product, namely the geometric structure of the theory is
beautiful, I will only stick to the latter and mention the fermions as
seldom as possible. Indeed, because the subject of this thesis is the study of BPS solutions, in which the fermions are fixed to be zero, they enter only in the determination of the bosonic lagrangian and the BPS conditions.
So let me start presenting the ``bestiary'' which characterizes the five dimensional supergravities.

\section{Supergravities in five dimension: an overview}
\label{d5}

I have already mentioned in the introductory chapter of the reasons for recent renewed interest in five--dimensional gauged supergravities.
However this theory has a long and
interesting history. The minimal theory (${\cal N}=2$ ), whose field content is given
by the metric $g_{\mu \nu }$, a doublet of pseudo Majorana gravitinos
$\psi_{i\mu}$ ($i=1,2$) and a vector boson $A_\mu$ was constructed
twenty years ago \cite{D'Auria:1981kq}  as the first non--trivial
example of a rheonomic construction\footnote{We leave aside pure
${\cal N}=1,d=4$ supergravity that from the rheonomic viewpoint
is a completely trivial case.}. This simple model remains to the
present day the unique example of a perfectly geometric theory where,
notwithstanding the presence of a gauge boson $A_\mu$, the action
can be written solely in terms of differential forms and wedge
products without introducing Hodge duals. This feature puts pure
$d=5$ supergravity into a selective club of few  ideal theories whose
other members are just pure gravity in arbitrary dimension and pure
${\cal N}=1$ supergravity in four dimensions. The miracle that
allows the boson $A_\mu$ to propagate without introducing its kinetic
term is due to the conspiracy of the first order formalism for the spin
connection $\omega^{ab}$ together with the presence of two
Chern--Simons terms. The first Chern Simons term is the standard gauge
one:
\begin{equation}
   {CS}_{gauge} = F\wedge F\wedge A
\label{purchsi}
\end{equation}
while the second  is a mixed, gravitational-gauge  Chern Simons that
reads as follows
\begin{equation}
  {CS}_{mixed} = T^a \wedge F\wedge V_a
\label{mixchsi}
\end{equation}
where $V^a$ is the vielbein and $T^a=\mathcal{D}V^a$ is its
\emph{curvature}, namely the torsion.
\par
The possible matter multiplets for ${\cal N}=2,d=5$ are the
\textbf{vector/tensor} multiplets and the \textbf{hypermultiplets}.
The field content of the first type of multiplets is the following
one:
\begin{equation}
    \left\{ \begin{array}{cccll}
    A^{\bar{I}}_\mu & \null & \null & (\bar{I}=1,\dots,n_V) & \mbox{vectors} \\
    \null & \lambda^{\tilde x}_i & \phi^{\tilde x} & (\tilde x=1,\dots,n_V+n_T \equiv n) & (i=1,2) \\
    B^M_{\mu \nu } & \null & \null &(M=1,\dots,n_T)&  \mbox{tensors} \
  \end{array} \right\}
\label{tensvect}
\end{equation}
where  by $n_V$ I have denoted the number of  vectors or gauge
$1$--forms $A^{\bar I}_\mu$,  $n_T$ being instead the number of tensors
or gauge $2$--forms $B^M_{\mu \nu }=-B^M_{\nu \mu}$. In ungauged supergravity, where everything is abelian, vectors and
tensors are equivalent since they can be dualised into each other by
means of the transformation:
\begin{equation}
  \partial _{[\mu} \, A _{\nu ]}=\epsilon _{\mu \nu }^{\phantom{\mu \nu
  }\lambda \rho \sigma } \, \partial _\lambda B_{\rho \sigma }
\label{dual2to3}
\end{equation}
but in gauged supergravity it is only the $1$--forms that can be
promoted to non--abelian gauge vectors, while the $2$--forms  describe
massive degrees of freedom. The other members of each vector/tensor
multiplet are a doublet of pseudo Majorana spin 1/2 fields:
\begin{equation}
\lambda^{\tilde x}_i=\epsilon^{ij} \, \mathcal{C} \, \left(
\overline{\lambda}^{j\tilde x}\right) ^T \, \quad ; \quad
\overline{\lambda}^{i\tilde x}= \left( \lambda^{\tilde x}_i\right) ^\dagger
\gamma_0 \quad ; \quad
 ~~~i,j=1,\dots,2\,.
\end{equation}
and a real scalar $\phi^{\tilde x}$.
The field content of hypermultiplets is the following:
\begin{equation}
  \mbox{hypermultiplets}=\left \{ q^X \,(X=1,\dots,4 \,n_H )\, ,
  \zeta^A \, (A = 1, \dots \, 2n_H) \right \}
\label{hypmulcont}
\end{equation}
where, having denoted $n_H$ the number of hypermultiplets, $q^X$ are $n_H$
quadruplets of real scalar fields and $\zeta^A$ are $n_H$ doublets
of pseudo Majorana spin 1/2 fields:
\begin{equation}
\zeta^A=\mathbb{C}^{A B } \, \mathcal{C} \, \left(
\overline{\zeta}_B\right) ^T \, \quad ; \quad
\overline{\zeta}_B = \left( \zeta^B \right) ^\dagger
\gamma_0 \quad ; \quad
 ~~~A ,B =1,\dots,2\,n_H
\end{equation}
the matrix $\mathbb{C}^T=-\mathbb{C}$, $\mathbb{C}^2 =-{\bf 1}$ being
the symplectic invariant metric of $\mathrm{Sp}(2m,\mathbb{R})$.
\par
In the middle of the eighties, Gunaydin, Sierra and Townsend \cite{GST1,GST2} considered
the general structure of $ {\cal N}=2,d=5$ supergravity coupled to an
arbitrary number $n=n_V+n_T$ of  vector/tensor multiplets. They discovered
that this is dictated by a peculiar geometric structure imposed by
supersymmetry on the scalar manifold $\mathcal{SV}_n$ that contains
the $\phi^{\tilde x}$ scalars. In  modern nomenclature this
peculiar geometry  is named \textbf{very special geometry} and $\mathcal{SV}_n$
are referred to as real \textbf{very special manifolds}.  The
characterizing property of very special geometry arises from the need
to reconcile the transformations of the scalar members of each
multiplet with those of the vectors in presence of the Chern-Simons
term (\ref{purchsi}), which generalizes to:
\begin{equation}
  \mathcal{L}^{CS} = \frac{1}{6\sqrt 6} \, C_{\tilde I \tilde J \tilde K}
  F^{\tilde I}_{\mu \nu } \, F^{\tilde J}_{\rho \sigma } \, A^{\tilde K}_\tau \, \epsilon ^{\mu \nu \rho \sigma \tau }
\label{Csmult}
\end{equation}
the symbol $ C_{\tilde I \tilde J \tilde K}$ denoting some appropriate
constant symmetric tensor and, having dualised all $2$--forms to
vectors, the range of the indices $\tilde I$, $\tilde J$, $\tilde K$ being:
\begin{equation}
  \tilde I= 1,\dots,n+1 \, = \left \{\underbrace{\, 0\, , \, \bar I }_I \,,
   , \, M\right\}
\label{Lamrang}
\end{equation}
Indeed, the total number of vector fields, including the graviphoton
that belongs to the graviton multiplet, is always $n_V+1$, $n_V$ being
the number of vector multiplets. It turns out that  very special geometry
is completely defined in terms of the constant tensors $C_{\tilde I \tilde J \tilde K} $,
that are further restricted by a condition ensuring positivity of
the energy. At the beginning of the nineties special manifolds were
classified and thoroughly studied by de Wit, Van Proeyen and some
other collaborators \cite{deWit:1992nm,deWit:1993wf,deWit:1995tf}.
They also explored the dimensional reduction from $d=5$ to $d=4$,
clarifying the way \emph{real very special geometry} is mapped into the
\emph{special K\"ahler geometry} featured by vector multiplets in $d=4$ and
generically related to Calabi--Yau moduli spaces.
\par
The $4n_H$ scalars  of the hypermultiplet sector have instead exactly
the same geometry in $d=4$ as in $d=5$ dimensions, namely they fill
a quaternionic manifold $\mathcal{QM}$.  These scalar geometries are
a crucial ingredient in the construction of both the ungauged and the
gauged supergravity lagrangians. Indeed, the  basic operations
involved by the gauging procedure are based on the specific
geometric structures of very special and quaternionic manifolds, in
particular the crucial existence of a moment--map (see sect.\ref{momentm1}).
For this reason the present section is devoted to a summary of these
geometries and to an illustration of the general form of the bosonic
$d=5$ lagrangians.

Independently from the number of supersymmetries we can write a
general form for the bosonic action of any $d=5$ \emph{ungauged
supergravity}, namely the following one:
\begin{eqnarray}
  \mathcal{L}^{(ungauged)}_{(d=5)}&=&\sqrt{-g} \, \left( \frac{1}{2}
  \,R \, - \, \frac{1}{4}\,a_{\tilde I\tilde J} F^{\tilde I}_{\mu \nu
  } \, F^{\tilde J \vert\mu \nu} + \frac{1}{2} \, g_{\Lambda \Sigma} \, \partial_\mu\varphi^{\Lambda} \, \partial ^\mu \, \varphi^{\Sigma} \right ) \nonumber\\
  &&+ \frac{1}{6\sqrt 6} C_{\tilde I \tilde J \tilde K} \, \epsilon ^{\mu \nu \rho \sigma \tau  } \,F^{\tilde I}_{\mu \nu } \, F^{\tilde J}_{\rho \sigma } \, A^{\tilde K}_\tau \, 
\label{genford5bos}
\end{eqnarray}
where $ g_{\Lambda \Sigma}$ is the metric of the scalar manifold $\mathcal{M}_{scalar}$ , $a_{\tilde I\tilde J}(\varphi)$ is a positive definite symmetric function of the
scalars that under the isometry group $\mathcal{G}_{iso}$ of $\mathcal{M}_{scalar}$ transforms in $\bigotimes^2_{sym} \mathbf{R}$, having denoted by $ \mathbf{R}$ a linear representation of $\mathcal{G}_{iso}$ to which the vector fields $A^\Gamma$ are assigned. Finally $C_{\tilde I \tilde J \tilde K}$ is a three--index symmetric tensor invariant with respect to the representation $\mathbf{R}$.


To see how the same items are realized in the case of an ${\cal N}=2$ theory
we have to introduce very special and quaternionic geometry. Before we elaborate this, it is worth nothing that also the supersymmetry
transformation rule of the gravitino field admits a general form
(once restricted to the purely bosonic terms), namely:
\begin{equation}
  \delta \psi_{i\mu} = \mathcal{D}_\mu \, \epsilon_i + i \frac{1}{4\sqrt 6} \,
  \mathcal{T}_{ij}^{\rho \sigma } \left( 4 g_{\mu \rho } \, \gamma _\sigma
  - \, \gamma_{\mu\rho \sigma }   \right) \, \epsilon ^j
\label{gensusrul}
\end{equation}
where the indices $i,j$ run in the fundamental representation of the
automorphism (R-symmetry) group $\mathrm{USp}({\cal N})$ and the tensor
$\mathcal{T}_{ij}^{\rho \sigma }$, antisymmetric both in $ij$ and in
$\rho \sigma $ and named the graviphoton field strength, is given by:
\begin{equation}
  \mathcal{T}_{ij}^{\rho \sigma } = \Phi^{\tilde I}_{ij}(\phi) \,
a_{\tilde I\tilde J}  F^{\tilde J\vert \rho \sigma }
\label{gravfotft}
\end{equation}
the scalar field dependent tensor $\Phi^{\tilde I}_{ij}(\phi)$ being
intrinsically defined as the coefficient of the term ${\bar \epsilon
}^i \, \psi^j_\mu$ in the supersymmetry transformation rule of the
vector field $A^{\tilde I}_\mu$, namely:
\begin{equation}
  \delta A^{\tilde I}_\mu = \dots - \, \frac {\sqrt 6}4 \mbox{i}\,
  \Phi^{\tilde I}_{ij}(\phi) \, {\bar \psi^i_\mu} \, \epsilon^j 
\label{CapFi}
\end{equation}
From its own definition it follows that under isometries of the scalar manifold
$\Phi^{\tilde I}_{ij}(\phi) $ must transform in the representation
$\mathbf{R}$ of $\mathcal{G}_{iso}$ times $\bigwedge^2 {\cal N}$
of the R-symmetry $\mathrm{USp}({\cal N})$. 
We see in the next subsection how this is generally
realized in an ${\cal N}=2$ theory via very special geometry.
\subsection[Very special geometry]{Very special geometry}\label{very}
\emph{Very special geometry} is the peculiar metric and associated
Riemannian  structure that can be constructed on a very special
manifold. By definition a \emph{very special manifold}
$\mathcal{VS}_n$ is a real manifold of dimension $n$ that can be
represented as the following algebraic locus in $\mathbb{R}^{n+1}$:
\begin{equation}
  1= \mathrm{N}(h) \, \equiv \, \left(\, C_{\tilde I \tilde J \tilde K} \,
  h^{\tilde I} \, h^{\tilde J} \, h^{\tilde K}\right)^{\frac 13} 
\label{defispec}
\end{equation}
where $h^{\tilde I}$ ($\tilde I=1,\dots,n+1$) are the coordinates of $\mathbb{R}^{n+1}$
while
\begin{equation}
C_{\tilde I \tilde J \tilde K}
\label{symtens}
\end{equation}
is a \textbf{constant symmetric tensor} fulfilling some additional
defining properties that I will recall later on.\\
A coordinate system $\phi^{\tilde x}$ on $\mathcal{VS}_n$ is provided by any
parametric solution of eq. (\ref{defispec}) such that:
\begin{equation}
  h^{\tilde I}  = h^{\tilde I} (\phi) \quad ; \quad \phi^{\tilde x} = \mbox{free} \quad ; \quad \tilde x=1,\dots , n
\label{coordspec}
\end{equation}
The \emph{very special metric} on the very special manifold is
nothing else but the pull--back
on the algebraic surface (\ref{defispec}) of the following $\mathbb{R}^{n+1}$ metric:
\begin{eqnarray}
  ds^2_{\mathbb{R}^{n+1}}&=&
  a_{\tilde I \tilde J} \, dh^{\tilde I}  \otimes dh^{\tilde J} 
  \label{rn+1met}\\
a_{\tilde I \tilde J}  & \equiv & - \partial _{\tilde I} \partial _{\tilde J} \, \ln \,
\mathrm{N}(X)
\label{scriptNsp}
\end{eqnarray}
In other words, in any coordinate frame the coefficients of the very
special metric are the following ones:
\begin{equation}
  g_{\tilde x \tilde y}(\phi) = a_{\tilde I \tilde J} \, h_{\tilde x}^{\tilde I} \,  h_{\tilde y}^{\tilde J}
\label{specmet}
\end{equation}
where we have introduced the new objects:
\begin{equation}
   h_{\tilde x}^{\tilde I} \, \equiv \, - \sqrt{\frac 32} \partial_{\tilde x} h^{\tilde I} = - \sqrt{\frac 32}\frac{\partial }{\partial
  \phi^{\tilde x}} h^{\tilde I} 
\label{fidefi}
\end{equation}
If we also define
\begin{equation}
  h_{\tilde I}= \frac{\partial }{\partial h^{\tilde I}} \, \ln \,
  \mathrm{N(X)} \quad ; \quad h_{\tilde I \tilde x}\equiv \, \sqrt{\frac 32} \partial _{\tilde x}h_{\tilde I} 
\label{Flam}
\end{equation}
and introduce the $2(n+1)$-vectors:
\begin{equation}
  U = \left( \begin{array}{c}
    h^{\tilde I} \\
    h_{\tilde I} \
  \end{array} \right)  \quad; \quad U_{\tilde x} = \partial_{\tilde x} U = \sqrt{\frac 23} \left( \begin{array}{c}
    -h^{\tilde I}_{\tilde x} \\
    h_{\tilde I\tilde x} \
  \end{array} \right)
\label{UiUvec}
\end{equation}
 it can be shown that taking a second covariant derivative the following identity is true:
\begin{equation}
  \nabla_{\tilde x} U_{\tilde y} = \frac{2}{3} \, g_{\tilde x\tilde y} U + \sqrt{\frac{2}{3}} \,
  T_{\tilde x\tilde y\tilde z} \, g^{\tilde z \tilde w} \, U_{\tilde w}
\label{dUide}
\end{equation}
where the world--index symmetric coordinate dependent tensor
$T_{\tilde x\tilde y\tilde z}$ is related to the constant tensor $C_{\tilde I \tilde J \tilde K}$
by:
\begin{equation}
  C_{\tilde I \tilde J \tilde K}=\frac 52  \, h_{\tilde I} \, h_{\tilde J}
  \,h_{\tilde K} \, -\, \frac 32 \, a_{(\tilde I \tilde J} \,
  h_{\tilde K)} + T_{\tilde x\tilde y\tilde z} \, 
  h_{\tilde I}^{\tilde x} \, h_{\tilde J}^{\tilde y} \,h_{\tilde K}^{\tilde z}
\label{Tidedide}
\end{equation}
The identity (\ref{dUide}) is the real counterpart of a completely
similar identity that holds true in special K\"ahler geometry and
also defines a symmetric $3$--index tensor.  In the use  of very
special geometry to construct a supersymmetric field theory the
essential property is the existence of the \emph{section}
$ h^{\tilde I}(\phi)$. Indeed it is this object that allows the
writing of the tensor $\Phi^\Lambda_{AB}(\phi)$ appearing in the
vector transformation rule (\ref{CapFi}). It is sufficient to set:
\begin{equation}
  \Phi^{\tilde I}_{ij}(\phi) =  h^{\tilde I}(\phi) \, \epsilon _{ij}
\label{Phiforn2}
\end{equation}
Why do we call it a section? Since it is just a section of a
\textbf{flat vector bundle} of rank $n+1$
\begin{equation}
  \mathrm{FB} \stackrel{\pi}{\rightarrow} \mathcal{SV}_n
\label{flatbundvsg}
\end{equation}
with base manifold the very special manifold and structural group
some subgroup of the $n+1$ dimensional linear group:
$\mathcal{G}_{iso}\subset \mathrm{GL}(n+1,\mathbb{R})$. The bundle is flat because
the transition functions from one local trivialization to another 
are constant matrices:
\begin{equation}
  \forall g \in \mathcal{G}_{iso} \quad : \quad  h^{\tilde I} (g \, \phi) = \left(
  M[g]\right )^{\tilde I}_{\phantom{\tilde I}\tilde J} \, h^{\tilde J} (\phi) \quad ;
  \quad M[g]=\mbox{constant matrix}
\label{urbano}
\end{equation}
The structural group $\mathcal{G}_{iso}$ is implicitly defined as the set of
matrices $M$ that leave the $C_{\tilde I \tilde J \tilde K}$ tensor
invariant:
\begin{equation}
M\in \mathcal{G}_{iso} \quad \Leftrightarrow \quad  M_{{\tilde I}_1}^{\phantom{\tilde I}{\tilde J}_1} \,
  M_{{\tilde I}_2}^{\phantom{\tilde I}{\tilde J}_2}\, M_{{\tilde I}_3}^{\phantom{\tilde I}{\tilde J}_3}
  \, C_{{\tilde I}_1 {\tilde I}_2 {\tilde I}_3 } \,=\, C_{{\tilde J}_1 {\tilde J}_2 {\tilde J}_3 }
\label{defiGs}
\end{equation}
Since the very special metric is defined by eq. (\ref{specmet}), it immediately
follows that $\mathcal{G}_{iso}$ is a subgroup of  the isometry group of  such a metric,\footnote{All the non linearly realized isometries of metric are not into  $ \mathcal{G}_{iso}$: explicit examples of spaces with this type of isometries exist.}  
 its action in any coordinate patch (\ref{coordspec}) being defined by the
action (\ref{urbano}) on the section $h^{\tilde I}$. This fact explains
the name given to this group.
\par
By this reasoning I have shown that the classification of
very special manifolds is partially reduced to the classification of the
constant tensors $C_{\tilde I \tilde J \tilde K}$ such that their group of invariances
acts transitively on the manifold $\mathcal{SV}_n$ defined by  eq. 
(\ref{defispec}) and that the special metric (\ref{specmet}) is positive definite.
This is the algebraic problem that was completely solved by de Wit
and Van Proeyen in \cite{deWit:1992nm}. They found all such tensors
and the corresponding manifolds. There is a large subclass of very
special manifolds that are homogeneous spaces but there are also infinite families
of manifolds that are not $\mathcal{G}/\mathcal{H}$ cosets.
\subsection{Quaternionic Geometry}
\label{hypgeosec}
Next I turn my attention to the hypermultiplet sector of an
${\cal N}=2$ supergravity. For these multiplets no distinction arises
between $d=4$ and $d=5$. Each hypermultiplet contains $4$ real scalar fields
and, at least locally, they can be regarded as the
four components of a quaternion. The locality caveat is, in this
case, very substantial because global quaternionic coordinates can be
constructed only occasionally, even on those manifolds that are
denominated quaternionic in the mathematical literature
\cite{alex}, \cite{gal}. Anyhow, what is important  is that, in
the hypermultiplet sector, the dimension of the scalar manifold $\mathcal{QM}$ is a multiple of four:
\begin{equation}
\mbox{dim}_{\bf R} \, \mathcal{QM} \, = \, 4 \, n_H \,\equiv \, 4 \, \# \,
\mbox{of hypermultiplets}
\label{quatdim}
\end{equation}
and, in some appropriate sense, it has a quaternionic structure.
\par
The geometry of the
hypermultiplet sector is called{\it Hypergeometry}, irrespectively whether we deal with global or
local ${\cal N}=2$ theories. Yet there are two kinds of hypergeometries.
Supersymmetry requires the existence
of a principal $\mathrm{SU}(2)$--bundle
\begin{equation}
{\cal SU} \, \longrightarrow \, \mathcal{QM} \label{su2bundle}
\end{equation}
The bundle ${\cal SU}$ is
{\bf flat} in the {\it rigid supersymmetry case} while its curvature is
proportional to the K\"ahler forms in the {\it local case}.
\par
These two versions of hypergeometry were already known in mathematics prior to
their use \cite{dWLVP}, \cite{skgsugra_13}, \cite{D'Auria:1991fj},
\cite{sabarwhal}, \cite{vanderseypen}  in the context of ${\cal N}=2$
supersymmetry and are identified as:
\begin{eqnarray}
\mbox{rigid hypergeometry} & \equiv & \mbox{HyperK\"ahler geometry.}
\nonumber\\ \mbox{local hypergeometry} & \equiv & \mbox{quaternionic
geometry} \label{picchio}
\end{eqnarray}
\subsubsection{Quaternionic versus HyperK\"ahler manifolds}
Both a quaternionic and a HyperK\"ahler manifold $\mathcal{QM}$
are a $4 n_H$-dimensional real manifold endowed with a metric $g$:
\begin{equation}
d s^2 = g_{X Y} (q) d q^X \otimes d q^Y   \quad ; \quad X,Y=1,\dots,
4  n_H \label{qmetrica}
\end{equation}
and three complex structures
\begin{equation}
(J^r) \,:~~ T(\mathcal{QM}) \, \longrightarrow \, T(\mathcal{QM}) \qquad
\quad (r=1,2,3)
\end{equation}
that satisfy the quaternionic algebra
\begin{equation}
J^r J^s = - \delta^{rs} \, \bfone \,  +  \, {\epsilon^{rs}}_t J^t
\label{quatalgebra}
\end{equation}
and with respect to which the metric is hermitian:
\begin{equation}
\forall   \mbox{\bf X} ,\mbox{\bf Y}  \in   T\mathcal{QM}   \,: \quad g
\left( J^r \mbox{\bf X}, J^r \mbox{\bf Y} \right )   = g \left(
\mbox{\bf X}, \mbox{\bf Y} \right ) \quad \quad
  (r=1,2,3)
\label{hermit}
\end{equation}
From eq.  (\ref{hermit}) it follows that one can introduce a triplet
of 2-forms
\begin{equation}
\begin{array}{ccccccc}
K^r& = &K^r_{X Y} d q^X \wedge d q^Y & ; & K^r_{XY} &=&   g_{XZ}
(J^r)^Z_Y \cr
\end{array}
\label{iperforme}
\end{equation}
that provides the generalization of the concept of K\"ahler form
occurring in  the complex case. The triplet $K^r$ is named the {\it
HyperK\"ahler} form. It is an $\mathrm{SU}(2)$ Lie--algebra valued
2--form  in the same way as the K\"ahler form is a $U(1)$
Lie--algebra valued 2--form. In the complex case the definition of a
K\"ahler manifold involves the statement that the K\"ahler 2--form is
closed. At the same time in Hodge--K\"ahler manifolds (those
appropriate to local supersymmetry in $d=4$) the K\"ahler 2--form can be
identified with the curvature of a line--bundle which in the case of
rigid supersymmetry is flat. Similar steps can be taken also here and
lead to two possibilities: either HyperK\"ahler or quaternionic
manifolds.
\par
Let us  introduce a principal $\mathrm{SU}(2)$--bundle ${\cal SU}$ as
defined in eq.  (\ref{su2bundle}). Let $p^r$ denote a connection
on such a bundle. To obtain either a HyperK\"ahler or a quaternionic
manifold we must impose the condition that the HyperK\"ahler 2--form
is covariantly closed with respect to the connection $p^r$:
\begin{equation}
\nabla K^r \equiv d K^r + 2 {\epsilon^r}_{s t} p^s \wedge K^t    \,
= \, 0 \label{closkform}
\end{equation}
The only difference between the two kinds of geometries resides in
the structure of the ${\cal SU}$--bundle.
\begin{definizione} A
HyperK\"ahler manifold is a $4 n_H$--dimensional manifold with the
structure described above and such that the ${\cal SU}$--bundle is
{\bf flat}
\end{definizione}
 Defining the ${\cal SU}$--curvature by:
\begin{equation}
R^r \, \equiv \, d p^r + 2 {\epsilon^r}_{st} p^s \wedge p^t \label{su2curv}
\end{equation}
in the HyperK\"ahler case we have:
\begin{equation}
R^r \, = \, 0 \label{piattello}
\end{equation}
Viceversa,
 \begin{definizione} A quaternionic manifold is a 
$4 n_H$--dimensional manifold with the structure described above and such
that the curvature of the ${\cal SU}$--bundle is proportional to the
HyperK\"ahler 2--form \end{definizione} Hence, in the quaternionic
case we can write:
\begin{equation}
R^r \, = \, { {\lambda}}\, K^x \label{piegatello}
\end{equation}
where $\lambda$ is a non vanishing real number.\footnote{In my notation $\lambda$ is equal to $\frac 12$.}
\par
As a consequence of the above structure the manifold $\mathcal{QM}$ has
a holonomy group of the following type:
\begin{eqnarray}
{\rm Hol}(\mathcal{QM})&=& \mathrm{SU}(2)\otimes {\cal H} \quad
(\mbox{quaternionic}) \nonumber \\ {\rm Hol}(\mathcal{QM})&=& \bfone
\otimes {\cal H} \quad (\mbox{HyperK\"ahler}) \nonumber \\ {\cal H} &
\subset & Sp (2n_H,\mathbb{R}) \label{olonomia}
\end{eqnarray}
In both cases, introducing flat indices $\{i,j,k= 1,2\}
\{A,B,C= 1,.., 2n_H\}$  that run , in the
fundamental representations of, respectively, $\mathrm{SU}(2)$ and
$\mathrm{Sp}(2n_H,\mathbb{R})$, we can find a vielbein 1-form
\begin{equation}
f^{iA} = f^{iA}_X (q) d q^X
\label{quatvielbein}
\end{equation}
such that
\begin{equation}
g_{XY} = f^{iA}_X f^{jB}_Y
\mathbb{C}_{AB}\epsilon_{ij} \label{quatmet}
\end{equation}
where $\mathbb{C}_{A B} = - \mathbb{C}_{B A}$ and $\epsilon_{ij}
= - \epsilon_{ji}$ are, respectively, the flat $\mathrm{Sp}(2n_H)$ and
$\mathrm{Sp}(2) \sim \mathrm{SU}(2)$ invariant metrics. The vielbein
$f^{iA}$ is covariantly closed with respect to the
$\mathrm{SU}(2)$-connection $p^r$ and to some
$\mathrm{Sp}(2n_H,\mathbb{R})$-Lie Algebra valued connection
$\Delta^{A B} = \Delta^{B A}$:
\begin{eqnarray}
\nabla f^{iA}& \equiv & df^{iA} + i p^r (\epsilon \sigma_r\epsilon^{-1})^i_{\phantom{i}j}
\wedge f^{jA} \nonumber\\ &+& \Delta^{AB} \wedge
f^{iD} \mathbb{C}_{B D} =0 \label{quattorsion}
\end{eqnarray}
\noindent where $(\sigma^r)_i^{\phantom{i}j}$ are the standard Pauli
matrices. Furthermore $f^{iA}$ satisfies  the reality
condition:
\begin{equation}
f_{iA} \equiv (f^{iA})^* = \epsilon_{ij}
\mathbb{C}_{AB} f^{jB} \label{quatreality}
\end{equation}
Eq. (\ref{quatreality})  defines  the  rule to lower the symplectic
indices by means   of  the  flat  symplectic   metrics
$\epsilon_{ij}$   and $\mathbb{C}_{AB}$. More specifically we can
write a stronger version of eq.  (\ref{quatmet}) \cite{witten:1983}:
\begin{eqnarray}
(f^{iA}_X f^{jB}_Y + f^{iA}_Y
f^{jB}_X)\mathbb{C}_{AB}&=& g_{XY} \epsilon^{ij}\nonumber\\
 \label{piuforte}
\end{eqnarray}
\noindent We have also the inverse vielbein $f^X_{iA}$
defined by the equation
\begin{equation}
f^X_{iA} f^{iA}_Y = \delta^X_Y \label{2.64}
\end{equation}
Flattening a pair of indices of the Riemann tensor ${\cal
R}^{XY}_{\phantom{XY}{ZW}}$ we obtain
\begin{equation}
{\cal R}^{XY}_{\phantom{XY}{ZW}} f^{iA}_X f^{jB}_Y = -\, i  R^r_{ZW} \epsilon^{ik} (\sigma_r)_k^{\phantom {k}j} \mathbb{C}^{AB}+
 \mathbb{R}^{AB}_{ZW}\epsilon^{ij}
\label{2.65}
\end{equation}
\noindent where $\mathbb{R}^{AB}_{ZW}$ is the field strength
of the $\mathrm{Sp}(2n_H) $ connection:
\begin{equation}
d \Delta^{AB} + \Delta^{AC} \wedge \Delta^{DB} \mathbb{C}_{CD} \equiv \mathbb{R}^{AB} =
\mathbb{R}^{AB}_{ZW} dq^Z \wedge dq^W \label{2.66}
\end{equation}
Eq.  (\ref{2.65}) is the explicit statement that the Levi Civita
connection associated with the metric $g$ has a holonomy group
contained in $\mathrm{SU}(2) \otimes \mathrm{Sp}(2n_H)$. Consider now
eq.s (\ref{quatalgebra}), (\ref{iperforme}) and (\ref{piegatello}).
We easily deduce the following relation:
\begin{equation}
g^{ZW} K^r_{XZ} K^s_{ZY} = -   \delta^{rs} g_{XY} +
  {\epsilon^{rs}}_t K^t_{XY}
\label{universala}
\end{equation}
that holds true both in the HyperK\"ahler and in the quaternionic
case. In the latter case, using eq.  (\ref{piegatello}), eq. 
(\ref{universala}) can be rewritten as follows:
\begin{equation}
g^{ZW} R^r_{XZ} R^s_{ZY} = - \lambda^2  \delta^{rs} g_{XY} +
\lambda  {\epsilon^{rs}}_t R^t_{XY}
 \label{2.67}
\end{equation}
Eq. (\ref{2.67}) implies that the intrinsic components of the curvature
 2-form $R^r$ yield a representation of the quaternion algebra.
In the HyperK\"ahler case such a representation is provided only by
the HyperK\"ahler form. In the quaternionic case we can write:
\begin{equation}
R^r_{iA, jB} \equiv R^r_{XY} f^X_{iA} f^Y_{jB} = - i \lambda \mathbb{C}_{AB}
(\sigma_r)_i^{\phantom {i}k}\epsilon _{kj} \label{2.68}
\end{equation}
\noindent Alternatively eq. (\ref{2.68}) can be rewritten in an
intrinsic form as
\begin{equation}
R^r =\,-{\rm i}\, \lambda \mathbb{C}_{AB} (\sigma_r)_i^{\phantom {i}k}
\epsilon _{kj} f^{iA} \wedge f^{jB} \label{2.69}
\end{equation}
\noindent whence we also get:
\begin{equation}
{i\over 2} R^r (\sigma_r)_i^{\phantom{i}j} = \lambda f_{iA} \wedge f^{jA} \label{2.70}
\end{equation}
\par
The quaternionic manifolds are not requested to be homogeneous
spaces, however there exists a subclass of quaternionic homogeneous
spaces that are displayed in Table~\ref{quatotable}.
\par
\begin{table}
\begin{center}
\caption{\sl Homogeneous symmetric quaternionic manifolds}
\label{quatotable}
\begin{tabular}{|c||c|}
\hline $n_H$  & $G/H$
\\
\hline ~~&~~\\ $m$ & $\frac {\mathrm{Sp}(2m+2)}{\mathrm{Sp}(2)\times
\mathrm{Sp}(2m)}$ \\ ~~&~~\\ \hline ~~&~~\\ $m$ & $\frac
{\mathrm{SU}(m,2)}{SU(m)\times \mathrm{SU}(2)\times U(1)}$  \\
~~&~~\\ \hline ~~&~~\\ $m$ & $\frac
{\mathrm{SO}(4,m)}{\mathrm{SO}(4)\times \mathrm{SO}(m)}$  \\ ~~&~~\\
\hline \hline ~~&~~\\ $2$ & $\frac {G_2}{\mathrm{SO}(4)}$ \\ ~~&~~\\
\hline ~~&~~\\ $7$ & $\frac {F_4}{\mathrm{Sp}(6)\times
\mathrm{Sp}(2)}$\\ ~~&~~~\\ \hline ~~&~~\\ $10$ & $\frac
{E_6}{\mathrm{SU}(6)\times U(1)}$\\ ~~&~~\\ \hline ~~&~~\\ $16$ &
$\frac {E_7}{S0(12)\times \mathrm{SU}(2)}$\\ ~~&~~\\ \hline ~~&~~\\
$28$ & $\frac {E_8}{E_7\times \mathrm{SU}(2)}$\\ ~~&~~\\ \hline
\end{tabular}
\end{center}
\end{table}

\subsection{${\cal N}=2$, $d=5$ supergravity before gauging}
\label{beforgau}
Relying on the geometric lore developed in the previous sections it
is now easy to state what is the bosonic Lagrangian of a general ${\cal N}=2$ theory
in five--dimensions. We just have to choose an $n$--dimensional very
special manifold and some quaternionic manifold $\mathcal{QM}$ of
quaternionic dimension $n_H$. Then recalling eq. (\ref{genford5bos}) we
can specialize it to:
\begin{eqnarray}
  \frac 1{\sqrt{-g}}\mathcal{L}^{(ungauged)}_{(d=5,{\cal N}=2)}&=& \, 
\frac{1}{2}
  \,R \, - \, \frac{1}{4}\, a_{\tilde I \tilde J} (\phi)  F^{\tilde I}_{\mu \nu
  } \, F^{\tilde J \vert\mu \nu} + \frac{1}{2} \, g_{\tilde x \tilde y}(\phi) \, \partial_\mu \phi^{\tilde x} \, \partial ^\mu \, \phi^{\tilde y}
 \cr
  &&   +\, \frac{1}{2} \, g_{XY}(q) \, \partial_\mu q^X \, \partial ^\mu \, q^Y 
+ \frac{1}{6\sqrt 6 \sqrt{-g}} C_{\tilde I \tilde J \tilde K} \, \epsilon ^{\mu \nu \rho \sigma \tau
  } \, F^{\tilde I}_{\mu \nu } \, F^{\tilde J}_{\rho \sigma } \, A^{\tilde K}_\tau \cr
&&\label{genn2d5ung}
\end{eqnarray}
where $g_{XY}(q)$ is the quaternionic metric on the quaternionic
manifold $\mathcal{QM}$, while $g_{\tilde x\tilde y}(\phi)$ is the very special metric
on the very special manifold. At the same time, the constant tensor
$C_{\tilde I \tilde J \tilde K}$ is that defining the cubic norm
(\ref{defispec}) while the kinetic metric $a$ is that
defined in eq. (\ref{scriptNsp}). The transformation rule of the
gravitino field takes the general form (\ref{gensusrul}) with the
graviphoton defined as in eq. (\ref{gravfotft}) and the tensor
$\Phi^{\tilde I}_{ij}$ given by eq. (\ref{Phiforn2}). In this respect it
is noteworthy that the gravitino supersymmetry transformation rule depends only on the vector multiplet scalars and it is independent
of the hypermultiplets. Such a situation will be changed by the
gauging, that introduces a gravitino mass-matrix depending also on the
hypermultiplets.




\section[Supergravity Gaugings]{Supergravity Gaugings}
\label{gaugchap}
\setcounter{equation}{0}
Before entering in the technical details of the gauging procedure for our specific case let me begin by recalling some very general aspects which are common to extended supergravities in different dimensions. Because of the fundamental property that the scalar potential is generated by the gauging, the discussion on these topics naturally involves spontaneous supersymmetry breaking and the super-Higgs mechanism. \footnote{They were codified in the literature of the
early  and middle eighties \cite{susyb1,deRoo:1985np,susyb2,susyb3}
(for a review see chapter II.8 of \cite{castdauriafre}) and were
further analyzed and extended in the middle nineties
\cite{Zinovev:1992mw,Ferrara:1996gu,Fre:1997js,Girardello:1997hf}.}

\subsection[Supersymmetry breaking in conventional vacua]{Supersymmetry breaking in
conventional vacua}\label{genaspect}
A conventional\footnote{I use the term ``conventional'' to distinguish them from the ones obtainable from Domain wall configurations (see Appendix \ref{domain wall}).}  vacuum of $d$--dimensional supergravity corresponds
to a space--time geometry with a maximally extended group of isometries, namely with
$\frac{1}{2} d (d+1)$ Killing vectors. This means that the metric
$ds^2 = g_{\mu\nu} dx^\mu \otimes dx^\nu$ necessarily has constant
curvature in $d$--dimensions and is one of the following three:
\begin{equation}
  \mathcal{M}_{space-time}=\left \{ \begin{array}{lcl}
  AdS_{d} & ;& \mbox{negative
  curvature}\\
  \mathrm{Minkowski_{d}}&;& \mbox{zero curvature}\\
\mathrm{dS_{d}} &;& \mbox{positive curvature}\ \end{array}\right.
\label{trecass}
\end{equation}
At the same time, in order to be consistent with this maximal
symmetry, the v.e.v.s of the scalar  fields, $< \varphi^\Lambda >=\varphi^\Lambda_0$,
must be constant and be extrema  of the scalar potential:
\begin{equation}
  \left.\frac{\partial \mathcal{V}}{\partial \varphi ^\Lambda}\right|_{\varphi=\varphi_0} =  0~,
\label{extrepot}
\end{equation}
Minkowski space occurs when $\mathcal{V}(\varphi_0)=0$,  anti de Sitter space AdS$_{d}$ occurs
when $\mathcal{V}(\varphi_0)< 0$ and finally de Sitter space dS$_{d}$ is generated by
$\mathcal{V}(\varphi_0) > 0$.
To be definite, I  focus on the $4$--dimensional case, that also historically was  the first to be analyzed, but all the mechanisms and properties I describe below have straightforward counterparts in higher
dimensions. Furthermore, as already underlined before, it presents a lot of similarities  with the $5$--dimensional case. So let me state that in relation
with the super-Higgs mechanism, there are just three
relevant items of the entire $d=4$ supergravity construction that  have
to be considered:
\begin{enumerate}
\item The \emph{gravitino mass matrix} $S_{ij}(\varphi)$ , namely the
non-derivative scalar field dependent term that appears in the
gravitino supersymmetry transformation rule:
\begin{equation}
  \delta \psi _{i\vert \mu} = \mathcal{D}_\mu \, \epsilon_i \, + \,
  S_{ij} \left( \varphi \right) \, \gamma_\mu \, \epsilon^j~ +\dots,
\label{gravsusy}
\end{equation}
and reappears as a mass term in the Lagrangian:
\begin{equation}
  \mathcal{L}^{\rm SUGRA} \, = \, \dots \, + \, \mbox{const} \,\left( \,
  S_{ij} (\varphi ) \, \psi^i_\mu \, \gamma^{\mu\nu} \, \psi^j_\nu \, +
  \,  S^{ij} (\varphi ) \, \psi_{i\vert\mu} \, \gamma^{\mu\nu} \,
  \psi_{j\vert\nu} \,\right)
\label{gravmasmat}
\end{equation}
\item The \emph{fermion shifts}, namely the non-derivative scalar field
  dependent terms in the supersymmetry transformation rule of the spin $\ft 12$
  fields :
\begin{eqnarray}
  \delta \, \lambda^{\Lambda}_R & =& \mbox{derivative terms} \, + \,
  \Sigma_{i}^{\phantom{i} \Lambda} \left( \varphi \right) \,  \epsilon^i~,
  \nonumber\\
  \delta \, \lambda^{\Lambda}_L & =& \mbox{derivative terms} \, + \,
  \Sigma^{i\vert \Lambda} \left( \varphi \right) \,  \epsilon_i~.
\label{fermioshif}
\end{eqnarray}
\item The scalar potential itself, $\mathcal{V}(\varphi)$.
\end{enumerate}
These three items are related by a general supersymmetry Ward
identity, firstly discovered in the context of gauged ${\cal N}=8$
supergravity \cite{dewit1} and later extended to all supergravities
\cite{susyb1,susyb2,susyb3}, that, in my conventions,\footnote{In fact I adopt the conventions of $d=5$  to avoid any confusion with the normalization and the role of the indices.}
reads as follows:
\begin{equation}
  6\,S_{ik} \, S^{kj} - 4 \, K_{\Lambda,\Gamma} \Sigma_{i}^{\phantom{i} \Lambda}  \Sigma^{j \vert \Gamma} \, = \, -\delta_{i}^{j} \,\mathcal{V}~,
\label{wardide}
\end{equation}
where $K_{\Lambda,\Gamma}$ is the kinetic matrix of the spin--1/2 fermions \cite{deRoo:1985np,wageman,deroowag,wagthesis}. The
numerical coefficients appearing in (\ref{wardide}) depend on the
normalization of the kinetic terms of the fermions, while $i,j,\dots
= 1,\dots,{\cal N}$ are $\mathrm{SU}({\cal N})$ indices that
enumerate the supersymmetry charges. We also follow the standard
convention that the upper or lower position of such indices denotes
definite chiral  projections of Majorana spinors, right or left
depending on the species of fermions considered\footnote{For
instance, we have $\gamma_5 \, \epsilon_i= \epsilon_i$ and $\gamma_5
\, \epsilon^i = - \epsilon^i$.}. The position denotes also the way of
transforming of the fermion with respect to
$\mathrm{SU}({\cal N})$, with lower indices in the fundamental and
upper indices in the fundamental bar. In this way we have $S^{ij} =
\left(S_{ij} \right) ^\star$ and $\Sigma_i^{\phantom{i} \Lambda}
=\left(\Sigma^{j \vert \Lambda} \right) ^\star$. Finally, the index $\Lambda$ is a collective index that enumerates all spin--1/2  fermions $ \lambda^\Lambda$ present in the theory\footnote{We denote by $ \lambda^\Lambda$ the right
handed chiral projection while $\lambda_\Lambda$ are the left handed ones.}.
\par
  The corresponding  fermion shifts are defined by
\begin{equation}
  \delta \, \lambda^\Lambda \, = \,\mbox{derivative terms} \, + \,
  \Sigma_{i}^{\phantom{i}\Lambda} \left( \varphi \right) \,  \epsilon^i~.
\label{gaushif}
\end{equation}
\par
A vacuum configuration $\varphi_0$ that preserves $\mathcal{N}_0$
supersymmetries is characterized by the existence of $\mathcal{N}_0$
vectors $\rho^i_{(\ell)}$ ($\ell=1,\ldots,\mathcal{N}_0$) of
$\mathrm{SU}({\cal N})$, such that
\begin{eqnarray}
\label{breakpat} S_{ij} \left( \varphi_0 \right) \, \rho^i_{(\ell)} & =
& e^{\rm i\theta}\, \sqrt{\ft{-\mathcal{V}(\varphi_0)}{6}} \, \rho_{i(\ell)}
~,\nonumber\\ \Sigma_i^{\phantom{i}\Lambda} \left( \varphi_0 \right) \,
\rho^i_{(\ell)} & = & 0~,
\end{eqnarray}
where $\theta$ is an irrelevant phase. Indeed, consider the spinor
\begin{equation}
  \epsilon ^i(x) = \sum_{\ell =1}^{\mathcal{N}_0} \, \rho^i_{(\ell)}
  \epsilon^{(\ell)}(x)~,
\label{Eadix}
\end{equation}
where $\epsilon^{(\ell)}(x)$ are $\mathcal{N}_0$ independent
solutions of the equation for covariantly
constant spinors in $\mathrm{AdS}_4$ (or Minkowski space) with 
$2\,e= \sqrt{-\mathcal{V}(\varphi_0)/{6}}$:
\begin{equation}
\label{susygravads4}
D_a^{(\mathrm{AdS})}\epsilon(x)\equiv
(\partial_a - {1\over 4}\omega^{bc}_{~~a} \gamma_{bc} -
2\, e\, \gamma_5\gamma_a)\epsilon(x)=0~,
\end{equation}
The integrability of eq. (\ref{susygravads4})
is guaranteed by the expression of the AdS$_4$ curvature, 
$R^{ab}_{~~cd}=-64\,e^2\,\delta^{ab}_{cd}$, that
corresponds to the Ricci tensor:
\begin{equation}
\label{cosmol4e7}
\mathcal{R}_{ab} = -96\, e^2\, \eta_{ab}~=\,\mathcal{V}(\varphi_0) \, \eta_{ab},\end{equation}
 Then it follows that under supersymmetry
transformations of the parameter (\ref{Eadix}), the chosen vacuum
configuration $\varphi=\varphi_0$ is invariant\footnote{As already
stressed, the v.e.v.s of all the fermions are zero and equation
(\ref{breakpat}) guarantees that they remain zero under supersymmetry
transformations of parameters (\ref{Eadix}).}. That such a
configuration is a true vacuum follows from another property proved,
for instance, in  \cite{susyb3}: all vacua that admit at least one
vector $\rho^i$ satisfying eq.  (\ref{breakpat}) are automatically
extrema  of the potential, namely they satisfy eq.  (\ref{extrepot}).
Furthermore, as one can immediately check, the four dimensional action 
 for constant scalar field configurations implies that the Ricci tensor must be 
 $\mathcal{R}_{\mu \nu
}=\,\mathcal{V}(\varphi_0)\, g_{\mu \nu }$, as in equation
(\ref{cosmol4e7}).
\par
The above integrability argument can be easily generalized to all
dimensions and to all numbers of supersymmetries ${\cal N}$.
Consider a supergravity action in $d$ dimensions that, once reduced to the
gravitational plus scalar field sector, has this general form:
\begin{equation}
  A_{grav+scal}^{[d]}=\int \, d^d x \, \sqrt{-g} \,  \frac{1}{2} \left[ \, R[g]
  +\alpha \, g_{\Lambda\Sigma} (\varphi) \, \partial^\mu \, \varphi^\Lambda \partial_\mu \varphi^\Lambda -
 2 \mathcal{V}(\varphi)\right]
\label{scalfielsect}
\end{equation}
where $\alpha$ is a normalization constant that can vary from case to
case, since it can always be reabsorbed into the definition of the
scalar metric. But the scalar potential  $\mathcal{V}$ has an
unambiguous and unique normalization with respect to the Einstein
term. For constant field configurations $\varphi_0$, the Einstein equations
derived from (\ref{scalfielsect}) imply that:
\begin{equation}
  R_{\mu \nu  }= \frac{2}{(d-2)} \, \mathcal{V}(\varphi_0) \, g_{\mu \nu }
\label{Ricciustens}
\end{equation}
Then the Riemann tensor of an anti de Sitter space $AdS_d$ consistent
with eq.  (\ref{Ricciustens}) is necessarily the following:
\begin{equation}
  R^{\rho \sigma }_{\mu \nu } = \frac{4}{(d-1)(d-2)}\, \mathcal{V}(\varphi_0) \,
  \delta^{[\rho}_{[\mu} \, \delta ^{\sigma]}_{\nu]}
\label{Riemantens}
\end{equation}
Consider next the equation for a covariantly constant  spinor in $AdS_d$. Its
general form is:
\begin{eqnarray}
D_\mu^{(\mathrm{AdS})}\epsilon &\equiv& \mathcal{D}_\mu \epsilon(x)-
\mu \, \gamma_\mu \, \epsilon
=(\partial_\mu - {1\over 4}\omega^{bc}_{~~\mu} \gamma_{bc} -
\mu \, \gamma_\mu)=0
\label{gencovconstspi}
\end{eqnarray}
where the parameter $\mu$ is fixed by integrability in terms of the
vacuum value of potential $\mathcal{V}(\varphi_0) $. Indeed, from the
condition $D_{[\mu}^{(\mathrm{AdS})} D_{\nu]}^{(\mathrm{AdS})} =0$ we
immediately get:
\begin{equation}
  |\mu |^2 = \, \frac{|\mathcal{V}(\varphi_0)|}{(d-1) (d-2)}
\label{modulmu}
\end{equation}
On the other hand the general form of the gravitino transformation
rule is, independently from the number of space--time dimensions,
that given in eq. (\ref{gravsusy}), so that, in a conventional vacuum
with an unbroken supersymmetry, $\mu$ is to be interpreted as
\textbf{eigenvalue} of the gravitino mass--matrix. So the general
conditions for the preservation of $\mathcal{N}_0$ supersymmetries
in $d$ dimensions are fully analogous to those in eq. (\ref{breakpat})
and correspond to the existence of $\mathcal{N}_0$ independent
vectors $\rho^i_{(\ell)}$ ($\ell=1,\dots,\mathcal{N}_0$), such that:
\begin{eqnarray}
\label{genbreakpat} S_{ij} \left( \varphi_0 \right) \, \rho^j_{(\ell)} & =
& e^{\rm i\theta}\, \sqrt{\ft{|\mathcal{V}(\varphi_0)|}{(d-2)(d-1)}} \, \rho_{i(\ell)}
~,\nonumber\\ \Sigma_i^{\phantom{i}i} \left( \varphi_0 \right) \,
\rho^i_{(\ell)} & = & 0~,
\end{eqnarray}
By extension of language, the vectors $\rho^i_{(\ell)}$ are named
\textbf{Killing spinors}.\\
In the next section we explicitly discuss these features for ${\cal N}=2$ theory in $d=5$.



\section{Gauged supergravities in five dimensions}
\label{gaugd5theo}
The general form of \emph{gauged} ${\cal N}=2$ supergravity in five--dimensions
has been obtained only recently. This occurred  through the contributions of two groups of authors. In a first step, G\"unaydin and Zagerman analyzed
the problem of gauging in the presence of an arbitrary number of vector
and tensor multiplets. In a series of papers \cite{Gunaydin:2000zx,Gunaydin:2000xk,Gunaydin:2001ph}
they established the key new features involved by the gauging procedure in this space-time dimension. 
In a subsequent step, Ceresole and Dall'Agata \cite{ceresole:2000}, utilizing the
general methods of the geometrical gauging
\cite{dewit1,Castellani:1986ka,D'Auria:1991fj,bertolo},  reconsidered
the problem and succeeded in including also the coupling to
hypermultiplets.
Since ${\cal N}=2$, that corresponds to $N_Q=8$ supercharges, is the minimal possible
number of supersymmetries in a five--dimensional space--time, it is clear
that this result is a relevant step for the construction of phenomenological models with minimal supersymmetry like the Randall Sundrum scenarios and more generically to understand better the supposed gravity/gauge theory correspondence.
I already pointed out that in maximal supergravities  the number of
available gauge vectors is fixed a priori and the possible gauge
algebras fill a discrete set. In matter coupled supergravities, on the other hand,
the number of vector multiplets varies and one has much more
possibilities. If the number of supercharges $N_Q$ is larger than
eight, the only available multiplets, beside the graviton multiplet,
are the vector or tensor multiplets; furthermore, given their number $n$,
the geometry of the scalar manifold is fixed and corresponds to a homogeneous space
$\mathcal{G}_n/\mathcal{H}_n$. At $N_Q=8$, instead, besides vector
or tensor multiplets (that can be dualised to vector multiplets), one
has also hypermultiplets so that the scalar manifold
$\mathcal{M}_{scalar}=\mathcal{M}_{vect. scal.}
\times \mathcal{M}_{hyp.scal.}$
is the tensor product of two submanifolds  containing the vector scalars
and the hyper scalars respectively.
As we have seen in sect.\ref{very}-\ref{hypgeosec}, although severely constrained, the geometry of these two submanifolds is not completely
fixed by supersymmetry and can vary within an
ample class that contains both homogeneous and non homogeneous
spaces. As I also recalled, there is a very
close structural relation between $N_Q=8$ supergravity in $d=4$
and in $d=5$ dimensions: the geometry of the hypermultiplet scalars
is the same in both theories, namely \emph{quaternionic geometry} (see
sect.\ref{hypgeosec}) while the vector scalars fill a \emph{special
K\"ahler} complex manifold in $d=4$ and a \emph{very special} real
manifold in $d=5$. Dimensional reduction on a circle maps $d=5$
theories into $d=4$ theories and provides a map from very special to
a subclass of special K\"ahler manifolds. Hence it is not
surprising that the \emph{gauging procedure} in $d=4$ and $d=5$
theories are extremely similar: yet there are some relevant differences
that had to be clarified before one could extend the
constructions of \cite{D'Auria:1991fj,bertolo} to one higher
space--time dimensions. These differences have essentially to do with
two specifically five--dimensional features:
\begin{enumerate}
  \item [a] Very special, differently from special K\"ahler manifolds
  are real and non--symplectic. So there is no notion of a moment-map
  for isometries
  \item [b] In the presence of gauging, vector and tensor multiplets
  become physically distinct and the vector fields that are in a non
  trivial non--adjoint representation of the gauge group have to be
  dualised to massive self--dual $2$--forms.
\end{enumerate}
\par
To accomplish this program  my first care is to discuss the general idea of the
moment map which constitutes an essential ingredient in the ${\cal N}=2$ case.
\subsection[The Moment Map]{The Moment Map}
\label{momentm1}
The moment map is a construction that applies to all manifolds
with a symplectic structure, in particular to K\"ahler, HyperK\"ahler
or quaternionic manifolds.
\par
I begin with the K\"ahler case, namely with the moment
map of holomorphic isometries which is the paradigm for all the other
cases. It is also the additional weapon one can use in gauging $d=4$ supergravity
while it is not available  for $d=5$ vector multiplets due to the real
structure of very special geometry.
The HyperK\"ahler and quaternionic cases correspond, instead,
to the moment map of triholomorphic isometries which equally applies to
$d=4$ and $d=5$ theories.
\subsubsection{The holomorphic moment map on K\"ahler manifolds}
I assume some basic knowledge of K\"ahler geometry which can be
retrieved from any standard textbook.
Let  $g_{i {j^\star}}$ be the K\"ahler metric of a K\"ahler
manifold ${\cal M}$ and let us assume that  $g_{i {j^\star}}$ admits
a non trivial group of continuous isometries ${\cal G}$
generated by Killing vectors $k_\mathbf{I}^i$ ($\mathbf{I}=1, \ldots, {\rm dim}
\,{\cal G} )$ that define the infinitesimal variation of the complex
coordinates $z^i$ under the group action:
\begin{equation}
z^i \to z^i + \epsilon^\mathbf{I} k_\mathbf{I}^i (z)
\end{equation}
Let $k^i_{\mathbf{I}} (z)$ be a basis of holomorphic Killing vectors for
the metric $g_{i{j^\star}}$.  Holomorphicity means the following
differential constraint:
\begin{equation}
\partial_{j^*} k^i_{\mathbf{I}} (z)=0
\leftrightarrow \partial_j k^{i^*}_{\mathbf{I}} (\bar z)=0 \label{holly}
\end{equation}
while the generic Killing equation (suppressing the
gauge index $\mathbf{I}$):
\begin{equation}
\nabla_\mu k_\nu +\nabla_\mu k_\nu=0
\end{equation}
in holomorphic indices reads as follows:
\begin{equation}
\begin{array}{ccccccc}
\nabla_i k_{j} + \nabla_j k_{i} &=&0 & ; &
\nabla_{i^*} k_{j} + \nabla_j k_{i^*} &=& 0
\label{killo}
\end{array}
\end{equation}
where the covariant components are defined as
$k_{j }=g_{j i^*} k^{i^*}$ (and similarly for
$k_{i^*}$).
\par
The vectors $k_{\mathbf{I}}^i$ are generators of infinitesimal
holomorphic coordinate transformations $\delta z^i = \epsilon^\mathbf{I} k^i_{\mathbf{I}} (z)$
which leave the metric invariant. In the same way as the metric is
the derivative of a more fundamental
object, the Killing vectors in a K\"ahler manifold are the
derivatives of suitable prepotentials. Indeed, the first of
eq.s (\ref{killo})  is automatically satisfied by holomorphic vectors
and the second equation reduces to the following one:
\begin{equation}
k^i_{\mathbf{I}}=i g^{i j^*} \partial_{j^*} P_{\mathbf{I}},
\quad P^*_{\mathbf{I}} = P_{\mathbf{I}}\label{killo1}
\end{equation}
In other words if we can find a real function $P^\mathbf{I}$ such
that the expression $i g^{i j^*} \partial_{j^*}
P_{(\mathbf{I})}$ is holomorphic, then eq. (\ref{killo1}) defines a
Killing vector.
\par
The construction of the Killing prepotential can be stated in a more
precise geometrical fashion through the notion of {\it moment map}.
Let us review this construction.
\par
Consider a K\"ahlerian manifold ${\cal M}$ of real dimension $2n$.
Consider a compact Lie group ${\cal G}$ acting on
 ${\cal M}$  by means of Killing vector
fields $\overrightarrow{X}$ which are holomorphic
with respect to the  complex structure
${ J}$ of ${\cal M}$; then these vector
fields preserve also the K\"ahler 2-form
\begin{equation}
\begin{array}{ccc}
\begin{matrix}
{\cal L}_{\scriptscriptstyle\overrightarrow{X}}g = 0 & \leftrightarrow &
\nabla_{(\mu}X_{\nu)}=0 \cr
{\cal L}_{\scriptscriptstyle\overrightarrow{X}}{  J}= 0 &\null &\null \cr 
\end{matrix}
  \Biggr \} & \Rightarrow &  0={\cal L}_{\scriptscriptstyle\overrightarrow{X}}
K = i_{\scriptscriptstyle\overrightarrow{X}}
dK+d(i_{\scriptscriptstyle\overrightarrow{X}}
K) = d(i_{\scriptscriptstyle\overrightarrow{X}}K) \cr
\end{array}
\label{holkillingvectors}
\end{equation}
Here ${\cal L}_{\scriptscriptstyle\overrightarrow{X}}$ and
$i_{\scriptscriptstyle\overrightarrow{X}}$
denote respectively the Lie derivative along
the vector field $\overrightarrow{X}$ and the contraction
(of forms) with it.
\par
If ${\cal M}$ is simply connected,
$d(i_{\overrightarrow{X}}K)=0$ implies the existence
of a function $P_{\overrightarrow{X}}$ such
that
\begin{equation}
-\frac{1}{2\pi}dP_{\overrightarrow{X}}=
i_{\scriptscriptstyle\overrightarrow{X}}K
\label{mmap}
\end{equation}
The function $P_{\overrightarrow{X}}$ is defined up to a constant,
which can be arranged so as to make it equivariant:
\begin{equation}
\overrightarrow{X} P_{\overrightarrow{Y}} =
P_{[\overrightarrow{X},\overrightarrow{Y}]}
\label{equivarianza}
\end{equation}
$P_{\overrightarrow{X}}$ constitutes then a {\it moment map}.
This can be regarded as a map
\begin{equation}
P: {\cal M} \, \longrightarrow \,
\mathbb{R} \otimes
{\mathbb{G} }^*
\end{equation}
where ${\mathbb{G}}^*$ denotes the dual of the Lie algebra
${\mathbb{G} }$ of the group ${\cal G}$.
Indeed, let $x\in {\mathbb{G} }$ be the Lie algebra element
corresponding to the Killing vector $\overrightarrow{X}$; then, for a given
$m\in {\cal M}$
\begin{equation}
\mu (m)\,  : \, x \, \longrightarrow \,  P_{\overrightarrow{X}}(m) \,
\in  \, \mathbb{R}
\end{equation}
is a linear functional on  ${\mathbb{G}}$.
If we expand
$\overrightarrow{X} = a^\mathbf{I} k_\mathbf{I}$ in a basis of Killing vectors
$k_\mathbf{I}$ such that
\begin{equation}
[k_\mathbf{I}, k_\mathbf{L}]= f_{\mathbf{I} \mathbf{L}}^{\ \ \mathbf{K}} k_\mathbf{K}
\label{blio}
\end{equation}
we  also have
\begin{equation}
P_{\overrightarrow{X}}\, = \, a^\mathbf{I} P_\mathbf{I}
\end{equation}
In the following we  use the
shorthand notation ${\cal L}_\mathbf{I}, i_\mathbf{I}$ for the Lie derivative
and the contraction along the chosen basis of Killing vectors $ k_\mathbf{I}$.
\par
From a geometrical point of view the prepotential,
or moment map, $P_\mathbf{I}$ is the Hamiltonian function providing the Poissonian
realization  of the Lie algebra on the K\"ahler manifold. This
is just another way of stating the already mentioned
{\it  equivariance}.
Indeed  the  very  existence  of the closed 2-form $K$ guarantees that
every K\"ahler space is a symplectic manifold and that we can define  a
Poisson bracket.
\par
Consider eqs.(\ref{killo1}). To every generator of the abstract  Lie algebra
${\mathbb{G}}$ we have associated a function  $P_\mathbf{I}$ on
${\cal M}$; the Poisson bracket of
$P_\mathbf{I}$ with $P_\mathbf{J}$ is defined as follows:
\begin{equation}
\{P_\mathbf{I} , P_\mathbf{J}\} \equiv 4\pi K
(\mathbf{I}, \mathbf{J})
\end{equation}
where $K(\mathbf{I}, \mathbf{J})
\equiv K (\vec k_\mathbf{I}, \vec k_\mathbf{J})$ is
the value of $K$ along the pair of Killing vectors.
\par
In reference \cite{D'Auria:1991fj} the following lemma is proved:
\begin{lemma}
{\it{The following identity is true}}:
\begin{equation}
\{P_\mathbf{I}, P_\mathbf{J}\}=f_{\mathbf{I}\mathbf{J}}^{\ \ \mathbf{L}} P_\mathbf{L} + C_{\mathbf{I} \mathbf{J}} \label{brack}
\end{equation}
{\it{where $C_{\mathbf{I} \mathbf{J}}$ is a constant fulfilling the
cocycle condition}}
\begin{equation}
f^{\ \ \mathbf{L}}_{\mathbf{I}\mathbf{M}} C_{\mathbf{L} \mathbf{J}} +
f^{\ \ \mathbf{L}}_{\mathbf{M}\mathbf{J}} C_{\mathbf{L} \mathbf{I}}+
f_{\mathbf{J}\mathbf{I}}^{\ \  \mathbf{L}} C_{\mathbf{L} \mathbf{M}}=0
\label{cocy}
\end{equation}
\end{lemma}
If the Lie algebra ${\mathbb{G}}$ has a trivial second cohomology group
$H^2({\mathbb{G}})=0$, then the cocycle $C_{\mathbf{I} \mathbf{J}}$ is a
coboundary; namely we have
\begin{equation}
C_{\mathbf{I} \mathbf{J}} = f^{\ \ \mathbf{L}}_{\mathbf{I} \mathbf{J}} C_\mathbf{L}
\end{equation}
where $C_\mathbf{L}$ are suitable constants. Hence, assuming
$H^2 (\mathbb{G})= 0$,
we can reabsorb $C_\mathbf{L}$ in  the definition of $P_\mathbf{I}$:
\begin{equation}
P_\mathbf{I} \rightarrow P_\mathbf{I}+ C_\mathbf{I}
\end{equation}
and we obtain the stronger equation
\begin{equation}
\{P_\mathbf{I}, P_\mathbf{J}\} =
f_{\mathbf{I}\mathbf{J}}^{\ \  \mathbf{L}} P_\mathbf{L}
\label{2.39}
\end{equation}
Note that $H^2({\mathbb{G}}) = 0$ is true for all semi-simple Lie
algebras.
Using eq. (\ref{brack}), eq. (\ref{2.39})
can be rewritten in components as follows:
\begin{equation}
{i\over 2} g_{ij^*}(k^i_\mathbf{I} k^{j^*}_\mathbf{J} -
k^i_\mathbf{J} k^{j^*}_\mathbf{I})=
{1\over 2} f_{\mathbf{I} \mathbf{J}}^{\  \  \mathbf{L}} P_\mathbf{L}
\label{2.40}
\end{equation}
Equation (\ref{2.40}) is identical with the equivariance condition
in eq. (\ref{equivarianza}).
\subsubsection{The triholomorphic moment map on quaternionic manifolds}
Next, closely following the original derivation of \cite{D'Auria:1991fj,mylecture},
I turn to a discussion of the triholomorphic isometries of the manifold $\mathcal{QM}$
associated with hypermultiplets.
Both in $d=4$ and in $d=5$ supergravity, $\mathcal{QM}$ is  quaternionic
and we can gauge only those of its isometries
that are  triholomorphic  and that  either generate an abelian group $\mathcal{G}$
or are \emph{suitably realized}  as isometries also on the special manifold $\mathcal{SV}$
\footnote{I anticipate the meaning of {\it suitably realized} to be
discussed in later sections. By definition the gauge vectors are in
the coadjoint representation of the gauge groups. The vectors
transform linearly under isometries as the sections
$h^{\tilde I}$ defining very special geometry. It follows that
under the gauge algebra, these latter must decompose in a
{\it coadjoint representation} plus, possibly, another representation
$R$. The vectors in the representation $R$ must be dualised to
massive self dual $2$--forms.}. This means  that on $\mathcal{QM}$ we
have Killing vectors:
\begin{equation}
\vec k_\mathbf{I} = k^X_\mathbf{I} {\vec \partial\over \partial q^X}
\label{2.71}
\end{equation}
\noindent
satisfying the same Lie algebra  as the corresponding Killing
vectors on ${\cal VM}$. In other words
\begin{equation}
\hat{\vec{k}}_\mathbf{I} =
k^x_\mathbf{I} \vec \partial_x + k_\mathbf{I}^X \vec\partial_X
\label{2.72}
\end{equation}
\noindent
is a Killing vector of the block diagonal metric:
\begin{equation}
\hat g = \left (
\begin{matrix} 
g_{xy} & \quad 0 \quad \cr \quad 0
\quad & g_{XY} \cr 
\end{matrix}
 \right )
\label{2.73}
\end{equation}
defined on the product manifold ${\cal VM}\otimes\mathcal{QM}$.
Let us first focus on the manifold $\mathcal{QM}$.
Triholomorphicity means that the Killing vector fields leave
the HyperK\"ahler structure invariant up to $\mathrm{SU(2)}$
rotations in the $\mathrm{SU(2)}$--bundle defined by eq. (\ref{su2bundle}).
Namely:
\begin{equation}
\begin{array}{ccccccc}
{\cal L}_\mathbf{I} K^r & = &\frac 12 \epsilon^{rst}K^s
\alpha^t_\mathbf{I} & ; &
{\cal L}_\mathbf{I} p^r&=& \frac 12 \nabla \alpha^r_\mathbf{I}
\end{array}
\label{cambicchio}
\end{equation}
where $\alpha^r_\mathbf{I}$ is an $\mathrm{SU(2)}$ compensator associated with the Killing vector $k^X_\mathbf{I}$.\footnote{The above relations find an effective application in this work for determining how gauge transformations act in presence of hypermultiplets couplings, see section \ref{geocla} below.} 
 The compensator $\alpha^r_\mathbf{I}$ necessarily
fulfills  the cocycle condition:
\begin{equation}
{\cal L}_\mathbf{I} \alpha^r_\mathbf{J} - {\cal L}_\mathbf{J} \alpha^r_\mathbf{I} + \epsilon^{rst}
\alpha^s_\mathbf{I} \alpha^t_\mathbf{J} = f_{\mathbf{I} \mathbf{J}}^{\cdot \cdot \mathbf{L}} \alpha^r_\mathbf{L}
\label{2.75}
\end{equation}
In the HyperK\"ahler case the $\mathrm{SU(2)}$--bundle is flat and the
compensator can be reabsorbed into the definition of the
HyperK\"ahler forms. In other words we can always find a
map
\begin{equation}
\mathcal{QM} \, \longrightarrow \, L^r_{\phantom{r}s} (q)
\, \in \, \mathrm{SO(3)}
\end{equation}
that trivializes the ${\cal SU}$--bundle globally. Redefining:
\begin{equation}
K^{r\prime} \, = \, L^r_{\phantom{r}s} (q) \, K^s
\label{enfantduparadis}
\end{equation}
the new HyperK\"ahler form  obeys the stronger equation:
\begin{equation}
{\cal L}_\mathbf{I} K^{r\prime} \, = \, 0
\label{noncambio}
\end{equation}
On the other hand, in the quaternionic case, the non--triviality of the
${\cal SU}$--bundle forbids to eliminate the $W$--compensator
completely. Due to the identification between HyperK\"ahler
forms and $\mathrm{SU(2)}$ curvatures, eq. (\ref{cambicchio}) is rewritten
as:
\begin{equation}
\begin{array}{ccccccc}
{\cal L}_\mathbf{I} R^r& = & \frac 12 \epsilon^{rst}R^s
\alpha^t_\mathbf{I} & ; &
{\cal L}_\mathbf{I} p^r&=& \frac 12\nabla \alpha^r_\mathbf{I}
\end{array}
\label{cambiacchio}
\end{equation}
In both cases, anyhow, and in full analogy with the case of
K\"ahler manifolds, to each Killing vector
we can associate a triplet $P^r_\mathbf{I} (q)$ of 0-form prepotentials.
Indeed, we can set:
\begin{equation}
{\bf i}_\mathbf{I}  K^r =
- \nabla P^r_\mathbf{I} \equiv -(d P^r_\mathbf{I} + 2 \epsilon^{rst} p^s P^t_\mathbf{I})
\label{2.76}
\end{equation}
where $\nabla$ denotes the $\mathrm{SU(2)}$ covariant exterior derivative.
\par
As in the K\"ahler case, eq. (\ref{2.76}) defines a moment map:
\begin{equation}
P: {\cal M} \, \longrightarrow \,
\mathbb{R}^3 \otimes
{\mathcal{G} }^*
\end{equation}
where ${\mathcal{G}}^*$ denotes the dual of the Lie algebra
${\mathcal{G} }$ of the group ${\cal G}$.
Indeed, let $x\in {\mathcal{G} }$ be the Lie algebra element
corresponding to the Killing
vector $\overrightarrow{\chi}$; then, for a given
$m\in {\cal M}$
\begin{equation}
\mu (m)\,  : \, x \, \longrightarrow \,  P_{\overrightarrow{\chi}}(m) \,
\in  \, \mathbb{R}^3
\end{equation}
is a linear functional on  ${\mathcal{G}}$. If we expand
$\overrightarrow{\chi} = a^\mathbf{I} k_\mathbf{I}$ on a basis of Killing vectors $k_\mathbf{I}$ such that
\begin{equation}
[k_\mathbf{I}, k_\mathbf{L}]= f_{\mathbf{I} \mathbf{L}}^{\ \ \mathbf{K}} k_\mathbf{K}
\label{blioprime}
\end{equation}
and we also choose a basis ${\bf i}_r \, (r=1,2,3)$ for $\mathbb{R}^3$
we get:
\begin{equation}
P_{\overrightarrow{\chi}}\, = \, a^\mathbf{I} P_\mathbf{I}^r \, {\bf i}_r
\end{equation}
Furthermore we need a generalization of the equivariance defined
by eq. (\ref{equivarianza})
\begin{equation}
\overrightarrow{\chi} \circ P_{\overrightarrow{\zeta}} \,=  \,
P_{[\overrightarrow{\chi},\overrightarrow{\zeta}]}
\label{equivarianzina}
\end{equation}
In the HyperK\"ahler case, the left--hand side of
eq. (\ref{equivarianzina})
is defined as the usual action of a vector field on a $0$--form:
\begin{equation}
\overrightarrow{\chi} \circ P_{\overrightarrow{\zeta}}\, =  \, {\bf i}_{\overrightarrow{\chi}} \, d P_{\overrightarrow{\zeta}}\, = \,
\chi^X \, {\frac{\partial}{\partial q^X}} \, P_{\overrightarrow{\zeta}}\,
\end{equation}
The equivariance condition   implies
that we can introduce a triholomorphic Poisson bracket defined
as follows:
\begin{equation}
\{P_\mathbf{I}, P_\mathbf{J}\}^r \equiv 2 K^r (\mathbf{I},
\mathbf{J})
\label{hykapesce}
\end{equation}
leading to the triholomorphic Poissonian realization of the Lie
algebra:
\begin{equation}
\left \{ P_\mathbf{I}, P_\mathbf{J} \right \}^r \, = \,
f^{\mathbf{K}}_{\phantom{\mathbf{K}}\mathbf{I}\mathbf{J}} \, P_\mathbf{K}^{r}
\label{hykapescespada}
\end{equation}
which in components reads:
\begin{equation}
K^r_{XY} \, k^X_\mathbf{I} \, k^Y_\mathbf{J} \, = \, {\frac{1}{2}} \,
f^{\mathbf{K}}_{\phantom{\mathbf{K}}\mathbf{I}\mathbf{J}}\, P_\mathbf{K}^{r}
\label{hykaide}
\end{equation}
In the quaternionic case, instead, the left--hand side of
eq. (\ref{equivarianzina})
is interpreted as follows:
\begin{equation}
\overrightarrow{\chi} \circ P_{\overrightarrow{\zeta}}\, =  \, {\bf i}_{\overrightarrow{\chi}}\,  \nabla
P_{\overrightarrow{\zeta}}\, = \,
\chi^X \, {\nabla_X} \, P_{\overrightarrow{\zeta}}\,
\end{equation}
where $\nabla$ is the $\mathrm{SU(2)}$--covariant differential.
Correspondingly, the triholomorphic Poisson bracket is defined
as follows:
\begin{equation}
\{P_\mathbf{I}, P_\mathbf{J}\}^r \equiv 2 K^r (\mathbf{I},
\mathbf{J})  - { {\lambda}} \, \varepsilon^{rst} \,
P_\mathbf{I}^s  \, P_\mathbf{J}^t
\label{quatpesce}
\end{equation}
and leads to the Poissonian realization of the Lie algebra
\begin{equation}
\left \{ P_\mathbf{I}, P_\mathbf{J} \right \}^r \, = \,
f^{\mathbf{K}}_{\phantom{\mathbf{K}}\mathbf{I}\mathbf{J}} \, P_\mathbf{K}^{r}
\label{quatpescespada}
\end{equation}
which in components reads:
\begin{equation}
K^r_{XY} \, k^X_\mathbf{I} \, k^Y_\mathbf{J} \, - \,
{ \frac{\lambda}{2}} \, \varepsilon^{rst} \,
P_\mathbf{I}^s  \, P_\mathbf{J}^t\,= \,  {\frac{1}{2}} \,
f^{\mathbf{K}}_{\phantom{\mathbf{K}}\mathbf{I}\mathbf{J}}\, P_\mathbf{K}^{r}
\label{quatide}
\end{equation}
Eq. (\ref{quatide}), which is the most convenient way of
expressing equivariance in a coordinate basis, was originally written
in \cite{D'Auria:1991fj} and has played a fundamental
role in the construction of  supersymmetric actions  for gauged ${\cal N}=2$ supergravity
both in $d=4$ \cite{D'Auria:1991fj,bertolo} and in $d=5$
\cite{ceresole:2000}.
\subsection{${\cal N}=2$ gaugings and the composite connections}
\label{gunasecti}
Equipped with the crucial geometric structure provided by the triholomorphic mo\-ment-map,
let us come to the problem of gauging a general ${\cal N}=2$ matter coupled supergravity
as described by the bosonic lagrangian (\ref{genn2d5ung}). To single out
a viable gauge group we have to go through a few steps that have been derived by G\"unaydin and Zagerman
in \cite{Gunaydin:2000zx,Gunaydin:2000xk,Gunaydin:2001ph}.
\par
The first thing we have to consider is the isometry group $\mathcal{G}_{iso}$ of the
special manifold $\mathcal{SV}_n$. Later we have to see how it might be
represented on the quaternionic manifold $\mathcal{QM}$.
\par
By definition, the vectors are in the representation $\mathbf{R}$ of $\mathcal{G}_{iso}$.
What we can gauge  is any  $n_V+1$--dimensional subgroup $G_g \subset
\mathcal{G}_{iso}$ such that certain conditions are satisfied. The
conditions are:
\begin{description}
\item[a)] The following branching must be true:
\begin{equation}
  \mathbf{R} \, \stackrel{G_g}{\Longrightarrow} \,
  \mathrm{Coadj}(G_g) \oplus \mathbf{D}_S
\label{brancio}
\end{equation}
where
\item [b)]
$\mathbf{D}_S$ denotes some reducible or irreducible
\textbf{symplectic representation} of the candidate gauge group $G_g$ different from the
coadjoint. By symplectic we mean the following.
Let us decompose the range of the index $\tilde I$ as in
eq. (\ref{Lamrang}), where
\begin{equation}
  \mathbf{I}=1,\dots, n_V+1 \equiv \mbox{dim}\, G_g
\label{dimGg}
\end{equation}
runs on the coadjoint representation of the $G_g$ Lie algebra and whose
generators we denote by $T_\mathbf{I}$,  with commutation relations
\begin{equation}
  \left[ T_\mathbf{I}\, , \, T_\mathbf{J} \right] =
  f^\mathbf{K}_{\mathbf{IJ}}\,T_\mathbf{K}
\label{Ggalge}
\end{equation}
and
\begin{equation}
  M=1,\dots ,n_T=\mbox{dim}\, \mathbf{D}_S
\label{D2repre}
\end{equation}
runs on a basis of the representation $\mathbf{D}_S$.
Let $\Lambda_{\mathbf{I} M}^N$ be the matrix
representing the generator $T_\mathbf{I}$ in $ \mathbf{D}_S$:
\begin{equation}
  T_\mathbf{I} \quad \rightarrow \quad \Lambda_\mathbf{I} \quad ;
  \quad \left[ \Lambda_\mathbf{I} \, , \, \Lambda_\mathbf{J}\right] = f^{\mathbf{K}}_{\mathbf{I}\mathbf{J}}
  \, \Lambda_\mathbf{K}
\label{repregenesym}
\end{equation}
In order for the representation to be symplectic there must exist an
antisymmetric  $n_T \times n_T$ matrix
$\Omega^T=-\Omega$ that
squares to minus the identity $\Omega^2=-{\bf 1}$ and such that:
\begin{equation}
  \forall T_\mathbf{I} \in G_g \quad \Omega  \, \Lambda_\mathbf{I} +
  \Lambda_\mathbf{I}^T \, \Omega =0
\label{omegazer}
\end{equation}
Indeed, this ensures that our algebra $G_g$ is a subalgebra of the
symplectic algebra $\mathrm{Sp}(n_T,\mathbb{R})$.
\item [c)]
The $C_{\tilde I \tilde J \tilde K}$ invariant tensor must
decompose under $G_g$ in the following way:
\begin{equation}
  C_{\tilde I \tilde J \tilde K} =\left\{ \begin{array}{rcl}
    C_{\mathbf{IJK}} & = & \mbox{Invariant tensor in the $\mathrm{Coadj}(G_g)$}  \\
    C_{MNP} & = & 0 \\
    C_{M \mathbf{IJ}} & = & 0 \\
    C_{MN \mathbf{I}} &  = & -\frac{\sqrt 6}{2} \Omega_{MP} \,
    \Lambda^P_{\mathbf{I}N}  \
  \end{array} \right.
\label{ergovit}
\end{equation}
\item [d)]  The group $G_g$, selected through the previous restriction,
  must act as a triholomorphic isometry on the quaternionic manifold
  $\mathcal{QM}$ \footnote{This last requirement is that spelled out by Ceresole and Dall'Agata in \cite{ceresole:2000}
  who have extended to the $d=5$ case the methods and procedures of the geometrical gaugings originally introduced
  in \cite{dewit1,Castellani:1986ka,D'Auria:1991fj,bertolo}.}.
\end{description}
The rationale for the above requirements is the following.
The reason for the requirement \textbf{a)}   is the same
as in four--dimensions. Since  the gauge vectors are by definition
in the coadjoint of the gauge algebra, it is necessary that the
representation to which the vectors are pre--assigned should contain
the coadjoint of what we want to gauge. Note also that for
semisimple groups,  adjoint and coadjoint representations are equivalent
but this is no longer true in the case of non semisimple gauge
algebras. An extreme possibility is provided by abelian algebras
\begin{equation}
  \mathcal{A}=\mathrm{U(1)}^{\ell} \otimes \mathbb{R}^m
\label{abelalg}
\end{equation}
where we were careful to distinguish compact from non compact
generators. In this case the coadjoint representation vanishes and
any set of $\ell +m$ vectors can be used to gauge an algebra $\mathcal{A}$
that has vanishing action on the very special manifold
$\mathcal{SV}_n$.
The rationale for the requirements \textbf{b)} and \textbf{c)}
is instead related to the consistent coupling of massive $2$--forms.
Gauge vectors that are in non trivial representations of a gauge
group different from the coadjoint representation are inconsistent
with their own gauge invariance. To cure this problem we have to
dualise them to massive self--dual $2$--forms, satisfying:
\begin{equation}
B^{M |\mu\nu} = m
\epsilon^{\mu\nu\rho\sigma\lambda}{\cal{D}}_\rho B^M_{\sigma\lambda}\,.
\label{selfconst}
\end{equation}
where the covariant derivative is :
\begin{equation}
  \mathcal{D}B^M \equiv dB^M
  + g  \Lambda^M_{\mathbf{I}N} A^\mathbf{I} \,
  \wedge \,
  B^N \equiv \mathcal{H}^M
\label{covderB}
\end{equation}
This is possible only if the part of
the Chern Simons term involving two $B$'s and one $A$
can be reabsorbed in  the kinetic term of the $2$--forms  that reads
as follows:
\begin{equation}
  \mathcal{L}^{kin}_{B} = \frac{1}{4g} \, \epsilon^{\mu \nu \rho \sigma \lambda
  } B^M_{\mu \nu } \,\mathcal{D}_\rho  B^N_{\sigma \lambda
  } \, \Omega_{MN}
\label{Bkinterm}
\end{equation}
\par
The last requirement \textbf{(d)} deals with the possible presence of
hypermultiplets. In particular, the action of $G_g$ on the quaternionic manifold
can be the identity action, which is certainly triholomorphic.
In this case the hypermultiplets are simply neutral with respect to the gauge group.
Alternatively, we can consider an abelian algebra, as in eq. (\ref{abelalg}),
that has no action on the very special manifold but acts by  non
trivial triholomorphic isometries on the quaternionic manifold.
Both of these are extreme cases that allow more freedom of choice
for one of the two manifolds. The general case corresponds to a
choice of $G_g$ that acts non trivially both on $\mathcal{SV}_n$ and
$\mathcal{QM}_m$ and respects conditions \textbf{a)-d)}.
\par
Assuming that the gauge algebra has been selected and satisfies the
above criteria, the gauging procedure becomes smooth and fully
parallel to the four--dimensional models we have already discussed.
The essential point is always the same, namely the \textbf{gauging} of the
\textbf{scalar vielbein} and of the \textbf{composite connections} acting on
the fermion fields. In the case of
maximal supersymmetry, where the scalar manifold is necessarily a
homogeneous space $\mathcal{G}/\mathcal{H}$, these two gaugings are
obtained in one stroke by gauging the Maurer--Cartan  $1$--forms. 
In the non maximal case we have
to do it separately and specifically for the different factors
occurring in the scalar manifold. These latter are not necessarily
coset manifolds but have  a sufficiently \emph{special} geometric
structure to allow the generic construction of those ingredients that
are necessary for the gauging of the composite connections, most
relevant being the role of the triholomorphic moment-map.
\par
Let us begin with the gauging of the scalar vielbein. This is
equivalent to replacing the ordinary derivatives (or differentials) of
the scalar fields with covariant ones, as follows:
\begin{equation}
  \begin{array}{rcl}
 \mathcal{D}\phi^{\tilde x} &=& d \phi^{\tilde x} + g A^\mathbf{I} k_\mathbf{I}^{\tilde x} (\phi)\\
\mathcal{D}q^{X} &=& d q^{X} + g A^\mathbf{I} k_\mathbf{I}^{X} (q)\\
\end{array}
\label{covdiffe}
\end{equation}
where $g$ is the gauge coupling constant and $k^{\tilde x}_\mathbf{I}(\phi),k_\mathbf{I}^X(q)$
are the killing vectors expressing the action of the gauge algebra generators
$T_\mathbf{I}$ on the two scalar manifolds:
\begin{eqnarray}
\delta_I \phi^{\tilde x} & = &  k_\mathbf{I}^{\tilde x} (\phi) \nonumber\\
\delta_Iq^{X} & = & k_\mathbf{I}^{X} (q)
\label{deltafiq}
\end{eqnarray}
Next we have the gauging of the composite connections. There are three of
them corresponding to the three vector bundles of which the fermions are
sections:
\begin{enumerate}
\item The Levi--Civita connection $ \Gamma^{\tilde x}_{\phantom{\tilde x}\tilde y }$ on the
tangent bundle to the very special manifold $T\mathcal{SV}_n$. This
enters because the gauginos $\lambda^{\tilde x}_i$ carry a world index of the
very special manifold, namely are sections of $T\mathcal{SV}_n$.
\item The $\mathrm{Sp}(2n_H,\mathbb{R})$ connection
$\Delta^{AB}$. This enters because the hyperinos
$\zeta^A$ are sections of the $\mathrm{Sp}(2n_H,\mathbb{R})$
bundle over the quaternionic manifold $\mathcal{QM}$. By definition
this latter has reduced holonomy, so that the structural group of the
tangent bundle $T\mathcal{QM}$ is $\mathrm{SU(2)}\times
\mathrm{Sp}(2n_H,\mathbb{R})$, as we know from section \ref{hypgeosec}.
\item The $\mathrm{SU_R(2)}$ connection of $R$--symmetry $p^{ij}$.
  This connection enters the game because both the
gravitino $\psi_i$ and the gauginos $\lambda^{\tilde x}_i$ are sections of the
$SU(2)_R$ vector bundle in the fundamental doublet representation
(the index $i=1,2$ denotes this fact). On the other hand the
$R$--symmetry bundle is identified with the
$\mathcal{SU} \rightarrow \mathcal{QM}$ bundle over the quaternionic
manifold and this means that the connection $p^{ij}$ is the
connection of the $\mathcal{SU}$ bundle described in section \ref{hypgeosec}.
\end{enumerate}
In terms of the Killing vectors and of the
the triholomorphic moment map $\mathcal{P}_\mathbf{I}^x(q) $
and just following the original recipe developed in
\cite{D'Auria:1991fj} and further clarified in \cite{bertolo}
the gauging of the connections is given by:
\begin{equation}
\begin{array}{cccccc}
{ T\mathcal{VS}} & : & \mbox{tangent bundle} &
 \Gamma^{\tilde x}_{\phantom{\tilde x}\tilde y}& \to &{\hat \Gamma}^{\tilde x}_{\phantom{\tilde x}\tilde y} =
 \Gamma^{\tilde x}_{\phantom{\tilde x}\tilde y} +
 g\, A^\mathbf{I}\, \partial_{\tilde y} k^{\tilde x}_\mathbf{I} \cr
{\cal SU} & : & \mbox{$\mathrm{SU(2)}$ bundle} &
p^r &\to &{\hat p}^r = p^r + g_R\, A^\mathbf{I}\, P^r_\mathbf{I} \cr
{\cal SU}^{-1}\otimes{ T\mathcal{QM}} & : & \mbox{$\mathrm{Sp}(2m,\mathbb{R})$ bundle} &
\Delta^{AB} &\to &{\hat  \Delta}^{AB}=
\Delta^{AB}  + g\, A^\mathbf{I}\,
 \partial_X k_\mathbf{I}^Y \, f^{X \vert i A}
 \, f^B_{Y \vert i} \cr
 \end{array}
\label{compogauging}
\end{equation}
where $g$ is the same gauge coupling constant as in
eq. (\ref{covdiffe}) while $g_R$ is an additional coupling constant
that allows to gauge or not to gauge the $R$-symmetry group $SU(2)$.
In the construction of the lagrangian and in checking the closure of
the supersymmetry algebra it turns out that $g$ and $g_R$ are
independent parameters \cite{ceresole:2000}.
\subsection{The Fermion shifts and gravitino mass-matrix}
Gauging the connections forces, through closure of the supersymmetry
algebra, the inclusion of new non--derivative terms in the \emph{susy}
rules of the fermions that are completely analogous to their
$4$--dimensional counterparts of eq.s (\ref{gravsusy}) and
(\ref{fermioshif}). Indeed as explained in sect.\ref{genaspect} the
gauging procedure of supergravity theories fits into a general and
uniform pattern for all space--time dimensions $d$ and for all number of supersymmetry
charges $N_Q$. The gravitino transformation rule (\ref{gensusrul})
becomes: 
\begin{equation}
\delta \psi_{i\mu} = \mathcal{D}_\mu \, \epsilon_i + i \frac{1}{4\sqrt 6} \,
  \mathcal{T}_{ij}^{\rho \sigma } \left( 4 g_{\mu \rho } \, \gamma _\sigma
  - \, \gamma_{\mu\rho \sigma }   \right) \, \epsilon ^j \, 
+ \,S_{ij} \, \epsilon ^j
\label{gaususrul}
\end{equation}
where the gravitino mass matrix is given by: 
\begin{equation}
  S_{ij} = \mbox{i}\,g_R \, \frac{1}{\sqrt 6} \, h^\mathbf{I} (\phi) \,
  P^r_\mathbf{I}(q) \, \left( \sigma_r \right)
  _i^{\phantom{i}k} \, \epsilon_{jk}
\label{ginmamatn2d5}
\end{equation}
while the transformation rules of the spin 1/2 fermions have been
determined by Ceresole and Dall'Agata  (see \cite{ceresole:2000}) to have, apart from some trivial choice of normalizations, an identical form to their
counterparts in $d=4$ ${\cal N}=2$ supergravity (see
\cite{D'Auria:1991fj,bertolo}).
Indeed one finds:
\begin{eqnarray}
\delta \zeta^A  & = & \mbox{derivative terms} \, + \,  \Sigma^{A \vert i} \,
\epsilon_i \nonumber\\
\delta \lambda^{\tilde x i}  & = & \mbox{derivative terms} \, + \,\Sigma^{\tilde x i \vert j}  \, \epsilon_j
\label{ceregashif}
\end{eqnarray}
where the fermion shifts take the following explicit form:
\begin{eqnarray}
\Sigma^{\tilde x i \vert j}&=& g \epsilon^{ij} W^{\tilde x} + g_R W^{\tilde x ij}\nonumber\\
W^{\tilde x} &=&\frac{\sqrt 6}{4}\,k_{\mathbf{I}}^{\tilde x}  h^\mathbf{I}\\
W^{\tilde x ij}&=&{\rm i}(\sigma_r)_{k}^{\phantom{k}j} \epsilon^{ki} P^r_{\mathbf{I}} g^{\tilde x \tilde y} h_{\tilde y}^{\mathbf{I}}\nonumber\\
\Sigma^{A \vert i} &=& g\, \epsilon ^{ij} \,\mathbf{N}_j^A \nonumber\\
N_j^A &=& - \frac{\sqrt 6}{4} \,f_{X A}^j \,k^X_{\mathbf{I}}\,h^{\mathbf{I}}
\label{pesamatrice}
\end{eqnarray}
Indeed, if one compares eq.s (\ref{ginmamatn2d5}),(\ref{pesamatrice}) with their
$4$--dimensional counterparts given in eq.s (8.23) of \cite{bertolo}
one sees that (apart from the overall normalization which can be
reabsorbed into the normalization of the corresponding fermionic
field)  the two sets of formulae are identical upon the
replacement of the complex section $L^\Lambda(z)$ of \emph{special K\"ahler
geometry} with the real section $h^\mathbf{I}(\phi)$ of \emph{very special
geometry}. The other noteworthy difference is that in $d=4$ the index
$\Lambda$ runs over the whole set of $n+1$ values, $n$ being the
dimension of the special K\"ahler manifold. In five dimensions,
instead, the index $ \mathbf{I}$ runs over the $n_V+1$ subset of values
corresponding to the gauged vectors while the total dimension of the
very special space is $n_V+n_T$. The remaining $n_T$ dimensions are,
as we know, associated with the massive self-dual $2$--forms.
\par
In five as in all other dimensions supersymmetry imposes a Ward
identity that is the straightforward generalization of
eq. (\ref{wardide}), namely:
\begin{equation}
  \alpha \, S^{ij} \, S_{jk} \, - \, \beta \, K_{\tilde x \tilde y} \Sigma^{\tilde x\vert i} \Sigma^{\tilde y \vert j} \, \epsilon_{jk} \, = \, - \,\delta^i_k \, \mathcal{V}
\label{d5wardide}
\end{equation}
where $K_{\tilde x \tilde y}$ is the kinetic matrix of the spin 1/2 fermions and $\alpha$ and $\beta$ are just numerical coefficients that differ from
their analogues in $4$--dimensions only because of the differences in
Lorentz algebra and $\gamma$-matrix manipulations. Verifying such an
identity  whose explicit form is not written down in their paper,
the authors of \cite{ceresole:2000} have proved the supersymmetry of the
gauged action and calculated the final form of the potential that
reads as follows:
\begin{equation}
  \mathcal{V}=  \, \frac 34 g^2  \left[   h^\mathbf{I} \,h^\mathbf{J}
  \, \left( k_\mathbf{I}^{\tilde x} k_\mathbf{J}^{\tilde y} \, g_{\tilde x \tilde y} + k_\mathbf{I}^X
  \,
  k_\mathbf{J}^Y \, g_{XY} \right) \right] - \,  g_R^2 \left[ \left(
  2 \, h^\mathbf{I} \, h^\mathbf{J} \, -\,g^{\tilde x \tilde y}
  h_{\tilde x}^\mathbf{I} \, h_{\tilde y}^\mathbf{J}\right) \, P^r_\mathbf{I} \, P^r_\mathbf{J}\right]
\label{ceregatpot}
\end{equation}
in terms of the moment-map (\ref{2.76}) and of the section
$h^\mathbf{I}$ of very special geometry and its derivative
$h^\mathbf{I}_{\tilde x} =- \sqrt{ \frac 32}\partial_{\tilde x} h^\mathbf{I}$ (see eq.s (\ref{fidefi})).
\subsection{The scalar potential and supersymmetry
breaking}
We can now summarize the results of the previous section writing the general
form of the  bosonic lagrangian for a general
\textbf{gauged} ${\cal N}=2,d=5$ supergravity. The ungauged action
(\ref{genn2d5ung}) is replaced by the following gauged one:
\begin{eqnarray}
  \mathcal{L}^{(gauged)}_{(d=5,{\cal N}=2)}&=& \frac 12 \sqrt{-g} \, \left( \,R \, - \, \frac 12 \,a_{\tilde I\tilde J} (\phi)  F^{\tilde I}_{\mu \nu
  } \, F^{\tilde J \vert\mu \nu} \right.\nonumber\\
  &&\left. 
+  \, g_{\tilde x \tilde y}(\phi) \, D_\mu\phi^{\tilde x} \, D^\mu \, \phi^{\tilde y}
  + \, g_{XY}(q) \, D_\mu q^X \, D^\mu \, q^Y - 2 \mathcal{V}(\phi,q)\right ) \nonumber\\
  &&+  \left(\, \frac{1}{6\sqrt 6} C_{\mathbf{IJK}} \,
   \, F^\mathbf{I}_{\mu \nu } \, F^\mathbf{J}_{\rho \sigma } \, A^\mathbf{K}_\tau
  + \frac{1}{4g}\,\Omega_{\mathcal{MN}} \, B^\mathcal{M}_{\mu \nu } \,
  \mathcal{D}_\rho
  \, B^N_{\sigma \tau } \,\right) \, \epsilon ^{\mu \nu \rho \sigma \tau}\cr
& &\label{genn2d5gau}
\end{eqnarray}
where the potential $\mathcal{V}(\phi,q)$ is that given in
eq.  (\ref{ceregatpot}).
\paragraph{General pattern of supersymmetry breaking in $d=5$}
Following the general discussion of eq.s (\ref{genbreakpat})
  in ${\cal N}$-extended $d=5$ supergravity, a conventional vacuum configuration
$\phi_0$ that preserves $\mathcal{N}_0$
supersymmetries is characterized by the existence of $\mathcal{N}_0$
vectors $\rho^i_{(\ell)}$ ($\ell=1,\ldots,\mathcal{N}_0$) of
$\mathrm{USp}({\cal N})$, such that
\begin{eqnarray}
\label{d5breakpat} S_{ij} \left( \phi_0 \right) \, \rho^j_{(\ell)} & =
& e^{\rm i\theta}\, \sqrt{\ft{|\mathcal{V}(\phi_0)|}{12}} \, \rho_{A(\ell)}
~,\nonumber\\ \Sigma_i^{\phantom{i}\tilde x} \left( \phi_0 \right) \,
\rho^i_{(\ell)} & = & 0~,
\end{eqnarray}
where $\theta$ is an irrelevant phase, $\mathcal{V}$ is the scalar
potential and $S_{ij}$ is the gravitino mass-matrix, uniformly defined
for all ${\cal N}$ by eq. (\ref{gaususrul}).


\subsubsection{Properties of the ${\cal N}=2$ potential and anti de Sitter vacua}
At this point we finally discuss the properties of the scalar potential and which type of vacua it admits. In fact a great part of the interest in gauged supergravity with matter couplings is due to existence of a ``tunable'' potential with the choice of the gauging. This feature is fundamental to study phenomenological applications as Randall--Sundrum scenarios or to explore possible extensions of the gravity/gauge field correspondence.

In view of these facts it is specifically interesting to survey the conditions for the existence of anti de Sitter vacua. According to our general discussion, following eq. (\ref{trecass}), we have anti de Sitter vacua if  $\mathcal{V}(q_0, \phi_0) < 0$ for $\mathcal{V}^{\prime} (q_0, \phi_0) = 0$. Thus it is  straightforward to see that the only contribution which can allow for such solutions is the term\footnote{Without loosing in generality we take $g=g_R$ for all the following part of the thesis.}
\begin{equation}\mathcal{V} =  - \, g^2 \, P^r \, P_r + \mbox{positive contributions} \label{pertica}
\end{equation}
coming from the $R$--symmetry gauging of the gravitinos. We have introduced the compact notation $P^r \equiv P_I^r h^I$ (see app.\ref{convention}). Indeed this is the only negative contribution to the potential. This implies that a simple Yang--Mills gauging, even in presence of both tensor and hypermultiplets, does not allow any anti de Sitter solution.
\par 
We can briefly analyze various cases.
\begin{description}
  \item[a)] If we set $n_H=0$ there are no hypermultiplets and the
  quaternionic manifold disappears. Correspondingly,
  as already noted in \cite{D'Auria:1991fj,bertolo} for the
  $4$-dimensional case, the killing vector $k_\mathbf{I}^X$ is zero
  while the triholomorphic moment maps are $\mathrm{SU(2)}$ Lie algebra valued
constants $\xi^r_\mathbf{I}$ that, because of eq. (\ref{quatide}), must satisfy the
condition:
\begin{equation}
  g \, f^\mathbf{K}_{\mathbf{IJ}} \, \xi_\mathbf{K}^r = \ft 1 2
  \, g_R \, \epsilon^{rst} \, \xi_\mathbf{I}^s \, \xi_\mathbf{J}^t
\label{identitaxi}
\end{equation}
Generically the $\xi_\mathbf{I}^r$ break $\mathrm{SU(2)} \to
\mathrm{U(1)}$. If the gauge group
$\mathcal{G}$ contains a subgroup $H\equiv SU(2)$, this can be identified
with the R--symmetry group setting $\xi^r_\mathbf{\hat I} =
\delta_\mathbf{\hat I}^r$, $\xi^r_\mathbf{\hat I}\in H$.
  \item[a1)] If at $n_H=0$ one makes the choice
$\xi_\mathbf{I} = (0, V_\mathbf{I}, 0)$, the condition (\ref{identitaxi})
reduces to
\begin{equation}
f^\mathbf{K}_{\mathbf{IJ}} V_\mathbf{K} = 0
\label{pirlon}
\end{equation}
As already noted in \cite{D'Auria:1991fj,bertolo} for the four--dimensional case
this is the Fayet--Iliopoulos
phenomenon which corresponds, in mathematical language to the
possibility of lifting the moment-maps to a non zero level for all
the generators belonging to the center of the gauge Lie algebra.
\item[b)] If we both set $n_H=0$, namely we include no hypermultiplet
but we also set $n_T=0$ namely we consider only vector fields in the
coadjoint representation of the gauge group (i.e. the symplectic
representation $\mathbf{D}_S$ of the massive two forms is deleted),
then one can easily prove that
\begin{equation}
  k^{\tilde x}\equiv h^\mathbf{I} \, k^{\tilde x}_\mathbf{I}=0
\label{ortogcond}
\end{equation}
This implies that the scalar potential \ref{ceregatpot} reduces to:
\begin{eqnarray}
  \mathcal{V} & = &-6 g^2  \left[ W^2 \, - \, \frac 34 g^{xy}
  \partial_x W \, \partial_y W \right ]\label{townsendo}\\
  W(\phi) & \equiv & \sqrt{\frac 23} V_\mathbf{I} h^\mathbf{I}(\phi)
\label{superpotential}
\end{eqnarray}
the constant coefficients $V_\mathbf{I}$ being those introduced above
and satisfying the consistency condition (\ref{pirlon}). The
interesting thing about the potential \\(\ref{townsendo}) is that it
follows from the general class of potentials of the form:
\begin{equation}
  \mathcal{V}  =  -  \alpha^2 \left[ (d-1) \, W^2 \, - \, (d-2) g^{\Lambda \Sigma}
  \partial_\Lambda W \, \partial_\Sigma W \right ]\label{townskend}
\end{equation}
where $W(\varphi)$ is a real function named the \emph{superpotential},
$d$ denotes the space--time dimensions and $g^{xy}$ is the positive
definite kinetic metric of the scalar fields. In \cite{townsupot}
Townsend has shown that the structure (\ref{townskend}) is precisely
that required for vacuum stability. We note that in the presence of an arbitrary number of hypermultiplets and vector multiplets to require this form for the potential  implies a condition on the phase of the prepotential $Q^r$ (see app.\ref{convention}). We will show  in chapter \ref{vet} that for the electrostatic spherically symmetric configurations, such condition is recovered as a consequence of the BPS requirement. The same happens for the flat domain wall solutions as observed in \cite{ceresole:2001}.

\item[c)] If we set $n_V+n_T=0$ there are no vector multiplets and
we have simply hypermultiplets. Then $h^0=1$ and there is just one gauge vector: the graviphoton whose action on the quaternionic manifold is
described by the triholomorphic Killing vector $k_\mathbf{0}^X$.
The potential is still non--zero and becomes
\begin{equation}
\mathcal{V} = \,\frac 32g^2 \, k_\mathbf{0}^X \, k_\mathbf{0}^X \, g_{XY}  -
\,4 g^2 \, P^r_\mathbf{0} \, P_{\mathbf{0}r}
  \label{novectors}
\end{equation}
which in principle can admit anti de Sitter vacua.
\item [d)] The minimal gauged  $5$--dimensional supergravity is retrieved as a
subcase of the above,  setting also the number of hypermultiplets to zero, with the scalar potential reducing to a cosmological term.
\end{description}
The topics described in this chapter will find a concrete application in the study of BPS solutions.

\chapter{Electrostatic spherically symmetric BPS solutions with Hypermultiplets}\label{black}

\section{Introduction}
As has been already underlined in the introductory chapter,
in recent years five-dimensional ${\cal N}=2$ gauged supergravity theories have received considerable attention primarily for their relevance  to the AdS/CFT correspondence \cite{maldapasto},\cite{ads}. In particular much 
interest has been directed toward the study of domain-wall supergravity solutions \cite{Behrndt:2000kz},\cite{flat-domainwall}, 
\cite{cardoso:2002},\cite{berhndt:2002},\cite{curved-domainwall} as duals of renormalization group (RG) flows in the corresponding  field theory 
\cite{RGF},\cite{stability}. Also a  strong motivation in this direction derives from phenomenological requests in 
brane-world scenarios obtained via M-theory compactifications and/or domain-wall type models \cite{RS2},\cite{RS1},\cite{phenom}.

Finding supersymmetric solutions of ${\cal N}=2$, $D=5$ supergravities is never an easy exercise \cite{Gauntlett:2002nw}; it 
becomes a quite difficult task if one considers general couplings to matter and general gaugings. Partial 
results have been obtained so far, i.e. for cases where only special vector or hypermultiplet gaugings 
have been considered \cite{partialsolut}, \cite{gutperle:2001}. Here we start a systematic program with the general aim to classify BPS solutions 
with vector, tensor and hypermultiplet couplings. 
The introduction of the hypermultiplets is crucial for widening the variety of solutions as compared to the case where only vector multiplets are present \cite{ceresole:2001},\cite{renandrei},\cite{Behrndt:2000tr}. In particular, 
aside for the special example analyzed in \cite{gutperle:2001},
the existence of BPS  black-hole  solutions has not been investigated systematically before. In this chapter I address this task, reporting on and discussing the results of my work with S. Cacciatori and D. Zanon \cite{noialtri}. Of course, black-hole solutions are especially relevant since, via the AdS/CFT correspondence, they could describe the RG flows between field theories in different dimensions \cite{maldacena:2000}.   

To treat the problem we restrict ourselves to the case of 
hypermultiplet couplings, with generic gauging, and a static $SO(4)$ symmetric ansatz for the metric. 
In this setting we study the 
integrability conditions that follow from the BPS equations and find a set of equations
for the functions in the ansatz. The quaternionic geometries give equations for the scalars  which 
are a generalization of the ones found in \cite{ceresole:2001}.
Then we analyze all these equations and check directly that they 
satisfy the equations of motion.  

The presentation is organized as follows: in the next section we introduce the model specializing the general formula present in the previous chapter. 
 We describe  the form of the solutions we are looking for: obviously this choice determines the {\em physics} contained in the solution. 
Then we focus on the derivation of 
the BPS equations and study their 
integrability conditions.
In section \ref{blackhole}
we find the set of independent first order differential equations that are equivalent to the BPS 
conditions. In section \ref{equamotion} we show that the family of solutions we have found satisfy 
the equations of motion. In section \ref{solution} we discuss the properties of the BPS solutions 
in the special case of the universal hypermultiplet \cite{witten:1983} and find an explicit result 
for a simple choice of the gauging.
 We conclude with some final remarks.
Our notations and conventions, which are the same as in \cite{noialtri}, are summarized in Appendix \ref{convention}. 

\section{The model and its BPS equations}\label{model}
We consider $\cal{N}$=2 gauged supergravities in five dimensions interacting with an arbitrary number of 
hypermultiplets (we postpone the study of vector multiplet coup\-lings to the next chapter.). 
The field content of 
the theory is the following
\begin{itemize}
\item the supergravity multiplet \eqn \{ e^a_\mu \ , \psi^{i}_\mu \ , A_\mu \} \feqn containing the 
{\em graviton} $e^a_\mu$, two {\em gravitini} $\psi^{i}_\mu$ and the {\em graviphoton} $A_\mu$,
which is the only (abelian) gauge field present in the theory; \item $n_H$ hypermultiplets 
\eqn \{ \zeta^A \ , q^X \} \feqn containing the {\em hyperini} $\zeta^A $ with ${\scriptstyle A}=1,2,\dots,2 n_H$, and the {\em scalars} $q^X$ with ${\scriptstyle X}=1,2,\dots,4 n_H$ which define a quaternionic K\"ahler manifold (see section \ref{hypgeosec})  with 
metric $g_{XY}$.
\end{itemize}

 \noindent The bosonic sector of the theory is described by the Lagrangian density presented in
\cite{ceresole:2000}
\eqn \label{lagr}        
&&{\cal {L}}_{BOS} =\ - \ \frac 12 e \left[ R+\frac 12 F_{\mu \nu} F^{\mu \nu} 
                  +g_{XY} D_\mu q^X D^\mu q^Y \right] \cr
&&~~~~~~~~~~~~
+ \frac 1{6\sqrt 6} \epsilon^{\mu\nu\rho\sigma\tau} F_{\mu\nu}           F_{\rho\sigma}A_{\tau} 
- e {\cal V} (q)  
\feqn
with 
$$
D_\mu q^X=\partial_{\mu} q^X + g A_\mu K^X(q)
$$
where $K^X(q)$ is a Killing vector on the quaternionic manifold and ${\cal V} (q)$ is the scalar potential as given in Appendix \ref{convention}.

We look for electrostatic spherical solutions that preserve half of the ${\cal N}=2$ supersymmetries. 
To this end we make the following ansatz for the supergravity fields: we choose a metric which is $SO(4)$ symmetric with all the other fields depending on the holographic space-time coordinate $r$ only. Moreover we fix the gauge for the graviphoton keeping only the $A_t$ component different from zero. 
 
Introducing  spherical coordinates $(t,r,\theta ,\phi, \psi)$ we write 
\cite{gutperle:2001}
 \eqn \label{metricansatz} ds^2 =-e^{2v} dt^2 +e^{2w}  dr^2 + r^2 
(d\theta^2 +\sin^2 \theta d\phi^2 +\cos^2 \theta d\psi^2 ) \feqn 
where the functions $v$ and $w$ depend on $r$ only. The variations of the fermionic fields under supersymmetry 
transformations give rise to the following BPS equations: for the gravitini we have 
\cite{vanproeyen:2001} 
\eqn 0 \ = \ \delta_\epsilon \psi_{\mu i} &=& \partial_\mu \epsilon_i +\frac 
14 \omega^{ab}_\mu \gamma_{ab} \epsilon_i -\partial_\mu q^X p_{Xi}^{\quad j} \epsilon_j -  g 
A_{\mu} P_{i}^{\ j} \epsilon_j \cr &+&\frac i{4 \sqrt 6} (\gamma_{\mu \nu \rho} -4g_{\mu \nu} 
\gamma_\rho )F^{\nu \rho } \epsilon_i - \frac i{\sqrt 6} g  P_{i}^{\ j} \gamma_\mu \epsilon_j  
\label{BPSgrav} \feqn 
We note that in ref.\cite{gutperle:2001}  the corresponding equation contains
an additional term. This extra term, which arises due to a incorrect interpretation of the covariant 
derivative acting on the spinor
$\epsilon$ as given in \cite{ceresole:2000},  should not be present. 

For the hyperini the equations $ \delta_\epsilon \zeta^A=0$ lead to
 \eqn \left[\frac i2 e^{-w} f^{Ai}_X 
q^{\prime X} \gamma_1 - i \frac g2 f^{Ai}_X K^X e^{-v} A_t \gamma_0 \right] \epsilon_i &=&
 \frac {\sqrt 6}4 g K^X f^{Ai}_X \epsilon_i
\label{BPShyper}  \feqn where we have set $q^{\prime X} =\partial_r q^X$

Without loss of generality it is convenient to parametrize the graviphoton as follows \footnote{Note that in \cite{gutperle:2001}  $a$ was chosen to be a constant.}
\eqn \label{elettrico} A_t =\sqrt {\frac 32} a(r)e^v \feqn 
This allows to write explicitly  the BPS equations for the gravitini 
\eqn & & \left\{
\partial_t \delta_l^{\ k} +\frac 12 v^\prime e^{v-w} \delta_l^{\
k} \gamma_0 \gamma_1 +\frac 1{\sqrt 6} g e^v P^r (\sigma_r )_l^{\
k} \gamma_0 +\frac i2 e^{v-w} (v^\prime a+ a^\prime)  \delta_l^{\ k}
\gamma_1 \right. \cr
& &~~~~~~~~\left. -i\sqrt{\frac 32}g a e^v  P^r  (\sigma_r )_l^{\ k} \right\}
\epsilon_k =0 \label{tBPSgrav} \\
&&~~~\cr
& & \left[ \partial_r \delta_i^{\ j} - iq^{\prime X} p_X^r (\sigma_r )_i^{\ j}
+\frac 12 (v^\prime a + a^\prime) (i\gamma_0 )\delta_i^{\ j}
+\frac {g} {\sqrt 6} e^w P^r \gamma_1 (\sigma_r )_i^{\ j} \right] \epsilon_j =0
\cr
&&\label{rBPSgrav} \\
&&~~~\cr
&&~~~\cr
& & \left[
\delta_i^{\ j} \partial_\theta -\frac 12 e^{-w}
\gamma_1 \gamma_2 \delta_i^{\ j} +i\frac {re^{-w}}4 (v^\prime a + a^\prime)
\gamma_{012}
\delta_i^{\ j} 
 +\frac {g r}{\sqrt 6} P^s (\sigma_s)_i^{\ j} \gamma_2
\right] \epsilon_j =0 \cr
&&\label{teta}  \\
&&~~~\cr
& & \left\{
\delta_i^{\ j} \left[ \partial_\phi -\frac 12 e^{-w}
\sin \theta \gamma_1 \gamma_3 -\frac 12 \cos \theta \gamma_2 \gamma_3
+i\frac {re^{-w}}4 (v^\prime a + a^\prime) \sin \theta \gamma_{013} \right] \right.
\cr
& & ~~~~~~~~~~~\left. +\frac {g r \sin \theta}{\sqrt 6} P^s (\sigma_s)_i^{\ j}\gamma_3 \right\} \epsilon_j =0 \label{fi}    \\
&&~~~\cr
&&~~~\cr
& & \left\{ \delta_i^{\ j} \left[ \partial_\psi -\frac 12 e^{-w} \cos \theta \gamma_1 
\gamma_4 +\frac 12 \sin \theta \gamma_2 \gamma_4 +i\frac {re^{-w}}4 (v^\prime a + a^\prime) \cos \theta 
\gamma_{014} \right] \right.  \cr 
& & ~~~~~~~~~~~\left. +\frac {g r \cos \theta}{\sqrt 6} P^s (\sigma_s)_i^{\ j} 
\gamma_4 \right\} \epsilon_j =0 \label{psi} \feqn 

At this point, using (\ref{fquadrato}) and the $SU(2)$ projection as in (\ref{pro}), we can rewrite the 
algebraic relations in (\ref{BPShyper})  as
\eqn  
&&\left( i e^{-w} q^{\prime X} \gamma_1 -i g\sqrt{\frac 32} a K^X
\gamma_0 - \sqrt{\frac 32} g K^X \right) \left(g_{ZX} \delta_j^{\ i} +
2 i R_{ZX}^s {(\sigma_s)}_j^{\ i}\right) \epsilon_i =0\cr
&&\label{1BPShyper}
\feqn
We will make use of the above expression in the following.

\subsection{Integrability conditions}

Now we want to discuss the integrability of the gravitini equations in order to ensure the existence of a 
Killing spinor (i.e of residual supersymmetry). The standard procedure is to impose the vanishing of the 
various commutators. In this way one obtains equations that combined with the hyperini ones determine the 
unknown functions in the ansatz and impose restrictions on the geometry (gauging). We find it useful to 
adopt the following notation: given the vector $P^s$  $s=1,2,3$ we introduce the phase $\vec Q$ 
so that $\vec Q \cdot \vec Q =1$ and use the decomposition of the vector into its norm and phase
\eqn
\vec P =\sqrt {\frac 32} W \vec Q
\ .
\label{1scomposizione}
\feqn
We find four independent\footnote{Symmetry arguments show that the angular 
equations lead to only one independent condition.} integrability conditions that we list below:\\
from the commutators between the BPS equations (\ref{teta}), (\ref{fi}) and (\ref{psi}) we obtain
\eqn
& &
\left\{ i\gamma_0 \left[ 1-e^{-2w}
 -\frac {r^2 e^{-2w}}4 (v^\prime a + a^\prime)^2  +{g}^2
 r^2 W^2 \right] \delta_i^{\ j} \right. \cr
& & \Bigl.
+\bigl[ re^{-2w}\left(v^\prime a + a^\prime\right) \bigr]
\delta_i^{\ j} +\gamma_1 \bigl[ g e^{-w} r^2  \left(v^\prime a + a^\prime\right)
W Q^s \bigr] (\sigma_s)_i^{\ j} \Bigr\} \epsilon_j =0\cr
&&~~~~~
\label{tetafi}
\feqn
The commutators between (\ref{tBPSgrav}) and
the angular components give the conditions
\eqn
& & \Bigl\{
\gamma_0 \left[ v^\prime e^{-w} -{g}^2 e^{w} r
W^2 \right] \delta_i^{\ j} -i g r (v^\prime a+ a^\prime) WQ^s
(\sigma_s)_i^{\ j} \gamma_1 
 -i(v^\prime a+ a^\prime) e^{-w} \delta_i^{\ j}
 \Bigr\} \epsilon_j =0 \cr
&&~~~~~
\label{tteta}
\feqn
The commutators between (\ref{rBPSgrav}) and the angular 
components give
\eqn
& & \left\{ -\frac i2 g (v^\prime a+ a^\prime) 
  rW Q^s (\sigma_s)_i^{\ j} \gamma_0 +\left[ \frac 12
-w^\prime e^{-w}  -\frac {{g}^2}2 e^{w} rW^2
\right] \gamma_1 \delta_i^{\ j} \right. \cr
& & \left. +\frac {g}2 r {q^\prime}^X D_X \left( WQ^s \right) (\sigma_s)_i^{\ j} 
+\frac i4 \partial_r \left[ re^{-w} (v^\prime a + a^\prime)
\right] \gamma_0 \gamma_1 \delta_i^{\ j}  \right\} \epsilon_j =0
~~~~~\label{rteta}
\feqn
Finally the commutator between (\ref{tBPSgrav}) and
(\ref{rBPSgrav}) gives
\eqn
& & \left\{  
ig{q^\prime}^X D_X \left(
\sqrt{\frac 32} e^v a WQ^s \right) (\sigma_s)_i^{\ j}
-\gamma_0 \frac {g}2 e^v {q^\prime}^X D_X(WQ^s ) (\sigma_s)_i^{\ j}
\right. \cr
& & +\gamma_1 \left[ i \frac {v^\prime}2  e^{v-w}
(v^\prime a + a^\prime) \delta_i^{\ j} 
-\frac i2 \partial_r \left(
e^{v-w} (v^\prime a + a^\prime) \right) \delta_i^{\ j} \right] \cr
& & \left. +\gamma_0 \gamma_1 \delta_i^{\ j} \left[ \frac 12 {g}^2
e^{v+w} W^2 +e^{v-w} \frac 12 (v^\prime a+ a^\prime)^2 -\frac 12
\partial_r (v^\prime e^{v-w}) \right] \right\} \epsilon_j =0 \cr
&&
\label{tierre}
\feqn

We begin with the study of the equation in (\ref{tetafi}). Unless all
coefficients vanish it can be written as 
\eqn \label{ansatz}
(if^0 \gamma_0 \delta_l^{\ k} +f^r (\sigma_r )_l^{\
k} \gamma_1 )\epsilon_k =\epsilon_l
\feqn
where
\eqn
& & f^r=-g e^w rWQ^r \label{fr} \\
& & \cr
& & f^0=-\frac {1-e^{-2w} -\frac
{r^2 e^{-2w}}4 (v^\prime a + a^\prime)^2 +{g}^2 r^2 W^2}{r
e^{-2w} (v^\prime a + a^\prime) } \label{fo}
\feqn
The Killing equation in (\ref{ansatz}), viewed as a projector equation, leads to the consistency requirement 
\eqn
(f^0 )^2 +\sum_r (f^r )^2 =1 \label{normal}
\feqn
Now we compare the above results with the content from the hyperini equation in (\ref{1BPShyper}). 
Starting from (\ref{1BPShyper}),  multiplying by the projector
$K_{\tilde Z} \delta_i^{\ j} -2iR^s_{\tilde Z X} K^X (\sigma_s)_i^{\ j}$
and symmetrizing in ${\scriptstyle Z, \  \tilde Z}$, we obtain
\eqn
& & \left[ \Bigl(K_Z K_{\tilde Z} +4K^X K^Y R^r_{ZX} R^s_{\tilde Z Y} \delta_{rs} \Bigr)
\Bigl[ a \gamma_0 \delta_i^{\ j} - i \delta_i^{\ j}\Bigr] 
-\left( \sqrt {\frac 23} \frac {e^{-w}}{g} q^{\prime
X} g_{X(Z} K_{\tilde Z )} \delta_i^{\ j} \right. \right. \cr
& & + 2 \sqrt{\frac 23} i
\frac {e^{-w}}{g} q^{\prime X} R^r_{(Z|X|} K_{\tilde Z )} (\sigma_r)_i^{\ j}
-2 \sqrt{\frac 23} i \frac {e^{-w}}{g} q^{\prime X} g_{X(Z}
R^r_{\tilde Z )Y} K^Y (\sigma_r)_i^{\ j} \cr
& & \left. \left. +4 \sqrt {\frac 23} \frac {e^{-w}}{g} q^{\prime X}
R^r_{(Z |Y|} R^s_{\tilde Z )X} K^Y \Bigl(\delta_{rs} \delta_i^{\ j}
+i \epsilon_{rst} (\sigma_r)_i^{\ j} \Bigr) \right) \gamma_1 
 \right] \epsilon_j =0 \label{w}
\feqn
The above equation is compatible with (\ref{ansatz}) only if the conditions
\eqn
q^{\prime X} (g_{X(Z} K_{\tilde Z )} -4R^r_{X(Z} R^s_{\tilde Z )Y} K^Y
\delta_{rs})=0 \label{conditions}
\feqn
are satisfied. 
If this is the case then (\ref{w}) becomes
\eqn
\left[ -ia \gamma_0 \delta_i^{\ j} - 2 \sqrt{\frac 23} e^{-w} q^{\prime X}
\frac {U^r_{XZ\tilde Z}}{g( K_Z K_{\tilde Z} +4K^T K^Y R^r_{ZT} R^s_{\tilde Z Y} \delta_{rs}) } \gamma_1 (\sigma^r)_i^{\ j} \right] \epsilon_j =\epsilon_j \cr
&& \label{confronto}
\feqn
with
\eqn
U^r_{XZ\tilde Z} \equiv R^r_{(Z|X|} K_{\tilde Z )} -g_{X(Z} R^r_{\tilde Z )Y} K^Y
+2 R^t_{X(Z} R^s_{\tilde Z )Y} K^Y \epsilon_{tsr}
\feqn
The second term in (\ref{confronto}) must be independent of $Z$ and $\tilde Z$, so that one obtains   
\eqn
q^{\prime X}\frac {U^r_{XZ\tilde Z}}{K_Z K_{\tilde Z} +4K^T K^Y R^t_{ZT}
R^s_{\tilde Z Y}\delta_{ts}} =\ -\ \frac{q^{\prime X} D_X P^r}{|K|^2}\ = \ 
-\sqrt{\frac 32} \frac {g e^w}2 f^r
\label{vincolo}
\feqn
The above equations have important consequences  for the geometry of the moduli space; we will discuss them in the next section.

At this point from the angular integrability condition and the hyperini supersymmetry variation we have
\eqn
\sqrt {\frac 32}~ g^2 e^{2w} rWQ^r
&=& -2\frac{q^{\prime X} D_X P^r}{|K|^2} \label{210} \\
f^0 &=& -a    \label{semplice}  
\feqn
The above result and the relation in (\ref{normal}) fix $f^r$ to be
\eqn
f^r =\pm \sqrt {1-a^2} Q^r  \label{parallela}
\feqn
In addition from the vector relation\footnote{$Q^rQ_r=1$ implies $(D_XQ^r)Q_r=0$.}  (\ref{210})
we have
\begin{gather}
q^{\prime X} D_X Q^r =0 \label{costante} \\
ge^w |K|^2 \sqrt {1-a^2} =\pm 2W^\prime  \label{corol}
\end{gather}
If we use (\ref{semplice}) and (\ref{parallela}) in (\ref{fo}) and
(\ref{fr}) then we find
\begin{gather}
\sqrt {1-a^2} =\mp g e^w rW  \label{1fr} \\
a= \frac {1-e^{-2w} -\frac
{r^2 e^{-2w}}4 (v^\prime a + a^\prime)^2 +{g}^2  r^2 W^2}{r 
e^{-2w} (v^\prime a + a^\prime)} \label{1fo}
\end{gather} 
Finally inserting (\ref {1fr}) into (\ref{1fo}) we obtain
\eqn
1=a^2 e^{-2w} \left[ 1 +\frac {r}2 
\left( v^\prime + \frac {a^\prime} a \right) \right]^2  \label{511}
\feqn

We postpone the discussion of the other consistency conditions and
analyze next the implications of what we have just found for the geometric structure of the moduli space. 

\subsection{Geometric restrictions}

Now we want to consider the equations in (\ref{conditions}) and (\ref{vincolo}) and show that they determine the space-time $r$-dependence of the scalars, i.e. of all the quantities that enter in the description of the quaternionic geometry like the prepotential $P^r$ and the Killing vectors.
In order to elucidate their meaning in a transparent manner it is convenient to proceed as follows. First a double contraction of (\ref{conditions}) with the Killing vector leads to ${q^\prime}^X K_X =0$. Then using this result and contracting (\ref{conditions}) with $K^Z$ one obtains
\eqn
{q^\prime}^Z =\pm 3ge^w \sqrt {1-a^2} \partial^Z W \label{erreeffe}
\feqn
The above equation is quite important: it describes the path in the moduli space associated 
to the BPS solution. It shows explicitly
 that, if we exclude the case $a^2=1$, the condition to have a fixed point is $\partial_Z W=0$ which corresponds to a local minimum of the potential as observed in \cite{ceresole:2001}. 

From (\ref{erreeffe})
using $q^{\prime Z} W_Z =W^\prime$ we obtain
\eqn
|q^\prime |^2 =\pm 3ge^w\sqrt {1-a^2} W^\prime  \label{1erreeffe}
\feqn
The contraction of (\ref{conditions}) with $q^{\prime Z}$ gives
\eqn \label{condq}
|q^\prime |^2 K_Z =2\sqrt 6 \delta_{rs} q^{\prime X} R^r_{XZ} Q^s W^\prime
\feqn
Acting now with $K^Z$ one obtains
\eqn
|K|^2 |q^\prime |^2 =6W^{\prime 2} \label{1condq}
\feqn
(\ref{1condq}) together with (\ref{1erreeffe}) gives again (cfr
\ref{corol})
\eqn
|q^\prime |^2 = \frac 32 |K|^2 g^2 e^{2w} (1-a^2 )   \label{2corol}
\feqn
which is in agreement with the fact that the Killing vector $K^X$ has to be null at the fixed point.
Using (\ref{2corol}) and (\ref{erreeffe}) in (\ref{condq})
one easily obtains
\eqn
K^Z =2\sqrt 6 \delta_{rs} Q^r R^{sXZ} \partial_X W  \label{3corol}
\feqn
Now we consider  (\ref{vincolo}) rewritten as
\eqn
|K|^2 q^{\prime X} U^r_{XZ\tilde Z} =
-\left[ K_Z K_{\tilde Z} +4K^T K^Y R^t_{ZT}
R^s_{\tilde Z Y}\delta_{ts} \right]
q^{\prime X} D_X P^r
\label{1vincolo}
\feqn
Contracting with $K^{\tilde Z}$ and using $K^X D_X P^s =0$ (as follows from
the definition (\ref{prepot})) we obtain
\eqn
K^2 q^{\prime X} R^s_{ZX} =-\sqrt {\frac 32} W^\prime Q^s K_Z -3WW^\prime
Q^t D_Z Q^r {\epsilon_{tr}}^s \label{importante}
\feqn
Moreover contracting (\ref{1vincolo}) with $q^{\prime \tilde Z}$ and using
(\ref{1condq})
we find
\eqn
|q^\prime |^2 \partial^Z W ={q^\prime}^Z W^\prime
\feqn
(which can be obtained also from (\ref{2corol}) and
(\ref{3corol})) and
\eqn
|q^\prime |^2 W D_Z Q^r +2R^t_{ZX} W^\prime q^{\prime X} Q^s
{\epsilon_{ts}}^r =0
\feqn
which after use of (\ref{importante}) gives 
(\ref{1condq}). \\
Note that (\ref{prepot}) gives
\eqn
K^Z =-\frac 43 R^{rZX} D_X P^r
\feqn
so that (\ref{3corol}) can be written as
\eqn \label{I3corol}
K_Z =\sqrt 6 W R^{rX}_{\quad Z} D_X Q_r
\feqn
Finally the contraction of (\ref{importante}) with $Q^r \epsilon_{srt}$ gives
\eqn
3WW^\prime D_Z Q_t =|K|^2 Q^r q^{\prime X} R^s_{ZX} \epsilon_{srt}
\label{final}
\feqn
Note that from (\ref{erreeffe}) and (\ref{1erreeffe}) we also have
\eqn
|\partial W|^2 =\frac {|K|^2}6  \label{ultima}
\feqn
Let us collect the main results:
\eqn
& & {q^\prime}^Z =\pm 3ge^w \sqrt {1-a^2} \partial^Z W \label{restrizione} \\
& & |K|^2 |q^\prime |^2 =6W^{\prime 2} \label{1restrizione} \\
& & |q^\prime |^2 K_Z =2\sqrt 6 \delta_{rs} q^{\prime X} R^r_{XZ} Q^s
W^\prime  \label{2restrizione} \\
& & K^Z =2\sqrt 6 \delta_{rs} Q^r R^{sXZ} \partial_X W \label{3restrizione} \\
& & K^2 q^{\prime X} R^s_{ZX} =-\sqrt {\frac 32} W^\prime Q^s K_Z -3WW^\prime
Q^t D_Z Q^r \epsilon_{tr}^{~ s}  \label{4restrizione}
\feqn
From the above relations it follows
\eqn
& & |q^\prime |^2 = \frac 32 |K|^2 g^2 e^{2w} (1-a^2 ) \label{5restrizione}\\
& & K_Z =\sqrt 6 W R^{rX}_{\quad Z} D_X Q_r \label{6restrizione} \\
& & |\partial W|^2 =\frac {|K|^2}6 \label{7restrizione} \\
& & 3WW^\prime D_Z Q_t =|K|^2 Q^r q^{\prime X} R^s_{ZX} \epsilon_{srt} \\
& & K_X q^{\prime X} =0 \label{ok}
\label{8restrizione}
\feqn
Now it is straightforward to verify that these conditions solve  (\ref{conditions}), (\ref{vincolo}) 
and (\ref{1BPShyper}) identically.
Finally we note that (\ref{8restrizione}) gives
\eqn
q^{\prime X} D_X Q^r =0 \label{1costante}
\feqn
It is interesting  and not at all obvious that (\ref{erreeffe}) and the other relations we 
have found in this section look like a simple generalization of those obtained for flat domain 
wall solutions (where the gauge fields are zero). This is suggestive of an underlying general 
structure, independent of the form of the space-time solution.

\subsection{Further restrictions}

The equations obtained in the previous subsection are quite general. 
Now we have to consider the other integrability conditions together with (\ref{ansatz}).
We start from (\ref{tteta}): it is easy to show that either all the coefficients vanish or it must be equivalent to (\ref{ansatz}). The first case reduces to the case in which all the
coefficients of (\ref{ansatz}) vanish.
The second case is verified when the following conditions are true:
\eqn
& & f^0 =-\frac {v^\prime  
-{g}^2 e^{2w} rW^2}{(v^\prime a + a^\prime) } \label{questa} \\
& & \cr
& & f^r=-g e^w rWQ^r 
\label{verificare}
\feqn
Other consequences of (\ref{tteta}) are the following: \\
from (\ref{questa}) and (\ref{fo}) we find
\eqn
1+2{g}^2 r^2 W^2 +\frac {r^2 e^{-2w}}4 \left[{v^\prime}^2-(v^\prime a + a^\prime)^2 \right]
={\left(1+\frac {r}2 v^\prime \right)}^2 e^{-2w} \label{serve?}
\feqn
while inserting (\ref{1fr}) into (\ref{questa}) we obtain
\eqn
a= \frac {rv^\prime -1+a^2 }{r(v^\prime a + a^\prime)} \label{a}
\feqn

Now we consider the integrability condition (\ref{rteta}): by means of (\ref{ansatz}) 
we obtain the equations\footnote{These two equations give again the condition (\ref{1costante}).}
\eqn
& & g r (v^\prime a + a^\prime) W +g a rW^\prime \mp \frac 12 
 \sqrt {1-a^2} \partial_r [re^{-w}(v^\prime a + a^\prime)]=0   \label{b} \\
& & \cr
& &  \mp g r(v^\prime a + a^\prime)  \sqrt {1-a^2} W -a
w^\prime e^{-w}  
-{g}^2 a e^{w} rW^2 +\frac 12 \partial_r [r
e^{-w}(v^\prime a + a^\prime)] =0 \cr
& & \label{c}
\feqn

Similarly from (\ref{tierre}) we have
\eqn
& & \mp \sqrt {1-a^2} Q^s \left[ \frac 12 {g}^2 e^{v+w} W^2 
+ \frac 12 e^{v-w} (v^\prime a + a^\prime)^2 
-\frac 12 \partial_r (v^\prime e^{v-w}) \right] \cr
& &~~~~~~~~~~~~~~~~~~~~~~
- g^2 a {q^\prime}^X D_X \left( \sqrt {\frac 32} e^v aP^s \right) +\frac {g}2 e^v W^\prime Q^s =0  \label{d}\\
& & \cr
& & \mp g \sqrt {1-a^2} e^v W^\prime +a e^v \partial_r \bigl[e^{-w}
\bigl(v^\prime a + a^\prime\bigr)\bigr] +{g}^2 e^{v+w} W^2 \cr
& &~~~~~~~~~~~~~~~~~~~~~~~~~~~~~~~~+e^{v-w} \bigl(v^\prime a + a^\prime\bigr)^2 -\partial_r 
\bigl(v^\prime e^{v-w}\bigr)=0  \label{e}
\feqn

In the next section we analyze the system of first order differential equations obtained above. 

\section{Static BPS configurations}\label{blackhole}
Let us begin with equation (\ref{a}) from which we easily obtain
\eqn
e^v =\frac {r}{r_0 \sqrt {1-a^2}}  \label{delta}
\feqn
where $r_0$ is an integration constant.
Using (\ref{1fr}) this can be rewritten as
\eqn
1=\mp g r_0 We^{v+w} \label{alfa}
\feqn

We focus on the equations derived in the previous section
to obtain a set of independent ones.
We start with the equation (\ref{b}). Using
(\ref{delta}) we find
\eqn
\partial_r \left( g a e^v W \mp \frac r{2r_0} e^{-w} (v^\prime a + a^\prime) \right) =0
\label{91}
\feqn
It is straightforward to verify that (\ref{91}) is satisfied by (\ref{alfa}) and (\ref{511}).
Thus (\ref{b}) is identically satisfied.

Now we turn to the analysis of the equation (\ref{c}). 
Using
\eqn
rv^\prime a +ra^\prime =a-\frac{ra^\prime}{a^2 -1} \label{lambda}
\feqn
it becomes
\eqn
\partial_r \left( ae^{-w} +\frac 12 r e^{-w} (v^\prime a + a^\prime)\right)=0
\feqn
which is solved by (\ref{511}). 
Thus we conclude that the equation  (\ref{c}) is identically satisfied.
We notice that also the equation (\ref{serve?}) is identically satisfied.
Indeed (\ref{511}) gives
\eqn
1-\frac {r^2}4 e^{-2w} (v^\prime a + a^\prime)^2 = a^2 e^{-2w}
+ae^{-2w} r(v^\prime a + a^\prime)
\feqn
Inserting this expression into (\ref{serve?}) we obtain
\eqn
2{g}^2  r^2  W^2 -e^{-2w} (1+rv^\prime ) +ae^{-2w}
r(v^\prime a + a^\prime) + a^2 e^{-2w}=0
\feqn
Using (\ref{1fr}) in the first term and multiplying by $e^{2w}$
we have
\eqn
1-a^2 - rv^\prime +a r (v^\prime a + a^\prime) =0 \label{eta}
\feqn
which is equivalent to (\ref{delta}).

In a similar way we can study the equation (\ref{e}). 
If we use (\ref{1fr}) in the first term of (\ref{e}) we have
\eqn
& & {g}^2 e^{v+w} rWW^\prime +{g}^2 e^{v+w} W^2 +a e^v \partial_r \bigl(e^{-w} \bigl(v^\prime a 
+ a^\prime\bigr)\bigr)  \cr
& &~~~~~~~~~~~~~~~~~~~~~~~~~~~~~~~~ + e^{v-w} \bigl(v^\prime a + a^\prime\bigr)^2 - \partial_r \bigl(v^\prime e^{v-w}\bigr) =0     \label{102}
\feqn
 
This can be rewritten in the form
\eqn
{g}^2 e^{v+w} rWW^\prime +{g}^2 e^{v+w} W^2 +\partial_r (a e^{v-w} (v^\prime a + a^\prime) -v^\prime 
e^{v-w}) =0
\feqn
If we now multiply by $r$ and then add and subtract the quantity
$ e^{v-w} (a(v^\prime a + a^\prime) - v^\prime)$ and finally use (\ref{eta}) 
we find
\eqn                          
{g}^2 \frac {e^{v+w}}2 \partial_r (r^2 W^2) - e^{v-w} (a(v^\prime a + a^\prime) - v^\prime) +\partial_r 
(e^{v-w} (a^2 -1)) =0
\feqn
Now we use $ g^2 r^2 W^2  = (1-a^2 )e^{-2w}$  and obtain
\eqn
\frac {e^{v+w}}2 \partial_r [(1-a^2 )e^{-2w}] - e^{v-w} (a(v^\prime a + a^\prime) - v^\prime) +\partial_r (e^{v-w} (a^2 -1)) =0
\feqn
This shows that (\ref{e}) is identically satisfied.

At the end we consider the equation (\ref{d}).
Using (\ref{1scomposizione}) we have
\eqn
& & \pm \sqrt {1-a^2} \left[ \frac 12 {g}^2 e^{v+w} W^2 + \frac 12 e^{v-w} 
{\left(v^\prime a + a^\prime\right)}^2 -\frac 12 \partial_r \left(v^\prime e^{v-w}\right) \right]\cr 
& &~~~~~~~~~~~~~~~~~~~~~~~~~~~~~~~+\frac 32 g a \partial_r \left(ae^v W\right) 
+\frac {g}2 e^v W^\prime =0
\feqn
By means of (\ref{e}) this can be written as
\eqn
-3ga\partial_r \left(ae^v W \right) + g a^2 e^v W^\prime  \pm a\sqrt {1- a^2} e^v \partial_r 
\left(e^{-w} \left(v^\prime a + a^\prime \right)\right)=0 \label{i}
\feqn
From (\ref{1fr}) we have $\mp g W =\sqrt {1-a^2} \frac {e^{-w}}r $
so that
\eqn
\pm g W^\prime =\frac {a a^\prime}{\sqrt {1-a^2}} \frac {e^{-w}}r 
+\frac {\sqrt {1-a^2}}{r^2} e^{-w} -\frac {\sqrt {1-a^2}}r
\partial_r e^{-w}
\feqn
and then
\newpage
\eqn
& & g ae^v W^\prime \pm \sqrt {1-a^2} e^v \partial_r (e^{-w} (v^\prime a + a^\prime)) = \cr
&&~~~~~~~~~~~~~~~\pm \frac {v^\prime a}r \sqrt {1-a^2} e^{v-w} \mp \frac ar \sqrt {1-a^2} e^v \partial_r e^{-w} 
\pm \sqrt {1-a^2} e^v \partial_r \left(e^{-w} \left(v^\prime a + a^\prime\right)\right) \cr
& &~~~~~~~~~~~~ = \pm \frac {\sqrt {1-a^2}}r \left( re^v \partial_r (e^{-w} (v^\prime
a + a^\prime)) +e^v v^\prime ae^{-w} -ae^v \partial_r e^{-w} \right) \cr
& &~~~~~~~~~~~~  = \pm \frac {\sqrt {1-a^2}}r e^v \left( -a \partial_r e^{-w} -e^{-w}
a^\prime + \partial_r (r e^{-w} (v^\prime a + a^\prime)) \right)  \cr
& &~~~~~~~~~~~~  = \mp \frac {\sqrt {1-a^2}}r e^v \partial_r \left( ae^{-w} -r e^{-w}
(v^\prime a + a^\prime) \right) =\mp \frac 3{r_0} \partial_r \left(ae^{-w}\right)
\feqn
where in the last step we have used (\ref{511}) and (\ref{delta}). \\
Then using $\frac{e^{-w}}{r_0} = \mp g_R e^v W$ we see that (\ref{i}) is identically satisfied.

In the next section we verify that BPS solutions we have obtained so far satisfy the equations of motion.

\section{Equations of motion}\label{equamotion}

The equations of motion of our system are given by
\eqn
& & -e^{v-w} \partial_r (v^\prime e^{v-w}) -3\frac {v^\prime}r  e^{2(v-w)}
+e^{2(v-w)} (v^\prime a + a^\prime)^2 + 4 g^2 e^{2v} W^2  \cr
& &~~~~~~~~~~~~~~~~~~~~~~~~~~~~~~~~~~~~~~~~~~~~~ -\frac 12 g^2 e^{2v} |K|^2 + \frac 32 g^2 e^{2v} 
|K|^2 a^2 =0  
\label{1} \\
& & \cr
& & e^{w-v} \partial_r (v^\prime e^{v-w}) -\frac 3r w^\prime +|q^\prime |^2
- (v^\prime a + a^\prime)^2 - \frac 12 e^{2w}(8g^2 W^2 + g^2 |K|^2) =0 \cr
&& \label{2} \\ 
& & re^{-2w} \bigl(v^\prime -w^\prime \bigr) -2\bigl(1-e^{-2w}\bigr) +\frac 12 r^2 e^{-2w} 
\bigl(v^\prime a + a^\prime \bigr)^2 \cr
& & ~~~~~~~~~~~~~~~~~~~~~~~~~~~~~~~~~~~~~~~~~~~~~
+ \frac 23 r^2 \left( -6g^2 W^2 +\frac 34 g^2 |K|^2
\right) =0 \label{3} \\
& & a e^v K_X q^{\prime X} =0 \label{4} \\
& & \cr
& & e^{-(v+w)} r^{-3} \partial_r \left(r^3 e^{v-w} {q^\prime}^Z \right)-\frac 12
\partial^Z g_{XY} e^{-2w} q^{\prime X} q^{\prime Y}  \cr
& &~~~~~~~~~~~~~~~~~~~~~~~~~~~~~~~~~ +\frac 34 g^2 a^2
\partial^Z |K|^2 - \partial^Z \left( -6g^2 W^2 +\frac 34 g^2 |K|^2 \right) = 0
\cr
&& \label{5} \\
& & 3e^{-2w} (v^\prime a + a^\prime) +r\partial_r (e^{-w} (v^\prime a + a^\prime))
= g r ae^w |K|^2  \label{6} \\
& & \cr
& & g e^{-2w} K_X q^{\prime X} =0  \label{7}
\feqn
where (\ref{1}), (\ref{2}), (\ref{3}) and (\ref{4}) are the Einstein equations, (\ref{5}) is
the equation for the
scalar fields and (\ref{6}), (\ref{7}) are the Maxwell equations.

First we observe that both (\ref{4}) and (\ref{7}) are an immediate consequence of (\ref{ok}). 
Then we consider the sum of (\ref{1}) and (\ref{2}). Multiplying by $e^{2(v-w)}$ and using (\ref{5restrizione})
we obtain
\eqn
g^2 |K|^2 = 2\frac {e^{-2w}}r (v^\prime +w^\prime ) \ . \label{8}
\feqn
This is solved by (\ref{alfa}), (\ref{1fr}) and (\ref{corol}).

It is straightforward to check that all the equations are indeed satisfied:\\
Equation (\ref{1}) is solved using (\ref{8}), (\ref{102}),
(\ref{alfa}), (\ref{1fr}), (\ref{511}) and (\ref{delta}). \\
Equation (\ref{3}) is solved by (\ref{8}) and (\ref{serve?}).\\
Equation (\ref{5}) is solved using (\ref{restrizione}), (\ref{8}),
(\ref{7restrizione}), (\ref{1fr}) and (\ref{delta}). \\
Finally (\ref{6}) is solved by (\ref{511}) and (\ref{8}).

\section{The Universal Hypermultiplet case}\label{solution}

Now we collect the set of first order differential equations obtained by the BPS conditions:
\begin{gather}
1=a^2 e^{-2w} \left[ 1 +\frac {r}2 
\left( v^\prime + \frac {a^\prime} a \right) \right]^2 \label{5.1} \\
e^v =\frac {r}{r_0 \sqrt {1-a^2}} \label{def_v}\\
\sqrt {1-a^2} =\mp g e^w rW \label{def_w}\\
{q^\prime}^Z =\pm 3ge^w \sqrt {1-a^2} \partial^Z W \label{5.4}
\end{gather}
It is easy to reduce the above system to
\eqn
\begin{cases}
{q^\prime}^Z = -3 \frac{1-a^2}r \partial^Z \ln W & \text{ }  \\
1-a^2 = \frac{g^2}4 {\left(3 a + r \frac{a^\prime}{1-a^2}\right)}^2 r^2 W^2
& \text{ }  \label{problem}
\end{cases}
\feqn
(\ref{def_v}) and (\ref{def_w}) simply define 
$v$ and $w$ respectively in terms of $a$ and of the scalars, hence as functions of $r$.

In this form our problem is analogous to the domain-wall case: the main difference
is that now the two differential equations are coupled equations and therefore finding an
explicit solution is more involved.
To solve (\ref{problem}) we have to specify $W$ as a function of the scalars
$q^X$ i.e we have to choose which isometry of the quaternionic manifold represents
the action of the $U(1)$ gauge symmetry.
As we have already argued in the previous sections the most interesting
configurations are obtained for isometries in the isotropy group of some point of
the quaternionic manifold. By definition this choice corresponds to a
fixed point solution i.e one with asymptotic constant scalars.
For (supersymmetric) black hole solutions it implies the existence of 
an horizon. The scale invariance appearing in the
near-horizon limit gives rise to the enhancement of the unbroken
supersymmetry associated to a fixed point.

In fact the configurations that have the most relevant role in the AdS/CFT correspondence
are the ones with two fixed points: this type of solutions should describe a
RG flow between two different CFT's defined on the boundary of the five dimensional
space-time. For black hole solutions this means that the space-time is maximally
symmetric in the $r \rightarrow \infty$ limit (for example AdS).
Since as shown in \cite{Alekseevsky:2001if}, 
in order to obtain such configurations one needs the introduction of vector multiplets, 
we postpone the study of black hole configurations to future work.\footnote{The results presented in the ch.\ref{vet} furnish the instruments to affort it.}

Here, as an example, we construct an explicit solution of the BPS equations in the case of a
$n_H =1$ hypermultiplet, i.e. the so called universal hypermultiplet.
However, this simple example is  important because 
this hypermultiplet (which contains the Calabi-Yau volume) appears in any Calabi-Yau compactification.
We adopt for this manifold the notations and the coordinate system of \cite{ceresole:2001}. The main properties of this space are reviewed in the appendix \ref{convention}. 
\noindent
For simplicity we make the following choices for the gauged isometry and the graviphoton: 
\begin{gather}
K\equiv \vec{k}_1 = \left(
\begin{array}{c}
0 \\ 1 \\ 0 \\ 0
\end{array}\right) \label{s1}\\  
a^\prime \equiv 0 \label{s2}
\end{gather}

The Killing vector $\vec{k}_1$ has a simple interpretation: it represents the translation of 
the axionic scalar ($\sigma$ in our notations). This solution has been already considered 
in  \cite{gutperle:2001}\footnote{As discussed in section \ref{model} in  \cite{gutperle:2001} 
a mistake affects the calculations; however the final solution has the right functional 
form.}. We remark that since $\vec{k}_1$ is not 
in the isotropy group of any point of the manifold, the presence of fixed points is excluded. 
In general it is easy to see that fixed point solutions are ruled out 
by the assumption $a^\prime=0$. 

We observe that, being the superpotential 
for $\vec{k}_1$,  $W_{\vec{k}_1}=\mp \sqrt{\frac 23} \frac 1{4V}$, the only dynamical scalar 
is $V$. The others are space-time constants and can be set consistently equal to zero.  
Imposing (\ref{s1}) and (\ref{s2}) the system (\ref{problem}) becomes
\eqn
\begin{cases}
V^\prime= 6 \frac{1-a^2}r V \label{p1}\\
\frac{1- a^2}{a^2}= \frac 3{32} {\left( \frac{g r}V \right)}^2 \label{p2}
\end{cases}
\feqn
Since the metric of the quaternionic manifold (\ref{quatmetric}) in our parametrization is singular  for $V=0$ 
we restrict ourselves to the branch $V>0$.  
From (\ref{p1}) it follows that the scalar $V$ has the form
\eqn
V={\cal C}\ r^\Lambda
\feqn
with  ${\cal C}$ and $\Lambda={6(1-a^2)}$ fixed,  by consistency with (\ref{p2}), to be
\begin{gather}
\Lambda=1\\
{\cal C}= \sqrt{\frac{3}{32}} \sqrt{\frac{a^2}{1- a^2}} g 
\end{gather}
that in particular gives $a=\sqrt{\frac 56}$.
At the end we find
\begin{gather}
V=\sqrt{\frac {15}{32}} g r 
\end{gather} 
and using (\ref{def_v}) and (\ref{def_w})
\begin{gather}
e^v=\sqrt 6 \frac r{r_0} \\
e^w=\sqrt{\frac {15}8}
\end{gather}
Rescaling the time coordinate $t$ by the constant $\frac{\sqrt 6}{r_0}$ the space-time metric  and 
the electrostatic potential become
\begin{gather}
ds^2 =- r^2 dt^2 + \frac {15}8 dr^2 + r^2 
(d\theta^2 +\sin^2 \theta d\phi^2 +\cos^2 \theta d\psi^2 ) \\
A_t= \sqrt{\frac 54} r
\end{gather}
We notice that as expected the gauging constant $g$ appears only in the 
quaternionic scalar while the other space-time quantities do not depend on it.

\section{Discussion and Outlook} \label{conclusion_b}

Here we have presented electrostatic spherical BPS (${\cal N}=1$) solutions in ${\cal N}=2$ 
gauged supergravity in five dimensions with hypermultiplet couplings. 
In particular we have discussed the possibility of finding (extreme) black-hole solutions. 
Although we study BPS solutions following the ``traditional'' way
i.e fixing an ansatz of interest we get general indications useful for the program of this thesis.
The main results we have obtained 
can be summarized as follows:
first of all we have derived the BPS equations in a general setting, 
going beyond special cases treated previously with restrictive  ansatz \cite{gutperle:2001}. 

Then we have discussed the structure of the integrability conditions and in particular 
of the hyperini equations. We have obtained relations 
which appear to be a generalization of those found in \cite{ceresole:2001} for  flat domain walls. 
This result is somewhat surprising since our ansatz is quite different from the one 
in \cite{ceresole:2001}.
In addition we have considered a 
configuration 
with a non-vanishing gauge field whose presence complicates considerably the structure of the equations. 

We have verified that our BPS solutions satisfy the equations of motion.

The above results leave much space for further studies.
First of all it would be interesting to explore if and under which conditions the structure found 
for the hyperini equation is maintained when more general ansatz are considered and 
vector multiplets are introduced. These two points will discussed in the following chapters. In these general settings one would like
to  explore  the existence of black hole solutions leading to nontrivial RG flows. 

Finally it would be interesting to consider such explicit solutions 
and extend them to non extreme black holes along the lines of what
has been done for the case of vector multiplets \cite{notextremal}.

These open problems are currently under investigation.

\chapter{The addition of vector multiplets}\label{vet}

\section{Introduction}

In this chapter we generalize what it has been done in \cite{noialtri}, with the introduction of an arbitrary number of vector multiplets considering only abelian gauge groups (${U(1)}^{n_V+1}$). This extention is necessary to consider configurations with two fixed points and it can be useful to understand which new features arise when considering charged solutions in presence of both hypermultiplet and vector multiplet couplings.  
We derive integrability conditions for this case using the same ansatz  of  \cite{noialtri} for the metric and for the gauge fields and study them together with the hyperini and gaugini equations. As for the flat domain configurations of \cite{ceresole:2001} we obtain that the BPS conditions ensure the stability of the potential as shown in \cite{stability},\cite{stab2} and that the supersymmetric flow equations are controlled by the superpotential $W$. The set of differential equations we get has  a structure  analogous to the sub-case $n_V=0$ treated in the previous chapter: in particular we find also for the scalars of very special geometry the behavior $\varphi^{\prime \Lambda}\propto \partial_\Lambda W$ (where with this notation we indicate generically all the scalars) as in \cite{ceresole:2001}.\footnote{As it will be emphasized later on, the factor of proportionality is not longer the same for vector multiplet and hypermultiplet scalars.}. We explicitly show that the BPS conditions satisfy the equations of motion (to not tire the reader 
the calculations are given in the appendix). At the end we present a preliminary discussion of the possible applications of our results.

\section{The model and its BPS equations}\label{model_v}
We consider $\cal{N}$=2 supergravity in five dimensions with an arbitrary number of hypermultiplets and vector multiplets. The field content of the theory is the following:
\begin{itemize}
\item the supergravity multiplet
\eqn
\{ e^a_\mu \ , \psi^{\alpha i}_\mu \ , A^0_\mu \}
\feqn
containing the {\em graviton} $e^a_\mu$, two {\em gravitini}
$\psi^{\alpha i}_\mu$ and the {\em graviphoton} $A^0_\mu$;
\item $n_H$ hypermultiplets
\eqn
\{ \zeta^A \ , q^X \}
\feqn
containing the {\em hyperini} $\zeta^A $ with $A=1,2,\dots,2n_H$, and the
{\em scalars} $q^X$ with $ X=1,2,\dots,4n_H$ which define a quaternionic
Kahler manifold with  metric $g_{XY}$;
\item the vector multiplet
\eqn
\{ A^{\hat{I}}_\mu\ , \lambda^{ai} \ , \phi^x \}
\feqn
containing $n_V$ {\em gaugini} $\lambda^{ia} \ , \quad a=1, \ldots ,n_V$ with
spin $\frac 12$, $n_V$ real scalars $\phi^x $,
$ \quad x=1, \ldots ,n_V$
which define a very special manifold and $n_V$ {\em gauge vectors}
$A^{\hat{ I}}_\mu $ , $\quad \hat{I} =1\ldots , n_V $. Usually the graviphoton is included
by taking $I=0\ldots n_V$. 
\end{itemize}

The bosonic sector of the gauged Lagrangian density is given by
\cite{ceresole:2000}
\eqn
{\cal {L}}_{BOS} &=& \frac 12 e\{ R-\frac 12 a_{IJ} F^I_{\mu \nu} F^{J \mu \nu}
-g_{XY} D_\mu q^X D^\mu q^Y -g_{x  y} D_\mu \phi^x D^\mu \phi^y
-2g^2 {\cal V} (q,\phi) \} \cr
& & +\frac 1{6\sqrt 6} \epsilon^{\mu\nu\rho\sigma\tau}
C_{IJK} F^I_{\mu\nu} F^J_{\rho\sigma}A^K_{\tau}
\label{lagr_v}
\feqn 
with 
\begin{gather*}
D_\mu q^X=\partial_{\mu} q^X + g A^I_\mu K_I^X(q) \\
D_\mu \phi^x=\partial_{\mu} \phi^x  + g A^I_\mu K_I^x(\phi)
\end{gather*}
where $K_I^X(q)$, $K_I^x(\phi)$ are the  Killing vector on the quaternionic and the very special real manifold respectively and ${\cal V} (q,\phi)$ is the scalar potential as given in Appendix.

At this point we concentrate our attention to abelian case: this implies (see sect. \ref{gunasecti}) that the action of the gauge group is non trivial only on the quaternionic manifold while scalars of vector multiplet are uncharged under it. This means that $D_\mu \phi^x \equiv \partial_\mu \phi^x$ and   the existence of any isometry for the very special geometry is not required. 
\noindent
Then the variations of the fermions for abelian gauge symmetry ${U(1)}^{n_V+1}$ reduce to:\\
 for the gravitini 
\eqn
\delta_\epsilon \psi_{\mu i} &=& \partial_\mu \epsilon_i
+\frac 14 \omega^{ab}_\mu \gamma_{ab} \epsilon_i -\partial_\mu q^X
p_{Xi}^{\quad j} \epsilon_j 
+ g A^I_\mu P_{Ii}^{\ j} \epsilon_j \cr &+&\frac i{4 \sqrt
6} (\gamma_{\mu \nu \rho} -4g_{\mu \nu} \gamma_\rho )h_I F^{I\nu
\rho } \epsilon_i - \frac i{\sqrt 6} g h^I P_{Ii}^{\ j} \gamma_\mu
\epsilon_j =0 \label{BPSgrav_v}
\feqn
for gaugini
\eqn
& & \left[ -i \phi^{\prime x} e^{-w} \gamma_1 \delta_i^{\ j}
-2 i g h^{Ix} P_I^s (\sigma_s)_i^{\ j} \right. \cr
& & \left. + \sqrt {\frac 32} e^{-w} h_I^x 
\left(v^\prime a^I + {a^I}^\prime \right) 
\gamma_{01} \delta_i^{\ j} \right]
\epsilon_j =0 \label{gaugini_v}
\feqn
and for hyperini
\eqn
f_{iX}^A \left[ - i q^{\prime X} e^{-w} \gamma_1 
+ i \sqrt {\frac 32} g a^I K_I^X \gamma_0 
+\sqrt {\frac 32}  g h^I K_I^X \right] \epsilon^i =0 
\label{hyperini_v}
\feqn
where we have set $\phi^{\prime x} :=\partial_r \phi^x$ and
$q^{\prime X} :=\partial_r q^X$.\footnote{This notation applies to all the quantities with the only exception of $a^I$ for which we explicitly define $a^\prime \equiv (\partial_r a^I) h_I$. This choice is motivated by the aim to be manifest the similarities with the case $n_V=0$.}


As already explained at the begining, we want to consider the direct generalization of the problem considered in the chap.\ref{black}
We look for electrostatic spherical solutions that preserve half of the ${\cal N}=2$ supersymmetries. 
We choose the same metric of the previous chapter, which is $SO(4)$ symmetric with all the other fields that only depend on the holographic space-time coordinate $r$. Moreover we fix the gauge for the gauge fields keeping only the $A^I_t$ component different from zero. 
 
Introducing  spherical coordinates $(t,r,\theta ,\phi, \psi)$ we write 
 \eqn \label{metricansatz_v} ds^2 =-e^{2v} dt^2 +e^{2w}  dr^2 + r^2 
(d\theta^2 +\sin^2 \theta d\phi^2 +\cos^2 \theta d\psi^2 ) \feqn 
where the functions $v$ and $w$ depend on $r$ only.

We parametrize the vector fields as
\eqn \label{elettrico_v}
A^I_t =\sqrt {\frac 32}
a^I(r)e^v 
\feqn 
so that \eqn A_t^{\prime I} =\sqrt {\frac 32} 
\left(v^\prime a^I+ {a^I}^\prime \right) e^v
\feqn

\subsection{Integrability conditions}\label{integ_v}
We now consider the BPS equations for the gravitini: their integrability
condition is the vanishing of their commutators; using the general
formulas in \cite{gen:to be published} one finds only four independent commutators
\eqn
& &\left\{ \left[ -\frac 12 \partial_r (v^\prime e^{v-w}) +\frac 12 e^{v-w}
(v^\prime a+ a^\prime )^2 +\frac {g^2}2 e^{v+w} W^2 \right] \gamma_0
\gamma_1 \delta_i^{\ j} \right. \cr
& & - \sqrt {\frac 32} ig   e^v (v^\prime a^I +{a^I}^\prime) P_I^s (\sigma_s )_i^{\ j}
- \sqrt {\frac 32} i g  e^v a^I \tilde{D}_r P_I^s (\sigma_s )_i^{\ j} -\frac g{\sqrt 6}
e^v \tilde{D}_r P^s (\sigma_s )_i^{\ j} \gamma_0 \cr
& & \left. -\left[ \frac i2 \partial_r (e^{-w} (v^\prime a+ a^\prime )) e^v
\delta_i^{\ j} +g^2 e^{v+w} a^I h^J P_I^r P_J^s \epsilon_{rs}^{\quad t}
(\sigma_t )_i^{\ j} \right] \gamma_1 \right\} \epsilon_j =0 \label{tr_v} \\
& & \cr
& & \cr
& & \left\{
\left[-\frac 12 v^\prime e^{v-2w} + \frac {g^2}2 r e^v W^2 \right] \gamma_0 \delta_i^{\ j}   +\frac {ig}{\sqrt 6} r e^{v-w} (v^\prime a+ a^\prime) P^s
(\sigma_s)_i^{\ j} \gamma_1 \right. \cr
& & + \left. \left[ \frac i2 e^{v-2w} (v^\prime a+ a^\prime) \delta_i^{\ j}
+g^2 r e^v a^I h^J P_I^r P_J^s \epsilon_{rst} (\sigma^t)_i^{\ j} \right]
\right\} \epsilon_j =0 \label{tteta_v}  \\
& & \cr
& & \cr
& & \left\{
\left[ \frac 12
\partial_r (e^{-w}) - \frac {g^2}2 r e^w W^2
\right] \gamma_1 \delta_i^{\ j}
+\frac i4 \partial_r \left[ r e^{-w} (v^\prime a+ a^\prime) \right] \gamma_0 \gamma_1  \delta_i^{\ j}
\right. \cr
& & \left.
-\frac {ig}{\sqrt 6} r (v^\prime a+ a^\prime) P^s (\sigma_s)_i^{\ j} \gamma_0 
+\frac {ig}{\sqrt 6} r \tilde{D}_r P^s (\sigma_s)_i^{\ j}
\right\} \epsilon_j =0
\label{rteta_v} \\
& & \cr
& & \cr
& &
\left\{ \frac 12 \left[ 1- e^{-2w} -\frac {r^2 e^{-2w}}4
(v^\prime a+ a^\prime)^2 +\frac 23 g^2 r^2 W^2 \right]
\delta_i^{\ j}  +\frac i2 r e^{-2w} (v^\prime a+ a^\prime) \delta_i^{\ j} \gamma_0 \right. \cr
& & \left. +\frac {ig}{\sqrt 6} r^2 e^{-w} (v^\prime a+ a^\prime) P^s (\sigma_s)_i^{\ j} \gamma_0 \gamma_1 \right\}
\sin \theta \epsilon_j =0 \label{tetafi_v}
\feqn
with the scalar derivative defined as
$$
\tilde{D}_\mu = \partial_\mu \phi^{\prime x} \partial_x +  \partial_\mu q^{\prime X} D_X
$$
Here we have defined $a :=h_I a^I$ and $a^\prime := h_I {a^I}^\prime$.

\subsection{Matter field conditions: Hyperini equation}\label{hypequa_v}
Now we compare the information coming from the integrability condition with the supersymmetric variation of the matter fermions. We consider first the equation for the hyperini. Assuming that
\eqn
re^{-2w} (v^\prime a+ a^\prime) \neq 0
\label{cond_v}
\feqn
we can rewrite (\ref{tetafi_v}) in the form
\eqn \label{ansatz_v}
(if^0 \gamma_0 \delta_l^{\ k} +f^r (\sigma_r )_l^{\
k} \gamma_1 )\epsilon_k =\epsilon_l
\feqn
with
\eqn
& & f^r= - g r e^w W Q^r \label{fr_v} \\
& & f^0=-\frac {1- e^{-2w} - {\left(\frac {r e^{-w}}2 (v^\prime a+ a^\prime)\right)}^2 
+g^2 r^2 W^2}{r e^{-2w} (v^\prime a+ a^\prime)} \label{fo_v}
\feqn
Also if we define $\Lambda =-f^0$ we have
\eqn
f^r =\pm \sqrt {1-\Lambda^2} Q^r  \label{parallela_v}
\feqn
Put in (\ref{hyperini_v}) and using (\ref{fquadrato}) we have
\eqn
[iA\delta_l^{\ k} +B^s (\sigma_s )_l^{\ k} ]\gamma_1 \epsilon_k =
[C \delta_l^{\ k} +iD^s (\sigma_s )_l^{\ k}] \epsilon_k
\feqn
with
\eqn
& & A=+\frac 12 q^{\prime Z} e^{-w} f^0 +\sqrt {\frac 32} ga^I D^Z P_I^s f_s \\
& & B^s =+\frac 12 \sqrt {\frac 32} ga^I K_I^Z f^s -q^{\prime X}
R^{sZ}_{\ \ \ X} e^{-w} f^0 -\sqrt {\frac 32} ga^I D^Z P_r f_t
\epsilon^{rts} \\
& & C=\frac 12 \sqrt {\frac 32} a^I K_I^Z g +\frac 12 \sqrt {\frac 32}f^0 gK^Z \\
& & D^s =\sqrt {\frac 32} ga^I D^Z P_I^s +g\sqrt {\frac 32} h^I D^Z P_I^s f^0
\feqn
It is now easy to see that this condition is not compatible with (\ref{ansatz_v})
so that one must put $A=B^s =C=D^s =0$ that is
\eqn
& & q^{\prime Z} e^{-w} \Lambda =\sqrt 6 ga^I D^Z P_I^s f_s \label{111_v}\\
& & q^{\prime X} R^{sZ}_{\ \ \ X} e^{-w} \Lambda = -\frac 12 \sqrt {\frac 32}
ga^I K_I^Z f^s +\sqrt {\frac 32} ga^I D^Z P_{I r} f_t \epsilon^{rts} \label{222_v}\\
& & a^I K_I^Z =K^Z \Lambda \label{333_v} \\
& & a^I D^Z P_I^s =h^I D^Z P_I^s \Lambda \label{444_v}
\feqn
Using (\ref{333_v}) and (\ref{444_v}) in (\ref{111_v}) and (\ref{222_v}) we find
\eqn
& & q^\prime_Z =\pm 3ge^w \sqrt {1-\Lambda^2} \partial_Z W \label{erreeffe_v}\\
& & q^{\prime X} R^{sZ}_{\ \ \ X} e^{-w} =
\mp \sqrt {1-\Lambda^2} \sqrt {\frac 32} g \left( \frac 12 K^Z Q^s
+\sqrt {\frac 32}W D^Z Q_r Q_t \epsilon^{rst} \right) \cr
&&\label{555_v}
\feqn
After contraction of (\ref{555_v}) with $K_Z$ we obtain
\eqn
\sqrt {\frac 32} g^2 r e^{2w} W Q^r
= -2\frac{q^{\prime X} D_X P^r}{{|K|}^2} \label{210_v} 
\feqn
which gives 
\eqn
q^{\prime X} D_X Q^r =0 \label{costante_v} 
\feqn
\eqn
ge^w {|K|}^2 \sqrt {1-\Lambda^2} =\pm 2q^{\prime X} \partial_X W \label{corol_v}
\feqn

Also (\ref{fo_v}) and (\ref{fr_v}) can be rewritten as
\eqn
& & \sqrt {1-\Lambda^2} =\mp g r e^w W  \label{1fr_v} \\
& & \Lambda=
\frac {1- e^{-2w} - {\left(\frac {r e^{-w}}2 (v^\prime a+ a^\prime)\right)}^2 
+g^2 r^2 W^2}{r e^{-2w} (v^\prime a+ a^\prime)}
\label{1fo_v}
\feqn
Using (\ref {1fr_v}) in (\ref{1fo_v}) we obtain
\eqn
1=\Lambda^2 e^{-2w}\left[ 1 +\frac {r}{2\Lambda} (v^\prime a+ a^\prime)\right]^2  \label{511_v}
\feqn
If we use (\ref{corol_v}) in (\ref{555_v}), the last one becomes
\eqn
K^2 q^{\prime X} R^s_{ZX} =- q^{\prime X} \partial_X W \left(\sqrt {\frac 32}  Q^s K_Z + 3W Q^t D_Z Q^r \epsilon_{tr}^{\quad s}\right)  \label{4restrizione_v}
\feqn

Many other relations, which will be useful to check the equations of motion,
follow from (\ref{erreeffe_v}), (\ref{costante_v}), (\ref{corol_v}) and
(\ref{4restrizione_v}):
\eqn
& & {|K|}^2 |q^\prime |^2 =6\left(q^{\prime X} \partial_X W\right)^2 \label{1restrizione_v} \\
& & |q^\prime |^2 K_Z =2\sqrt 6 \delta_{rs} q^{\prime X} R^r_{XZ} Q^s
q^{\prime Y} \partial_Y W  \label{2restrizione_v} \\
& & K^Z =2\sqrt 6 \delta_{rs} Q^r R^{sXZ} \partial_X W \label{3restrizione_v} \\& & |q^\prime |^2 = \frac 32 {|K|}^2 g^2 e^{2w} (1-\Lambda^2 ) \label{5restrizione_v}\\
& & K_Z =\sqrt 6 W R^{rX}_{\quad Z} D_X Q_r \label{6restrizione_v} \\
& & |\partial W|^2 =\frac {{|K|}^2}6 \label{7restrizione_v} \\
& & 3W q^{\prime X} \partial_X W D_Z Q_t ={|K|}^2 Q^r q^{\prime X} R^s_{ZX} \epsilon_{srt}
\label{8restrizione_v}
\feqn

\subsection{Matter field conditions: Gaugini equation}\label{gauequa_v}
Next let us consider gaugini: using (\ref{ansatz_v}) to replace $\gamma_0\epsilon$ in (\ref{gaugini_v}) one easily obtains
\eqn
& & \Lambda \phi^{\prime x}  + \sqrt {\frac 32} h_I^x
 (v^\prime a^I+ a^{\prime I}) =0 \label{1_v} \\
& & 2g \Lambda h^{x I} P_I^s - \sqrt {\frac 32} e^{-w} h_I^x
 (v^\prime a^I+ a^{I\prime}) f^s =0 \label{3_v}
\feqn
which gives
\eqn
\phi^{\prime x} f^s  =-2 g e^w  h^{x I} P_I^s 
\feqn
and using
\eqn
h_x^{I} P_I^s =-\frac 32 \partial_x (W Q^s )
\feqn
one finally has
\eqn
& & \partial_x Q^s =0 \label{chiq_v} \\
& & \pm \sqrt {1-\Lambda^2} \phi^{\prime x} =3 g e^w  g^{x y}
\partial_y W \label{strana_v} \\
& & \sqrt 6 g\Lambda \partial_x W =\mp e^{-w} h_I^x (v^\prime a^I
+a^{\prime I}) \sqrt {1-\Lambda^2} \label{bene_v}
\feqn

We can rewrite the information on the scalars in a compact way defining
$$
\varphi^\Sigma = 
\begin{cases}
\phi^x & \text{for } \Sigma=1,...,n_V \\
q^X       & \text{for } \Sigma=n_V+1,...,n_V+4n_H
\end{cases}
$$
as
\begin{gather}
\tilde{D}_r Q^s =0 \label{genq_v} \\
\varphi^{\prime \Lambda} =\pm 3 g e^w \left(1-\Lambda^2\right)^{\frac 12 
\Delta} g^{\Lambda \Sigma} \partial_\Sigma W \label{genscalar_v}\\
\intertext{with}\\
\Delta=
\begin{cases}
- 1 & \text{for } \Lambda=1,...,n_V \\
  1 & \text{for } \Lambda=n_V+1,...,n_V+4n_H
\end{cases}
\end{gather}
where $g^{\Lambda \Sigma}$ is simply the product metric.

Let me discuss the consequences of the above relations. First of all a strong similarity with the domain wall case \cite{ceresole:2001} emerges again: this observation is a not trivial  because the two configurations are quite different and it suggests that it should be possible to obtain a very general insight on BPS solutions in presence of generic matter couplings.
To be more specific in the two situations it happens that the phase of prepotential $Q^r$ do not depend on the vector multiplet scalars: under this condition the potential ${\cal V}(q,\phi)$ reduces the form that has been put forward for gravitational stability
\eqn
{\cal V}= -6W^2 + \frac 92 g^{\Lambda \Sigma} \partial_\Lambda W \partial_\Sigma W
\feqn
It is easy to see that in this case critical points of W are also critical points of ${\cal V}$.
Furthermore we find that $\varphi^\Lambda \propto \partial_\Lambda W$ but now the gauge interaction distinguishes between charged $q^X$ and uncharged $\phi^x$ via the factor $1-\Lambda^2$.
At we end we want to underline the importance of (\ref{1_v}) that practically
gives the component of the field strength on $h_x^I$ and with (\ref{a})
determines it as a vector of special geometry. This information will be
crucial to check whether BPS solutions satisfy the equations of motion.

\subsection{Further restrictions}

As usual we have to compare the previous information with the information coming from the other integrability conditions.
Let us consider
equation (\ref{tteta_v}): it is easy to show that or all the coefficients
vanish or it must be equivalent to (\ref{ansatz_v}). 
The first case reduces to the case in which all the
coefficients of (\ref{tetafi_v}) vanish.
The second case occurs when the following conditions are true:
\eqn
& & f^0 =-\frac {v^\prime  -g^2 r e^{2w} W^2}{v^\prime a+ a^\prime} \label{questa_v} \\
& & f^r=- gre^w W Q^r
\label{verificare_v}
\feqn
with \underline{$a^I P_I^r$  parallel to $h^J P_J^r$} 
\eqn
a^I P_I^r =\beta(r)h^J P_J^r  \label{parallelismo_v}
\feqn
for some function $\beta$.

 From the properties of very special geometry reviewed in the section \ref{very} the modulus of vector $h_I$ can
be chosen $h_I h^I = 1$ so the set $(h^I,h_x^I)$ is a base for the
$n_V+1-$dimensional space with $h_I h_x^I=0 $. Then this relation holds
\begin{gather}
a^I = a h^I + l^x h_x^I \label{decomp_v}
\end{gather}

Using the above decomposition in (\ref{444_v}) and (\ref{parallelismo_v}) we get
\begin{gather}
\begin{cases}
(a-\Lambda)D_Z P^r = - l^x D_Z P_x^r \\
(a-\beta) P^r      = - l^x P_x^r
\end{cases}
\end{gather}
remembering the BPS demand $\partial_x Q^r = 0$ gives
\begin{gather}
\begin{cases}
\beta=\Lambda\\
a-\Lambda = \sqrt{\frac 32} l^x\partial_x \ln W = -\frac r{\sqrt{6}} l_x \phi^{\prime x}\\
a-\Lambda = \sqrt{\frac 32} l^x\partial_x \ln \partial_Z W
\end{cases}
\label{set_v}
\end{gather} 

We continue to derive the other equations from integrability conditions.

Equation(\ref{questa_v}) together with (\ref{fo_v}) gives
\eqn
1+2g^2 r^2 W^2 +\frac {r^2 e^{-2w}}4 \left[v^{\prime 2} -(v^\prime a+ a^\prime)^2 \right]
= e^{-2w}(1+\frac {r}2 v^\prime )^2 \label{serve?_v}
\feqn
If we substitute (\ref{1fr_v}) into (\ref{questa_v})
\eqn
\Lambda= \frac {rv^\prime - 1 + \Lambda^2}{r(v^\prime a+ a^\prime)} \label{a_v}
\feqn
Using (\ref{ansatz_v}) in (\ref{rteta_v}) we obtain the equations
\eqn
& & g r (v^\prime a+ a^\prime) W +g r \Lambda W^\prime \mp \frac 12 \partial_r
\left[r e^{-w}(v^\prime a+ a^\prime)\right]\sqrt {1-\Lambda^2}=0  
\label{b_v} \\
& & \mp g r (v^\prime a+ a^\prime) W \sqrt {1-\Lambda^2} +\Lambda\partial_r
(e^{-w}) - \Lambda g^2 r e^w W^2 \cr
& & +\frac 12 \partial_r \left[r e^{-w} (v^\prime a+ a^\prime)\right] =0  
\label{c_v} 
\feqn
Similarly from (\ref{tr_v}) we have
\eqn
& & \Lambda g\tilde{D}_r \left( \sqrt {\frac 32} e^v a^I P_I^s \right) +\frac g2 e^v
W^\prime Q^s
\mp \sqrt {1-\Lambda^2} Q^s \left[ \frac 12 g^2 e^{v+w} W^2 + \frac 12 e^{v-w}
\left(v^\prime a + a^\prime \right)^2 \right. \cr
& & \left. -\frac 12 \partial_r (v^\prime e^{v-w}) \right] =0 \label{d_v}
\feqn
\newpage
\eqn
& & \mp  \sqrt {1-\Lambda^2} ge^v W^\prime +\Lambda e^v \partial_r \left[e^{-w}
\left(v^\prime a + a^\prime \right) \right] +g^2 e^{v+w} W^2 +e^{v-w} \left(v^\prime a + a^\prime \right)^2
\cr
& & -\partial_r (v^\prime e^{v-w})=0 \label{e_v}
\feqn

We observe that all the relations already derived reduce to those in
\cite{noialtri} if we take $\Lambda = a$. So it is easy to conclude that
$l^x \equiv 0$ is compatible with
BPS conditions and reduces to the set of equations (\ref{5.1})-(\ref{5.4}) plus the one for the scalars of the vector multiplets.

\section{Static BPS configurations}

In this section we derive the independent set of equations that characterizes
BPS configurations.

We start by considering integrability conditions (\ref{b_v}) and (\ref{c_v}):
subtracting from (\ref{b_v}) the equation (\ref{c_v}) multiplied by $\mp \sqrt{1-\Lambda}$
we get
\eqn
gr \Lambda (v^\prime a + a^\prime) W + gr W^\prime \mp \sqrt{1-\Lambda^2} w^\prime e^{-w} \mp g^2 r \sqrt{1-\Lambda^2} e^w W^2 = 0
\feqn
Using (\ref{a_v}) and (\ref{1fr_v}) it gives
\eqn
W^\prime = - (v^\prime + w^\prime) W
\label{b-c_v}
\feqn
that implies $gr_0 W= \mp e^{-(v+w)}$ where $r_0$ is a constant.
This last expression can be rewritten considering again (\ref{1fr_v}) as 
\eqn
e^v= \mp \frac r{r_0 \sqrt{1-\Lambda^2}}
\label{delta_v}
\feqn  
which is fundamental to demonstrate the compatibility of BPS conditions. Indeed taking the derivative with respect to $r$ and comparing  with (\ref{a_v}) we obtain
\eqn
v^\prime a + a^\prime = v^\prime \Lambda + \Lambda^\prime
\label{key_v}
\feqn
This means that the integrability conditions and consequently the BPS
equations for the metric ($w$ and $v$) and for the scalars of the hypermultiplets have the same form as the ones in \cite{noialtri}: hence their consistency is ensured and the only change is the replacement of $a$ by
$\Lambda$ . The new ingredients here, due to
the introduction of vector multiplets, are then the equation for $\phi^x$ and the relations between $a$ and $\Lambda$ (\ref{set_v}).

Note that here $a'$ is not the derivative of a with respect to $r$. Indeed
we have defined define $a' \equiv h_I a^{\prime I}$. Using (\ref{decomp_v}) it
is easy to find
\eqn
a' =\partial_r a -\sqrt {\frac 23} l_x \phi^{\prime x}
\feqn
Substituting it in (\ref{key_v}) and using the second eq. of (\ref{set_v}) we get
\eqn
v^\prime a + \partial_r a +\frac 2r (a-\Lambda) = v^\prime \Lambda +
\Lambda^\prime
\feqn
which gives
\eqn
\Lambda=a + \frac {\mu}{r^2} e^{-v}
\feqn
where $\mu$ is an integration constant. Using (\ref{set_v}) the last expression can be rewritten to show in a transparent manner the relation between $\mu$ and $l^x$as 
\eqn
\frac {\sqrt 6 \mu}{3 r^3} e^{-v} = l_x \phi^{\prime x}
\feqn
The implication of this expression on the existence of fixed points still has to be clarified.
Now it is not so difficult to show that the BPS conditions we have derived satisfy the equations of motion. We refer the reader to the appendix \ref{app:motion} for the technical details.

To summarize how it has been obtained, we conclude this section presenting a set of independent BPS equations
\begin{gather}
1={\Lambda}^2 e^{-2w} \left[ 1 +\frac {r}2 
\left( v^\prime + \frac {{\Lambda}^\prime} {\Lambda} \right) \right]^2 \label{5.1_v} \\
e^v =\frac {r}{r_0 \sqrt {1-{\Lambda}^2}} \label{def_v_v}\\
\sqrt {1-{\Lambda}^2} =\mp g e^w rW \label{def_w_v}\\
{q^\prime}^Z =\pm 3ge^w \sqrt {1-{\Lambda}^2} \partial^Z W \label{5.4_v}\\
\phi^{\prime x} =\pm 3ge^w \frac 1{\sqrt {1-{\Lambda}^2}} \partial^x W \label{5.5_v}\\
\Lambda=a + \frac {\mu}{r^2} e^{-v}\\
\frac {\sqrt 6 \mu}{3 r^3} e^{-v} = l_x \phi^{\prime x}
\end{gather}


\section{Discussion}

In this section we want to recall the results already obtained and to point out which topics deserve more study. First of all we have derived BPS equations for this case, studying in the line of \cite{noialtri}, the relations from the  hyperini and the gaugini and integrability conditions for the gravitini. We observe that the former ones have the same structure manifested in the domain wall case \cite{ceresole:2001}: this suggests the possibility of determining some properties of BPS solutions without starting from the specific ansatz. The importance of a similar study is evident: for example this could permit us to give a definitive answer in the quest for a realistic cosmological model in gauged supergravity. 
These features are currently under investigation \cite{gen:to be published}, and a part of the results, for hypermultiplet couplings only, will be presented in the next chapter.

Furthermore we have derived the system of differential equations that is a  gene\-ralization of the one in \cite{noialtri}: as we stated it should admit solutions with two fixed points. It remains however to be studied, hopefully leading to construction of an explicit example.



\chapter{Bps solutions with hypermultiplets}\label{gen}

\section{Introduction}

In this last chapter we discuss the properties of BPS solutions in the presence of matter couplings. We want to trascend the limitations due to the choice of a special configuration to study and, instead, try to investigate why there are some features common to different families of solutions. There we will concentrate on the Maxwell--Einstein supergravity theory with hypermultiplets charged under the $U(1)$ gauge group.

To further this aim we adopt a double strategy. First we consider the integrability conditions and the hyperini equation for a generic space--time metric $g_{\mu\nu}$ and graviphoton $A_\mu$ with an arbitrary number of hypermultiplets $n_H$, without selecting any particular ansatz. We show that it is possible, although with a lot of calculations, to obtain general results in this way: in particular we demonstrate that integrability conditions satisfy the Einstein equations for the metric (imposing only that the Maxwell--Einstein equation holds for the graviphoton) and that the hyperini equation can be reduced to a more manageable form which explains the geometrical meaning of the gauge covariant derivative of the scalars $D_\mu q^X$ in terms of the killing vector $K^X$ and the prepotential $P^r$. 

The second ``weapon'' we use is given by the geometric method, first introduced by  Gauntlett et al in \cite{Gauntlett:2002nw}, to classify BPS solutions by determining the group structure of the base space (see sect.\ref{geocla}).
The strength of this method is that it 
is possible to derive the general form of the BPS solutions by assuming only the existence of a killing spinor.\footnote{This is at the same time also a limitation, because the resulting coordinate system is in most of the cases not ``natural''.}    
Let us start by reminding of the main topic of the theory, reviewed in the chapter \ref{teo}.

\section{The Model}
The field content of the $\cal{N}$=2 supergravity in five dimensions with  hypermultiplet couplings is the following
\begin{itemize}
\item the supergravity multiplets \eqn \{ e^a_\mu \ , \psi^{i}_\mu \ , A_\mu \} \feqn containing the {\em graviton} $e^a_\mu$,
two {\em gravitini} $\psi^{i}_\mu$ and the {\em
graviphoton} $A_\mu$; 
\item a hypermultiplet 
\eqn 
\{ \zeta^A \ , q^X \} 
\feqn 
containing $2n_H$ {\em hyperini} $\zeta^A \ , \quad
A=1, \ldots ,2n_H$ with spin $\frac 12$ and $4n_H$ {\em scalars} $q^X$ which
define a quaternionic manifold with metric $g_{XY}$;
The bosonic Lagrangian density is \cite{ceresole:2000} 
\eqn 
{\cal
{L}}_{BOS} &=& -\frac 12 e\{ R+\frac 12 F_{\mu \nu}
F^{ \mu \nu}
+g_{XY} D_\mu q^X D^\mu q^Y 
\cr
&+&   2 {\cal V} (q,\phi) \} + \frac 1{6\sqrt 6} \epsilon^{\mu\nu\rho\sigma\tau} F_{\mu\nu} F_{\rho\sigma}A_{\tau}
\feqn 
${\cal V} (q,\phi)$
being the potential as given in appendix.
\end{itemize}
\section{BPS equations and equations of motion}
The supersymmetry transformations of the fermions are:\\
\smallskip
for the gravitini 
\eqn
\delta_\epsilon \psi_{\mu i} &=& \hat{\DD}_\mu \epsilon_i \\
                             &=& \partial_\mu \epsilon_i
+\frac 14 \omega^{ab}_\mu \gamma_{ab} \epsilon_i -\partial_\mu q^X
p_{Xi}^{\quad j} \epsilon_j - g A_\mu P_{i}^{\ j} \epsilon_j \cr 
&+&\frac i{4 \sqrt
6} (\gamma_{\mu \nu \rho} -4g_{\mu \nu} \gamma_\rho )F^{\tilde I
\nu \rho } \epsilon_i - \frac i{\sqrt 6} g P_{i}^{\ j} \gamma_\mu
\epsilon_j =0 \label{BPSgrav_g} \feqn

for the hyperini 
\eqn
\delta_\epsilon \zeta^A =\left[ -\frac i2 \gamma^a \partial_a q^X
-\frac i2 \gamma^a g A_a K^X +g \frac {\sqrt 6}4 K^X \right]
f_{Xi}^A \epsilon^i =0 \label{hyperini}
\feqn 

\subsection{Integrability conditions}
We now consider the BPS equations for the gravitini; their integrability conditions are given by the commutators $[\hat{\DD}_\mu,\hat{\DD}_\nu] = 0$ which we decompose in the sum of $15$ non trivial pieces\footnote{This relations, 
that here are specialized to the abelian case, can be easily extended also to the non-abelian case.}:
\eqn
& & \left[ \partial_\mu +\frac 14 \omega_\mu^{ab} \gamma_{ab} ,
\partial_\nu +\frac 14 \omega_\nu^{cd} \gamma_{cd} \right] \epsilon_i =
\frac 14 \Omega_{\mu \nu}^{\quad ab} \gamma_{ab} \epsilon_i  \label{1_g} \\
&&\cr
& &-\left[ \partial_\mu +\frac 14 \omega_\mu^{ab} \gamma_{ab} ,
\partial_\nu q^X p_{Xi}^{\quad j} \right] \epsilon_j  
-(\mu \leftrightarrow \nu ) =
-\left[ \partial_\nu q^X \partial_\mu p_{Xi}^{\quad j} 
-\partial_\mu q^X \partial_\nu p_{Xi}^{\quad j} \right]  \epsilon_j
\cr
&&\label{2_g} \\
& & \left[ \partial_\mu +\frac 14 \omega_\mu^{ab} \gamma_{ab} ,
-g A_\nu^I P_{Ii}^{\quad j} \right] -(\mu \leftrightarrow \nu ) =
-g \left( \partial_\mu \left[ A_\nu P_{i}^{\quad j} \right]
-\partial_\nu \left[ A_\mu P_{i}^{\quad j} \right] \right) \epsilon_j
\cr
&&\label{4_g} \\
& & \left[ \partial_\mu +\frac 14 \omega_\mu^{ab} \gamma_{ab} ,
\frac i{4 \sqrt 6} (\gamma_{\nu \tau \rho} -4g_{\nu \tau} \gamma_\rho ) 
F^{\tau \rho } \right] \epsilon_i -(\mu \leftrightarrow \nu ) \cr
& & =\frac i{4 \sqrt 6} \left[ (\gamma_\nu \gamma_a -4e_{\nu a})\gamma_b
\DD_\mu F^{ab}
-(\gamma_\mu \gamma_a -4e_{\mu a})\gamma_b
\DD_\nu F^{ab} \right] \epsilon_j \label{5_g} \\
&&\cr
& & \left[ \partial_\mu +\frac 14 \omega_\mu^{ab} \gamma_{ab} ,
-\frac i{\sqrt 6} g P_{i}^{\quad j} \gamma_\nu \right] \epsilon_j
-(\mu \leftrightarrow \nu ) =\frac {ig}{\sqrt 6} \left( 
\gamma_\mu \partial_\nu P^{\ j} -\gamma_\nu \partial_\mu P_i^{\ j}
\right) \epsilon_j \label{6_g} \\
&&\cr
& & \left[ i\partial_\mu q^X p_X^r (\sigma_r)_i^{\ j} ,
i\partial_\nu q^Y p_Y^s (\sigma_s)_i^{\ j} \right] \epsilon_j
= -i2 \partial_\mu q^X \partial_\nu q^Y p_X^r p_Y^s \epsilon_{rs}^{\quad t}
(\sigma_t)_i^{\ j} \epsilon_j \label{7_g} \\
&&\cr
& & \left[ i\partial_\mu q^X p_X^r (\sigma_r)_i^{\ j} ,
ig A_\nu P^s (\sigma_s)_i^{\ j} \right] \epsilon_j
-(\mu \leftrightarrow \nu ) =
-4ig \partial_{[\mu} q^X p_X^r A_{\nu ]} P^s \epsilon_{rs}^{\quad t}
(\sigma_t)_i^{\ j} \epsilon_j \cr
&& \label{9} \\
& & \left[ i\partial_\mu q^X p_X^r (\sigma_r)_i^{\ j} ,
\frac i{4 \sqrt 6} (\gamma_{\mu \nu \rho} -4g_{\mu \nu} \gamma_\rho ) 
F^{\nu \rho } \right] \epsilon_j -(\mu \leftrightarrow \nu ) =0 \label{10} \\
&&\cr
& & -\left[ i\partial_\mu q^X p_X^r (\sigma_r)_i^{\ j} ,
\frac {g}{\sqrt 6}  P^s \gamma_\nu (\sigma_s)_i^{\ j} \right] \epsilon_j
-(\mu \leftrightarrow \nu ) \cr
& & =\frac 4{\sqrt 6} g\partial_{[\mu} q^X p_X^r P^s e_{\nu ]}^a
\gamma_a \epsilon_{rs}^{\quad t} (\sigma_t)_i^{\ j} \epsilon_j \label{11} \\
&&\cr
& & \left[ ig A_\mu P^r (\sigma_r)_i^{\ j} ,
ig A_\nu P^s (\sigma_s)_i^{\ j} \right] \epsilon_j =
-2i g^2 A_\mu A_\nu P^r P^s \epsilon_{rs}^{\quad t} (\sigma_t)_i^{\ j} 
\epsilon_j =0 \label{16} \\
&&\cr
& & -\left[ ig A_\mu P^r (\sigma_r)_i^{\ j} ,
\frac i{4 \sqrt 6} (\gamma_{\mu \nu \rho} -4g_{\mu \nu} \gamma_\rho ) 
F^{\nu \rho } \right] \epsilon_j -(\mu \leftrightarrow \nu ) =0 \label{17} \\
& &\cr
& & -\left[ ig A_\mu P^r (\sigma_r)_i^{\ j} ,
\frac {g}{\sqrt 6} P^s \gamma_\nu (\sigma_s)_i^{\ j} \right] \epsilon_j
-(\mu \leftrightarrow \nu ) =
\frac 4{\sqrt 6} g^2 A_{[\mu} P^r P^s \gamma_{\nu ]}
\epsilon_{rs}^{\quad t} (\sigma_t)_i^{\ j} \epsilon_j =0 \cr 
&& \label{18} \\
& & -\frac 1{96} \left[
(\gamma_{\mu ab} -4g_{\mu a} \gamma_b ) F^{ab } ,
(\gamma_{\nu cd} -4g_{\nu c} \gamma_d ) F^{cd } \right] \epsilon_i \cr
& & =\left\{ \frac 1{24} F^2 \gamma_{\mu \nu}
+\frac 1{12} \left[ F_{\ \nu}^{2 \ d} \gamma_{d\mu}
-F_{\ \mu}^{2 \ d} \gamma_{d\nu} \right] 
-\frac 14 F_\mu^{\ \ a} F_\nu^{\ \ b} \gamma_{ab} 
-\frac 16 F_{[\mu}^{\ \ b} F^{cd} \gamma_{\nu ] bcd} \right\} \epsilon_i \cr
&&\label{19} 
\feqn
\eqn
& & \frac 1{24} \left[
(\gamma_{\mu ab} -4g_{\mu a} \gamma_b ) F^{ab } ,
g P_i^{\ j} \gamma_{\nu ]} \right] \epsilon_j 
-(\mu \leftrightarrow \nu )~~~~~~~~~~~~~~~~~~~~~~~~~~~~~~~~~~~~~ \cr
& & ~~~~~~~~~~~~~~~~~~~~~~~~~~~~=\frac 16 g \gamma_{\mu \nu ab} F^{ab} P_i^{\ j} \epsilon_j 
+g P_i^{\ j} \frac 23 
F_{[\mu}^{\ \ b} \gamma_{\nu ]b} \epsilon_j  \label{20} \\
&&\cr
& & \left[ \frac i{\sqrt 6} g P_i^{\ j} \gamma_\mu ,
\frac i{\sqrt 6} g P_i^{\ j} \gamma_\nu \right] \epsilon_j
=\frac {g^2} 3 \gamma_{\mu \nu} |P|^2 \epsilon_i \label{21}
\feqn
Difficulties in derivations of the above commutators arise
only in Eq.(\ref{5_g}) and Eq.(\ref{19}) and are explicitly shown in
appendix \ref{computation}. There it is also shown how the terms can be collected to give 
the following result
\eqn
& & \left\{ \frac 14 \Omega_{\mu \nu}^{\quad ab} \gamma_{ab} \delta_i^{\ j}
-iR^r_{\mu \nu} (\sigma_r)_i^{\ j} -igP^r F_{\mu \nu} (\sigma_r)_i^{\ j}
\right. \cr
& & +\frac i{4\sqrt 6} \left[ \gamma_{\nu ab} \DD_\mu F^{ab} 
-\gamma_{\mu ab} \DD_\nu F^{ab} \right]\delta_i^{\ j} 
+\frac i{\sqrt 6}
\left[ e_{\mu a} \gamma_b \DD_\nu F^{ab} -e_{\nu a} \gamma_b \DD_\mu 
F^{ab} \right] \delta_i^{\ j} \cr
& & -\frac {g}{\sqrt 6} (\gamma_\mu D_\nu P^r -\gamma_\nu D_\mu P^r )
(\sigma_r)_i^{\ j} +\frac 1{24} F^2 \gamma_{\mu \nu} \delta_i^{\ j}
+\frac 1{12} \left[ F_{\ \nu}^{2 \ d} \gamma_{d\mu} -F_{\ \mu}^{2 \ d} 
\gamma_{d\nu}  \right] \delta_i^{\ j} \cr
& & -\frac 14 F_{\mu}^{\ a} F_{\nu}^{\ b} \gamma_{ab} \delta_i^{\ j}
-\frac 16 F_{[\mu}^{\quad b} F^{cd} \gamma_{\nu]bcd} \delta_i^{\ j}
+\frac 16 g\gamma_{\mu \nu ab} F^{ab} P_i^{\ j} +\frac 23 g
F_{[\mu}^{\quad b} \gamma_{\nu]b} P_i^{\ j} \cr
& & \left. +\frac {g^2}3 \gamma_{\mu \nu} |P|^2 \delta_i^{\ j}
\right\} \epsilon_j =0 \label{integral}
\feqn
where we have defined
\eqn
& & F^{2ab} := \eta_{cd} F^{ca} F^{db} \ , \\
& & F^2 := F_{ab} F^{ab} \ , \\
& & |P|^2 =\delta_{rs} P^r P^s \ , \\
& & B_{*i}^{\ \ j} := iB_*^{\ r} (\sigma_r)_i^{\ j} \ , \\
& & R_{\mu \nu}^r := D_\mu q^X D_\nu q^Y R^r_{XY} \ , \\
& & R^r_{\mu \nu} := D_\mu q^X D_\nu q^Y R^r_{XY} \ , \\
& & D_\nu P^r := D_\nu q^X D_X P^r
\feqn

\subsection{Some manipulations}
Let us try to get some information out of (\ref{integral}) by means of contractions. 
If one multiplies (\ref{integral}) by $\gamma^\alpha$ and then contracts $\alpha$ 
with $\nu$, using the first Bianchi identity for $\Omega$ one obtains
\eqn
& & \left\{ -\delta_i^{\ j} \frac 12 \RR_{\mu b}\gamma^b -iR^r_{\mu a} \gamma^a
(\sigma_r )i^{\ j} -\frac i{2\sqrt 6} \delta_i^{\ j} \gamma_{ab} \DD_\mu F^{ab} \right. \cr
& & +\left[ \frac i{\sqrt 6} \gamma^{cb} \DD_c F_{\mu b} -\frac i{\sqrt 6} \DD_a
F^a_{\ \mu} +\frac i{2\sqrt 6} \DD_a F^{ab} \gamma_{\mu b}-\frac 1{4\sqrt 6} e_{d\mu} 
\epsilon^{dc}_{\quad abe} \gamma^e
\DD_c F^{ab} \right]\delta_i^{\ j} \cr 
& & +\left( \frac {4g}{\sqrt 6} D_\mu P^r 
+\frac {g}{\sqrt 6} \gamma_\mu^{\ a} D_a P^r \right) (\sigma_r )_i^{\ j}
+\left[ -\frac 1{12} F^2 \gamma_\mu +\frac 12 F_\mu^{2b} \gamma_b 
+\frac 16 F_\mu^{\ b} F^{cd} \gamma_{bcd} \right. \cr
& & \left. \left. -\frac i{12} e_\mu^f F^{ab} F^{cd} \epsilon_{abcdf} \right] \delta_i^{\ j}
-\frac 43 g^2 \gamma_\mu |P|^2 \delta_i^{\ j} 
\right\} \epsilon_j =0 \label{contratta}
\feqn
Here 
$\RR_{\mu\nu} =\Omega^a_{\ \mu a \nu}$ is the Ricci tensor and $R$ the Ricci Scalar.
\subsection{Equations of motion}
The Einstein equations are
\eqn
\RR_{\mu\nu} +F_{\mu a} F_\nu^{\ a} +g_{XY} D_\mu q^X D_\nu q^Y 
-\frac 16 |F|^2 g_{\mu \nu}
+\frac 23 \VV g_{\mu \nu} =0 \label{einstein}
\feqn
from which it follows in particular
\eqn
\frac 32 R +\frac 14 |F|^2 +\frac 32 |q^\prime |^2 -5 \VV =0  \label{particular}
\feqn
Variation with respect to the gauge fields gives
\eqn
\DD_a  F^{af} +\frac 1{2 \sqrt 6} \epsilon^{abcdf}
F_{ab} F_{cd} -gK_{Y} D^t q^Y =0 \label{maxwell}
\feqn
Finally the equations for the scalars are
\eqn
& & \hat D_\mu D^\mu q^W +gA^{\mu I} D_\mu K_I^W = g^{WX} \partial_X 
\VV \label{qq} 
\feqn
Here $\tilde D$ indicates a totally covariant derivative, i.e with respect
to all the indices. So for example
\eqn
\hat D_\mu \DD^\mu q^X =D_\mu D^\mu q^X 
+\Gamma^X_{YZ} D_\mu q^Y D^\mu q^Z \nonumber
\feqn
and in general
\eqn
D_\mu f^* (q)= D_\mu q^X \partial_X f^* \ .
\feqn

\subsection{Compatibility}

In this section we analyze the compatibility between the equations of motion and the BPS equations. In particular we will assume that the BPS equations together with the equations of motions and the Bianchi Identities for the gauge fields
are satisfied. 
\subsection*{Gravitini}
To simplify the analysis of (\ref{contratta}), let us organize it in different steps:
\begin{enumerate}
\item the term $a=-iR^r_{\mu a} \gamma^a (\sigma_r )_i^{\ j}$ \\
by mean of (\ref{1hyperini}) one readily finds
\eqn
-iR^r_{\mu a} \gamma^a (\sigma_r )_i^{\ j} \epsilon_j = & & 
\left\{ \frac 12 \gamma^a D_a q^X g_{XZ} D_\mu q^Z {\delta_i}^j 
+\frac i2 g\sqrt {\frac 32} K_Z D_\mu q^Z {\delta_i}^j \right. \cr
& & \left. -g\sqrt {\frac 32} D_\mu q^Z D_Z P^r (\sigma_r )_i^{\ j} \right\}
\epsilon_j
\feqn
\item the term $c=-\frac i{\sqrt 6} \DD_a F^a_{\ \mu} -\frac i{12} e_\mu^f F^{ab} F^{cd} 
\epsilon_{abcdf}$ \\
from (\ref{maxwell}), which we express in the form $\M^b=0$, it is equal to\footnote{For simplicity we omit $\epsilon_i$ but it is obvious that the following relations are true only when applied to the killing spinor.}
\eqn
c &=& 
-\frac {ig}{\sqrt 6} K_Z D_\mu q^Z \delta_i^{\ j} -\frac i{\sqrt 6} \M_\mu \delta_i^{\ j}
\feqn
\item for the term $d=\frac i{2\sqrt 6} D_a F^{ab} \gamma_{\mu b}$,
first we use (\ref{maxwell}) to give
\eqn
& & \frac i{2\sqrt 6} \M^b \gamma_{\mu b} \delta_i^{\ j}
-\frac i{24} \epsilon^{abcde} F_{ab} F_{cd} \gamma_{\mu e} \delta_i^{\ j} 
-\frac i{2\sqrt 6} gK_Y D_\mu q^Y \delta_i^{\ j}~~~~~~~~~~~~~~~~~~~~\cr
&&+\frac i{2\sqrt 6} gK_Y D^e q^Y \gamma_\mu \gamma_e \delta_i^{\ j}
\feqn
in the last term we can use (\ref{1hyperini}) contracted with $K^Z$ to give
\eqn
d &=& \frac i{2\sqrt 6} \M^b \gamma_{\mu b} \delta_i^{\ j}
-\frac i{24} \epsilon^{abcde} F_{ab} F_{cd} \gamma_{\mu e} \delta_i^{\ j}
\cr
& & -\frac i{2\sqrt 6} gD_\mu q^Y K_Y -\frac g{\sqrt 6} \gamma_\mu \gamma^a D_a q^X
D_X P^r (\sigma_r )_i^{\ j} +\frac {g^2}4 K^X K_X \gamma_\mu \delta_i^{\ j}\cr
&&  
\feqn
\item the term $e=\left( \frac {4g}{\sqrt 6} D_\mu P^r + \frac {g}{\sqrt 6} 
\gamma_\mu^{\ a} D_a P^r \right) (\sigma_r )_i^{\ j}$ \\
This is
\eqn
e= \frac {4g}{\sqrt 6} D_\mu q^X D_X P^r (\sigma_r )_i^{\ j} 
+\frac g{\sqrt 6} \gamma_{\mu a}
D^a q^X D_X P^r (\sigma_r )_i^{\ j} 
\feqn
\item the term $f=-\frac i{2\sqrt 6} \delta_i^{\ j} \gamma_{ab} \DD_\mu F^{ab}
+ \frac i{\sqrt 6} \gamma^{cb} \DD_c F_{\mu b} \delta_i^{\ j} $
can be rewritten as
\eqn
f=\frac i{2\sqrt 6} \gamma^{bc} [\DD_b F_{\mu c} +\DD_c F_{b\mu} +\DD_\mu F_{cb}]
\delta_i^{\ j} =:\frac i{2\sqrt 6} \gamma^{bc} \BB_{\mu cb}\delta_i^{\ j}
\feqn
$\BB_{\mu cb} =0$ being the Bianchi identity.
\item note that 
\eqn
\frac 16 F_\mu^{\ b} F^{cd} \gamma_{bcd} = 
\frac i{24} \epsilon^{abcde} F_{ab} F_{cd} \gamma_{\mu e} \delta_i^{\ j}
\feqn
and that
\eqn
\frac {g^2}4 K^X K_X 
-\frac 43 g^2 |P|^2 =\frac 13 \VV 
\feqn
\item the other terms are
\eqn
& &  -\delta_i^{\ j} \frac 12 \RR_{\mu b}\gamma^b \\
& & -\frac 1{4\sqrt 6} e_{d\mu} \epsilon^{dc}_{\quad abe} \gamma^e
\DD_c F^{ab}  \delta_i^{\ j} \\
& & -\frac 1{12} F^2 \gamma_\mu +\frac 12 F_\mu^{2b} \gamma_b \delta_i^{\ j}
\feqn
\end{enumerate}

If we now write the Einstein equations (\ref{einstein}) in the form $E_{\mu \nu} =0$
we then see that all the terms can be collected to give
\eqn
& & \frac 12 E_{\mu b} \gamma^b \epsilon_i -\frac i{\sqrt 6} \M_\mu \epsilon_i 
+\frac i{2\sqrt 6} \M^b \gamma_{\mu b} \epsilon_i \frac i{2\sqrt 6} \gamma^{bc} \BB_{\mu cb}
\epsilon_i=0 
\feqn
Let us now impose the equation of motion and the Bianchi identities for the Maxwell field. Then
\eqn
E_{\mu b} \gamma^b \epsilon_i =0
\feqn
In the time-like case\footnote{The meaning of this statement will be clarified in the section \ref{geocla}} this ensures that the Einstein equations are automatically satisfied.
In the light-like case, the $E_{++}$ components of the Einstein equations must be imposed
to vanish.

The derivation of the equation of motion for the scalars starting from the hyperini equation is still under investigation.

\section{Hypergeometric restrictions}
In this section we analyze the projector content of  Eq.(\ref{hyperini}) deriving the general form that holds for a generic BPS configuration. Then we discuss some particular subcases of interest.
Let us now consider the hyperini BPS equations rewritten in the form
\eqn
\left[ -i \gamma^a D_a q^X
+g {\sqrt \frac 32} K^X \right]
\left( g_{ZX}\delta_i^{\ j} +2R_{ZX}^r {(\sigma_r)_i}^j \right) \epsilon^i =0
\label{1hyperini}
\feqn 
where $K^X=h^I k_I^X$ and $D_a q^X$ is the gauge covariant derivative in flat index. We want to study it for $K\neq 0$ which means that we are away from the fixed point: this should be equivalent to requiring  that (\ref{1hyperini}) and all the other BPS conditions are not trivial.

Without loss of generality we can decompose $D_a q^X$ as
\begin{gather}
D_a q^X = M_a K^X + \Psi_a^X
\label{proje}
\end{gather}
with $\Psi_a^X K_X = 0$.
Now multiplying by\footnote{This vector is always different from zero if $K=0$.} $K_{\tilde Z} \delta_i^{\ j} -2i R^s_{\tilde Z X} K^X (\sigma_s)_i^{\ j}$  
and symmetrizing in $Z$, $\tilde Z$ we end up, using (\ref{proje}) and the properties of the $SU(2)$-curvature, with
\eqn
& &\left\{ \sqrt{\frac 32} g C_{Z \tilde Z} {\delta_i}^j - i \gamma^a 
\left[ \left( M_a C_{Z \tilde Z} + \Psi_a^X B_{X Z \tilde Z} \right) {\delta_i}^j + 2 i \Psi_a^X U^r_{X Z \tilde Z} {(\sigma_r)_i}^j \right] \right\} \epsilon_j =0 \cr
& & \label{proje1}
\feqn
where 
\eqn
C_{Z \tilde Z} & = & K_Z K_{\tilde Z} + 4 K^X K^Y R^r_{Z X} R^s_{\tilde Z Y} \delta_{rs}\\
B_{X Z \tilde Z} & = & g_{X(Z} K_{\tilde Z)} + 4 R^r_{(Z|X|} R^s_{\tilde Z)Y} K^Y\delta_{rs}\\
U^r_{X Z \tilde Z} & = & R^r_{(Z|X|} K_{\tilde Z} 
                       - g_{X(Z} R^r_{\tilde Z) Y} K^Y + 2 R^s_{(\tilde Z|Y|} R^t_{Z)X}  K^Y{\epsilon_{st}}^r
\label{tensor}
\feqn
At this point we observe that the four vectors $K^X$, ${R^{rX}}_Y K^Y\equiv D^X P^r$, are ortho\-gonal to each other and that they generate a subspace $V_K$ of the quaternionic tangent space $T({\cal M})$. Moreover the existence of the triple complex structure ${R^{rX}}_Y$ allows us to construct a local basis by an iterative process that decomposes $T({\cal M})$ in $n_H$ four dimensional subspaces: in fact if we choose a generic vector $u$ out of $V_K$ and orthogonal to the vectors therein (it is always possible to reduce to this case) also the vectors $u^r$ defined as $u^{rX}\equiv 2 {R^{rX}}_Y u^Y$ have the same properties so we can define a new subspace $V_u$ and so on. In this basis the matrices defined in (\ref{tensor}) can be written in a compact way as symmetric tensor products on $T({\cal M})$:
\eqn
C &=& B(K) =  K \otimes K + v^r \otimes v^s \delta_{rs} \\
B(u)& =& u^X B_X  =  K \otimes u + u^r\otimes v^s \delta_{rs} \\
U^r(u)& =&  u^X U^r_X  =  \frac 12  \left[ u^r\otimes K - u\otimes v^r + v^s \otimes u^t {\epsilon_{st}}^r  \right]
\label{tensor1} 
\feqn
where  $u$ is a generic vector, $u^r$ is $u^{rX}= 2 {R^{rX}}_Y u^Y= (R^r(u))^X$ and   $v^r= R^r(K)$. As consequence of the property of compositions of complex structures  ${R^r_Z}^X R^{sZY} = \frac 14 (g^{XY} \delta^{rs} + 2 {\epsilon^{rs}}_t R^{tXY}) $ some useful relations hold:
\begin{align}
B(v^r)&= 0 \\
U^r(u^s) =-\frac 12 B(u) \delta^{rs} &  + u^{[r} \otimes v^{s]} + \frac 12 {\epsilon^{rs}}_t \left( v^t \otimes u - K \otimes u^t \right)\\
\intertext{from which follows}
U^r(v^s) &= - \frac 12 C \delta^{rs}
\end{align}
As a consequence it is easy to see that for the consistency of projectors, $\Psi_a^X$ must be of the form
\eqn
\Psi_a^X & = & v_{ar} v^r
\label{psi_g}
\feqn
or in other words $D_a q^X \in V_K$.
Equations (\ref{proje1}) and (\ref{psi_g}) can be translated into
\eqn
& & \Psi_a^Y -\frac 6{|K|^2} \Psi_a^Y D_Z (WQ^s) D_Y (WQ_s) =0 \\
& & \left[ \sqrt {\frac 32} ig \delta_i^{\ j} +\gamma^a
M_a  \delta_i^{\ j} -
2i\Psi_a^X \frac {D_X P^r}{|K|^2} \gamma^a (\sigma_r)_i^{\ j}
\right] \epsilon_j =0  
\label{Ma}
\feqn
 where we write explicitly $v^{rX}$ as $2 D^X P^r$.

Now it remains to be demonstrated that the relations above are not only a necessary consequence of (\ref{1hyperini}) but also a sufficient condition for it.  

This fact can be easily checked by direct substitution  of (\ref{Ma}) in (\ref{1hyperini}) using the properties of the complex structure and gamma matrix algebra.

We then have the
following
\begin{prop}
Generically the BPS hyperini equations are equivalent to the system
\eqn
& &\partial_a q^X +gA_a^I K_I^X =M_a K^X + 2 v_{ar} D^X P^r \nonumber\\                  & &\sqrt {\frac 32} ig \epsilon_i +\gamma^a M_a  \epsilon_i  
+ i v_a^r \gamma^a \epsilon^j(\sigma_r)_{j i} =0 
\label{hy-sys} 
\feqn
These expressions contain all the constraints imposed on the scalars of hypermultiplets due to the quaternionic geometry.
\end{prop}

They explain the structure of the equations found for example in 
\cite{noialtri,ceresole:2001}.
Surely further restrictions arise  out from the BPS equation for the gravitini
but they will depend on the specific configuration of the
Einstein-Maxwell fields. We want to emphasize that this result holds also for a  non abelian gauge group with a generic number of vector multiplets. \\

\section{Geometric classification: results}\label{geocla}

In this section we present the results obtained in the presence of only hypermultiplets, following the analysis in \cite{0304064}. We discuss the new features arising with respect to the minimal gauged case studied there.

We start by assuming the existence of a Killing spinor and constructing the bosonic quantities from it:\footnote{We choose different conventions than in \cite{0304064}; the dictionary between our and their (primed) quantities is:
\begin{eqnarray*}
& & \eta_{ab}^\prime = - \eta_{ab}~~~~~~~~~\gamma_a^\prime = -i  \gamma_a \\
& & F^\prime = \frac 1{\sqrt 2} F~~~~~~~~~\chi \delta^{r1} = 2 \sqrt 2 g P^r
\end{eqnarray*}
where they choose the following representation for the Pauli matrices
\begin{eqnarray*}
\sigma_1 = \left( \begin{array}{cc} 0 & i \\ -i & 0 \end{array} \right) \ ,
\qquad
\sigma_2 = \left( \begin{array}{cc} 0 & 1 \\ 1 & 0 \end{array} \right) \ ,
\qquad
\sigma_3 = \left( \begin{array}{cc} 1 & 0 \\ 0 & -1 \end{array} \right) \ .
\end{eqnarray*}
Because we take $f^\prime=f$, ${V^\prime}_a=V_a$ and ${\phi^\prime}_{ab}^{ij}=\phi_{ab}^{ij}= i \chi^r_{ab} {(\sigma_r)_i}^j$ their definitions (\ref{bos_def}) in term of the spinor change with respect to those in \cite{0304064}.}
\begin{gather}
i f \varepsilon^{ij} = \bar{\epsilon}^i \epsilon^j \nonumber\\
- V_a \varepsilon^{ij} = \bar{\epsilon}^i \gamma_a \epsilon^j \label{bos_def}\\
\chi^r_{ab} (\sigma_r)^{ij}= \bar{\epsilon}^i \gamma_{ab} \epsilon^j \nonumber
\end{gather}
These are not independent and a lot of relations can be derived from the Fierz identities. In particular
\eqn
& & i_V \chi^r =0 \nonumber\\
& & V^aV_a = -f^2\nonumber\\
& & i_V(*\chi^r) = f \chi^r\label{f_id}\\
& & V^a\gamma_a \epsilon_i = f \epsilon_i\nonumber  
\feqn
Using (\ref{BPSgrav_g}) we get:
\eqn
& & df=\sqrt{\frac 23} i_V F \label{df} \\
& &\DD V = dV= -\sqrt{\frac 23} *(F\wedge V) + 2 \sqrt{\frac 23} f F + 2 \sqrt{\frac 23}g P_r \chi^r \label{dV}\\
& &\DD_a \chi^r_{bc} = \partial_a \chi^r_{bc} = \frac 1{\sqrt 6} \left[2 F_a^d \left(*\chi^r\right)_{dbc} + \eta_{a[b} \left(*\chi^r\right)_{c]de} F^{de} 
- 2 F_{[b}^d \left(*\chi^r\right)_{c]ad} + 4 g \eta_{a[b}V_{c]} P^r\right] 
\cr 
& &~~~~~~~~~~~~~~~~~~~~~~~- 2 {\varepsilon^r}_{st}\left[(\partial_a q^X p_X^s + g A_a P^s)\chi^t_{bc} + \frac g{\sqrt 6} P^s \left(*\chi^r\right)_{abc}\right]
\label{dchi}
\feqn
this imply 
\eqn
& & {\cal L}_V F=0\\
& & d\chi^r= -2{\varepsilon^r}_{st}\left[{\cal A}^s\wedge\chi^t + \sqrt{\frac 32} g P^s\left(*\chi^t\right)\right]
\feqn
where ${\cal A}^s$ is 
\eqn
& &{\cal A}^s= dq^Xp^s_X + g A P^s 
\feqn
As a consequence of (\ref{dchi}) and of the hyperini equations (\ref{hy-sys})
we obtain 
\eqn
{\cal L}_V \chi^r & = &(i_V \circ d) \chi^r = \cr
                  &   & -2 g (\sqrt{\frac 32}f + i_V A) {\varepsilon^r}_{st} (P^s-K^s)\chi^t
\feqn
It is possible, like in  \cite{0304064}, to use the gauge freedom to have ${\cal L}_V\chi^r= 0$ choosing $i_V A = - \sqrt{\frac 32} f$.

It would be very interesting to study the consequences of gauge invariance of the BPS equations on general grounds and in particular to determine  the space time dependence of the embedding of the gauge group U(1) in the R-symmetry group SU(2). Here we present a preliminary analysis in the case of an infinitesimal transformation:
\eqn
& & q^X \longmapsto q^X + g \Lambda K^X \\
& & A_\mu \longmapsto A_\mu - \partial_\mu \Lambda \\
& & \epsilon_i \longmapsto  {R_i}^j \epsilon_j~~~~~~ R=e^{\frac i2\alpha^r \sigma_r}\simeq \unity + \frac i2 \alpha^r \sigma_r 
\feqn
and as a consequence 
\eqn  
& & p^r \longmapsto p^r + g \Lambda {\cal L}_K p^r = p^r + g \Lambda \left(-DP^r+DK^r\right); ~~~~~~~K^r\equiv K^X p^r_X 
\cr & &\\
& & P^r \longmapsto P^r + g \Lambda {\cal L}_K P^r = P^r + 2 g \Lambda {\epsilon^r}_{st} P^s K^t
\feqn
Using the above relations in (\ref{BPSgrav_g}) we obtain an equation for the compensator $\alpha^r$ (see sect. \ref{momentm1})
\eqn
& &\left\{D_a \left[\alpha^r - 2 g \Lambda (K^r-P^r)\right] + 2g\left(A_a + \frac i{\sqrt 6} \gamma_a \right) {\varepsilon^r}_{st} P^s \left(\alpha^t - 2 g \Lambda (K^t-P^t)\right)\right\}\epsilon_i = 0
\cr
& &\label{gauge_inv}
\feqn

The main information we obtain from the equation above is that $V$ is a Killing vector and it generates a symmetry of all the solution (including scalars, if we choose $i_V A = - \sqrt{\frac 32} f$). To use it we have now to distinguish two cases: that $V$ is time-like or that it is light-like, meaning $f=0$.

\subsection{The time-like case}

As in  \cite{0304064} we take $V=-\partial_t$ and we write the spacetime metric as
\eqn
& & ds^2 =   - f^2 (dt+w)^2 + f^{-1} h_{mn} dx^m dx^n \label{metrica}
\feqn
We take $f>0$\footnote{From the Fierz identities this corresponds to taking $\gamma_0 \epsilon = \epsilon$.} , $w$ and $h$ not depending  on $t$, $e^0=f(dt+w)$ and $fdw = G^+ +G^-$. 
Specializing  the Fierz identities to this case we find that $w$ is anti-self-dual as a 4d 2-form on the base space $B$. An important relation is
\eqn
& & {\chi^r_m}^p{\chi^s_p}^n = -\delta^{rs} {\delta_m}^n + {\epsilon^{rs}}_t {\chi^t_m}^n\label{chi_algebra}
\feqn
where $m$, $n$, $p$ are curl indices of $B$. From (\ref{df}) and (\ref{dV}) we derive $F$
\eqn
& & F= H -2 g f^{-1} P_r \chi^r
\feqn
where $H = \sqrt{\frac 32} de^0 - \sqrt{\frac 23}G^+$.
Following the calculations of \cite{0304064} we express $\DD_a\chi^r_{bc}$ in terms of $H$ (in this way we isolate the part contributing to the pull-back on B)
\eqn
& & \DD_a \chi^r_{bc} = \frac 1{\sqrt 6} \left[2 H_a^d \left(*\chi^r\right)_{dbc} + \eta_{a[b} \left(*\chi^r\right)_{c]de} H^{de} 
- 2 H_{[b}^d \left(*\chi^r\right)_{c]ad}\right] 
\cr 
& &~~~~~~~~~~~~~~~~~~~~~~~-2 {\varepsilon^r}_{st}\left[(\partial_a q^X p_X^s + g A_a P^s)- \sqrt{ \frac 32} g f^{-1} P^s V_a\right] \chi^t_{bc}
\cr 
& &\label{dchi1}
\feqn
where we use $*\chi^r =- f^{-1} V\wedge\chi^r$.
Writing (\ref{dchi1}) with respect to the metric $h$ of $B$ we end up with 
\eqn
& &\nabla_m \chi^r_{pq} = -2  {\varepsilon^r}_{st}{\cal A}^s_m \chi^t_{pq}\label{base_str}\\
& & {\cal A}^s_m = \partial_m q^Xp^s_X + g \left(A_m-\sqrt{\frac 32} e^0_m\right)P^s\ =\ \partial_m q^Xp^s_X + g \left(A_m- \sqrt{\frac 32} f w_m\right)P^s
\cr
& &
\feqn
where $\nabla_m$ is the Levi--Civita connection on $B$ with respect to $h$.
Eq. (\ref{base_str}) needs some comments: it appears to be the direct generalization of the minimal gauged case,\footnote{This can be easily obtained in our framework as the sub--case $n_H$ choosing a constant prepotential $P^r$ in the $\sigma_1$ direction with $W$ proportional to the cosmological constant.} where a preferred (constant) direction exists, with the inclusion of a term coming from the scalars of the hypermultiplets. It is natural to suppose that the convenient quantity to perform a geometric classification of the solution is not directly $\chi^r$ but its combination with a two form depending on the matter that links together the two quaternionic manifolds.  This point deserves more study.   

To obtain some further insight, note that to invert (\ref{base_str}) and solve for ${\cal A}^s_m$ we need to know the value of ${\cal A}^r_m \chi^s_{pq} \delta_{rs}$. In order to evaluate this term we study systematically the relation between $D_m q^X$ and $f$, $V$ and $\chi^r$ imposed by the hyperini equations. All the information can be extracted by contracting (\ref{hy-sys}) with $\bar{\epsilon}^k$ and $\bar{\epsilon}^k \gamma_b$. This gives
\eqn
& & i_V v^r = 0\\
& & M_a = \frac 1f \left[ \sqrt{\frac 32} g V_a + v^b_r \chi^r_{ba}\right]\\
& & D_m q^X = v^n_r\left[\chi^r_{nm} K^X + 2 h_{nm} D^X P^r\right] + \sqrt{\frac 32} g f w_m K^X
\feqn
from which it follows that $\partial_0 q^X =D_0 q^X - g A_0 K^X=0 $\footnote{This is a consequence of the gauge choice and at the same time elucidates its meaning.}.
We note that the equations above, together with (\ref{chi_algebra}), satisfy   (\ref{hy-sys}) identically, hence they are the only requisites for the compatibility of hyperini equation with the gravitini one.

We can rewrite ${\cal A}_m^s$ as
\eqn
{\cal A}_m^s =  v^n_r\left[\chi^r_{nm} K^X + 2 h_{nm} D^X P^r\right] p_X^s +
               g \left(A_m- \sqrt{\frac 32} f w_m\right)\left(P^s-K^s\right)
\feqn 

Now we return to the discussion of gauge invariance: contracting (\ref{gauge_inv}) by $\bar{\epsilon}^j$ we find 
\eqn
& &D_a \left[\alpha^r - 2 g \Lambda (K^r-P^r)\right] + 2g\left(A_a -\frac 1{\sqrt 6 f} V_a \right) {\varepsilon^r}_{st} P^s \left(\alpha^t - 2 g \Lambda (K^t-P^t)\right)= 0\cr
& &\label{gauge_inv1}
\feqn
It is interesting to study this expression in the gauge $i_V A= -\sqrt{\frac 32} f$ and consider the residual gauge transformations i.e. with $\partial_0 \Lambda=0$, which in the following  we indicate as a four dimensional gauge transformation. In this case, using that $\partial_0 q^X = 0$ the above relation implies\footnote{It is easy to check using (\ref{alpha}) that  (\ref{base_str}) is four dimensional gauge covariant.}
\eqn
& & \alpha^r = 2g \Lambda (K^r-P^r) + \beta P^r \label{alpha}
\feqn
where $\beta$ is determined by the condition 
\eqn
& & D_{\bar{a}}(\beta P^r) = \partial_{\bar{a}} \beta ~P^r + \beta \partial_{\bar{a}} q^X D_X P^r =0 ~~~~~~~~~~~~~        \bar{a}=1,...,4
\feqn
This condition has a nice interpretation: it is equivalent to requiring either that $\beta=0$ or 
\eqn
& & v^r_{\bar{a}}= -\frac 2{K^2} (\partial_{\bar{a}} \log \beta) ~P^r \label{splitting}  
\feqn
The second possibility occurs for example in the electro-static spherically symmetric configurations studied in \cite{noialtri} and in the flat domain walls \cite{ceresole:2001}: as a consequence in these cases the BPS equation for the scalar (\ref{hy-sys}) results $\partial q^X \propto \partial^X W$ where $W$ is the prepotential. In general we obtain 
\eqn
& &\partial_{\bar{a}} q^X = \left(\frac {\sqrt{6}}{K^2} f^{-1}W\chi_{\bar{a}}^{r\bar{b}}\partial_{\bar{b}} \log \beta ~Q_r - g A_{\bar{a}}\right) K^X - 2 \frac {\sqrt{6}}{K^2}\partial_{\bar{a}} \log \beta ~W\partial^X W 
\cr
& &
\label{q_split}
\feqn

\section{Discussion}

Now we want to point out which are the main goals and which questions remain open in this first approach to the classification of BPS solutions with matter couplings. First of all, we think it is already an important result to have shown that a general discussion on BPS solutions of gauged supergravity is possible also in presence of non trivially coupled hypermultiplets. Indeed we have proved that, without performing any restrictive choice on the form of the metric and of the other fields, the integrability conditions (\ref{integral}) of the gravitini are sufficient to satisfy the Einstein equations for the metric.\footnote{We imposed as usual \cite{Gauntlett:2002nw} that the Maxwell equations are satisfied.} Furthermore we have demonstrated that the hyperini equation (\ref{hyperini}) can be analyzed in full generality clarifying its geometrical meaning. This can be used to obtain some informations \emph{ a priori} on the form of the BPS equations for the scalars and on the existence of fixed point solutions. By way of this analysis we are also near to demonstrating (but this has not been presented here) the relation between the hyperini equation and the equation of motion for the scalar fields. We want to remark that these results are relevant, not only to investigate the properties of BPS solutions, but also for the usual ansatz-based construction. 
\par 
Another important contribution, although incomplete, is given by the application of the Gauntlett procedure. As we have explained in section \ref{geocla} the method works in presence of matter quite similarly than in the minimal gauged theory \cite{0304064}.
The principal difficulty arises in the expression (\ref{base_str}) which gives the covariant derivative of the 2--form $\chi^r$ on the four dimensional base space $B$. In fact eq. (\ref{base_str}) generalizes quite naturally the corresponding relation of \cite{0304064} (that can be obtained as a subcase choosing the constant prepotential in the $\sigma_1$ direction) in the sense that the phase of triplet 1--form ${\cal A}^s_m$, as for the  prepotential $P^r$, is not a constant due to the gauging of the R--symmetry $SU(2)$.
The role of (\ref{base_str}) is fundamental because it is necessary to determine the base space $B$ and to demonstrate the consistency of the method. For example in the minimal gauged  case it gives immediately that $B$ is a K\"ahler space where $\chi^1$ is the K\"ahler form and that the graviphoton $A_\mu$ satisfies the Maxwell equation. 
We do not know the exact meaning of (\ref{base_str}) yet. At first reading it indicates only that the base space is quaternionic. However we suspect that 
the K\"ahler structure is not totally destroyed here but only deformed by the presence of hypermultiplets.
In particular we believe that the gauge transformations constitute the key to understanding  the content of (\ref{base_str}). It could happen that this relation reduces to the one in absence of matter via a gauge fixing. Our idea is that the gauge transformation might ``freeze'' the tree dimensional vector $\nabla \chi^r$ on a plane determining the existence of a K\"ahler form.
The problem of this conjecture is that its checking requires the knowledge of the finite gauge transformations hence of the compensator $\alpha^r$ as function of $\Lambda$ (not only at the first order as in (\ref{alpha})). This is equivalent to determine the exact embedding of the gauge group $U(1)$ in local R--symmetry $SU(2)$ which characterizes the theory when the hypermultiplets are not trivially coupled. As a last comment, another element that suggests we are on the right track is given by the equation (\ref{q_split}) which explains and generalizes the behavior observed and discussed in chap. \ref{black}.

\addcontentsline{toc}{chapter}{Conclusion}
\chapter*{Conclusion}\label{conclu}

At this point I want to summarize what it has been presented in this thesis. I will try to convince the reader that the aim of the work has been, at least partially, achieved. 

Rather than writing a detailed list of all the results, 
which could represent a repetition of what  
has already been presented at the end of each chapter, 
I will attempt to organize them in a more general discussion.
  
We have started with the study of charged configurations in presence of matter. This has been fruitful on the one hand because it allows to construct  solutions interesting for themselves (like black holes), on the other hand because of the new insights gained in the general structure of BPS solutions.
Hence the need of substantiating these general indications arises quite naturally.
In this light what has been treated in the last chapter represents at the same time the arrival point (in which the simple ``observations'' of the previous chapters become definite properties, e.g see the hyperini equation (\ref{hy-sys}))
and a starting point for an actual classification of BPS solutions in the presence of matter. 

With this aim, the synergy between an approach based on a brute force handling 
of the BPS conditions and the geometric approach due to Gauntlett appears to 
be fundamental in order to find out the aforementioned classification.  
If the former is limited by the evident difficulty to manipulate such 
tricky expressions, the latter gives rise to results which have still to be 
interpreted from a physical point of view.  

For example, it is not trivial to associate the explicit solution discussed in 
chapter \ref{black} to solutions obtained by means of Gauntlett's method. 
It is even not known if this solution is time--like or light--like 
in Gauntlett's framework. 


In order to explain these difficulties, let us consider the time--like case (the light--like one does not present new features). 
As already noted in \cite{Gauntlett:2002nw}, the time, as it appears in the metric (\ref{metrica}),  
is associated with supersymmetry and it does not correspond in general to the ``physical''  time which is related to the ADM mass (if any). 
 It means that, also when the solution has been determined with respect to the geometry of the base space, there remains however a lot of work to be done. 
This will require to follow both the above approaches.

As far as the perspectives in next future are concerned, we are going to overcome the technical obstacles  (like eq. (\ref{base_str})) already discussed, and we are confident  that this goal can be achieved on the grounds of our conjecture about the role of the residual gauge freedom. Then, supported by our analysis in chapter \ref{vet}, we hope to extend the results of the classification to the inclusion of the vector multiplet couplings.  

A more ambitious task would consist in finding a link with M--theory, in particular with the results of classification which has been started in \cite{0212008}. However this requires a deeper understanding of the compactification down to five dimensions.


\addcontentsline{toc}{chapter}{Acknowledgments}
\chapter*{Acknowledgments}

I would like to thank my advisor Daniela Zanon for her guidance over the last nearly three years. She gave me plenty of space to learn on my own, pushed me when I needed it, and supported me to the end.

A special note of appreciation goes to Sergio Cacciatori, who has been my principal collaborator  during my Ph.D, for his fundamental contribution to the results presented in this work. My gratitude to him is not only due to the great number of things I have learned about physics from  our discussions, but to the fact that he is, first of all, a friend. I hope that our collaboration will fruitfully continue in the future.

I would like to thank Antoine Van Proeyen for the interest showed in my research activity and for the useful indications and suggestions I have received from him.

I would also like to thank Bert, Geert and Joris that have been very kind and patient reading my thesis and helping me to clean up all the bugs and typos. I have appreciated a lot how they, as other people,  have welcomed me in Leuven.

Non posso non ricordare Francesco per aiuto, in campo scientifico e non, ricevuto in questi anni e per la disponibilità che ha sempre dimostrato. Il suo autorevole parere  è stato sempre molto gradito, richiesto e ascoltato da me, Federica e Luca, con i quali  non ho solo condiviso tanti pranzi in ufficio ma l'intera avventura del dottorato. A Luca in particolare va il mio grazie più sincero per il vicendevole sostegno su cui ho sempre potuto contare nonché per le interessanti discussioni a tutto campo. Ringrazio e saluto tutti coloro che ho avuto occasione di conoscere a Milano e con i quali conto di rimanere in contatto.

Da ultimo un saluto e un grazie del tutto speciale va alla mia famiglia: ai miei genitori, che hanno sempre dato fiducia alle mie scelte sostendole con ogni sforzo, a mia nonna e mia sorella, che ricordo con tanto affetto, e infine ad Antonella, luce dei miei occhi, che ha avuto il coraggio e la sventatezza di legare la sua vita alla mia, condividendone nel bene e nel male tutte le vicende.

\addcontentsline{toc}{chapter}{Appendix}

\appendix

\chapter{Conventions}\label{convention}

In this appendix we present some definitions and properties that we use in our work.
With 
\eqn \label{quatscalars} q^X \, \qquad {\scriptstyle X}=1, \ldots , 4 n_H 
\feqn 
we denote the scalars of the hypermultiplets which are the coordinates of a quaternionic manifold.
We introduce the $4n_H$beins as 
\eqn \label{fielbein} f^{iA}_X (q^Y ) \ ,\qquad i=1,2 \in SU(2) \ 
, A=1,\ldots,2n_H \in Sp(2n_H) \feqn

The splitting of the flat indices
in $i$ and $A$ reflects the factorization of the holonomy group in $USp(2)(\simeq SU(2))\otimes USp(2n_H)$ which is the main feature of those spaces. 
The indices as a consequence of  the symplectic structure are highered and lowered with the antisymmetric 
matrices \eqn
&& \epsilon_{ij} \ , \qquad \mathbb{C}_{AB} \\
&& \epsilon_{ij}=\epsilon^{ij} \ , \qquad \epsilon_{12} =1 \\
&& \mathbb{C}_{AB} \mathbb{C}^{CB} =\delta_A^{\quad C} \ , \qquad \mathbb{C}^{AB} =(\mathbb{C}_{AB})^* \ .
\feqn
following the NW-SE convention \cite{ceresole:2001}. 

The important relation 
\eqn 
f_{XiC} f_{Yj}^{\ \ C} =\frac 12 \epsilon_{ij} g_{XY} + R_{XYij} 
\label{fquadrato} 
\feqn
can be viewed as  a definition for the quaternionic metric $g_{XY}$ and for the $SU(2)$ curvature $R_{XYij}$.

We use the symbols $p_{Xi}^{\ \ \ j}$ for the $SU(2)$ spin connection
whereas $\omega_\mu^{ab}$ denotes the usual Lorentz spin connection.
The covariant derivative which appears in the gravitini supersymmetry variation  (\ref{gaususrul}) acts on the  symplectic Maiorana spinors $\epsilon_i$ as
\eqn
\DD_\mu \epsilon_i  &=& \partial_\mu \epsilon_i +\frac 14
\omega^{ab}_\mu \gamma_{ab} \epsilon_i -\partial_\mu q^X p_{Xi}^{\ \ j}
\epsilon_j 
- g A^I_\mu {P_I}_i^{\ j} \epsilon_j
\feqn
where  the generalized spin connection receives the following contributions: the first term represents 
the Lorentz
action while the others can be identified  with the $SU(2)$ action plus a term due  
to the $SU(2)$ R-symmetry  gauging. $A^I_{\mu}$ are $(n_V+1)$ $1-$forms and $P_I^r$ are the prepotentials while $g$ is the gauge coupling. We adopt the convention to define for the quantities with a $I$ index the corresponding ``dressed'' ones like $P^r\equiv P_I^r h^I$ or $F^{\mu\nu}\equiv F^{\mu\nu}_I h^I$. We note that in this notation the subcase $n_V=0$ is recovered in a natural way being $I=0$ and $h^I=h^0=1$.\footnote{This implies, from the definition of very special geometry (\ref{defispec}), that the normalization of $C_{I J K}$ is given by $C_{000}=1$.} 

It is useful to introduce the projection on the Pauli matrices for quantities in the adjoint representation of $SU(2)$, for example
\eqn
R_{XYi}^{\qquad j}=R_{XY}^r (i\sigma_r)_i^{\ j} \label{pro}
\feqn
where $(\sigma_r)_i^{\ j}$ are the usual Pauli matrices
\eqn
\sigma_1 = \left( \begin{array}{cc} 0 & 1 \\ 1 & 0 \end{array} \right) \ ,
\qquad
\sigma_2 = \left( \begin{array}{cc} 0 & -i \\ i & 0 \end{array} \right) \ ,
\qquad
\sigma_3 = \left( \begin{array}{cc} 1 & 0 \\ 0 & -1 \end{array} \right) \ .
\feqn
which satisfy
\eqn
& & (\sigma_r)_i^{\ j} (\sigma_s)_j^{\ k} =\delta_{rs} \delta_i^{\ k}
+i\epsilon_{rs}^{\quad t} (\sigma_t)_i^{\ k} \label{prodotto} \\
& & \left[\sigma_r ,\sigma_s \right] =2i
\epsilon_{rst} \sigma^t \label{parentesi}
\feqn
The prepotentials are defined by the relation
\eqn
& & R_{XY}^r K^Y = D_X P^r \label{prepot} \\
& & D_X P^r := \partial_X P^r +2\epsilon^{rst} p_X^s P^t 
\feqn 
where $D_X$ is the $SU(2)$ covariant derivative.
They can be expressed in terms of the Killing 
vectors 
\eqn P^r =\frac 1{2n_H} D_X K_Y R^{rXY} \label{killpot} 
\feqn 



The scalar potential (\ref{ceregatpot}) can be expressed for a generic number
of hypermultiplets and vector multiplets as
\begin{equation}
{\cal V}= g^2 [- P_r P^r +2 P_{xr} P^r_y g^{xy} + 2 N_{Ai}N^{Ai}] 
\label{scalarpot}
\end{equation} 
with
\eqn
& & N^{Ai}= \frac {\sqrt 6}4 h^I K_I^X f^{Ai}_X = 
\frac 2{\sqrt 6} f^{Ai}_X R^{rYX} D_Y P^r \ , \\
& & P^r_x \equiv -\sqrt{\frac 32} \partial_x P^r = h_x^I P_I^r
\feqn

Defining the superpotential $W$ by $P^r= \sqrt{\frac 32} W Q^r$ with $Q^rQ_r=1$ the potential becomes 
\eqn
{\cal V} = -6 g^2 W^2  + \frac92 g^2 [ g^{\Lambda \Sigma} \partial_\Lambda W \partial_\Sigma W + W^2 g^{xy}\partial_x Q^r \partial_y Q_r ]
\feqn
where $\Lambda$ is the curl index of the entire $n_V+4n_H$--dimensional scalar manifold. From the above relation it follows that the requirement on ${\cal V}$ to be of the form (\ref{townskend}), which ensures the gravitational stability, is 
$$\partial_x Q^r=0$$ as found in the sect. \ref{gauequa_v}.

The universal hypermultiplet ($n_H=1$) corresponds to the quaternionic K\"ahler space $\frac{SU(2,1)}{SU(2)\times
U(1)}$. A significant parametrization, from a M-theory point of view, is \cite{ceresole:2001}
$$q^X= \{V,\sigma, \theta, \tau\}$$ with the metric
\begin{equation}
\rmd  s^2 = \frac{\rmd V^2}{2V^2} + \frac{1}{2V^2}\left( \rmd \sigma + 2
\theta \, \rmd  \tau - 2 \tau \, \rmd  \theta\right)^2 + \frac{2}{V} \,
\left(\rmd \tau^2 + \rmd  \theta^2\right) \,. \label{quatmetric}
\end{equation}
Using the general properties of quaternionic geometry it is possible from (\ref{quatmetric}) to derive explicitly all the quantities presented above, in particular the Killing vectors and the prepotentials of the eight isometries of manifold. For the axionic shift we have:
\begin{equation}
\vec{k}_1 = \left(
\begin{array}{c}
0 \\ 1 \\ 0 \\ 0
\end{array}
\right)\,
~~~~~~~~~~~~~~~~~~~~~~~~~~~~~~~~~~~~~
\vec{P}_1 = \left(
\begin{array}{c}  0 \\ 0  \\ -\frac{1}{4V} \end{array}
\right)\,
\end{equation}

The Fierz identity is given by: 
\eqn
  \bar{\epsilon}_1 \epsilon_2 \bar{\epsilon}_3 \epsilon_4 = \frac{1}{4}
\left( \bar{\epsilon}_1 \epsilon_4 \bar{\epsilon}_3 \epsilon_2 +
\bar{\epsilon}_1 \gamma_a \epsilon_4 \bar{\epsilon}_3
\gamma^a \epsilon_2 - \frac{1}{2} \bar{\epsilon}_1
\gamma_{a b} \epsilon_4 \bar{\epsilon}_3
\gamma^{ab} \epsilon_2 \right) 
\feqn
 Most of the algebraic identities we recorded in section \ref{geocla} were obtained by using the Fierz identity with  $\bar{\epsilon}_1 = \bar{\epsilon}^i$,
$\epsilon_2 = \epsilon^l$, $\bar{\epsilon}_3 = \bar{\epsilon}^k$
and then setting in turn $\epsilon_4 = \epsilon^j$, $\gamma_a
\epsilon^j$ and $\gamma_{a b} \epsilon^j$. The remaining
identites were obtained using $\bar{\epsilon}_1 =
\bar{\epsilon}^i$, $\epsilon_2 = \gamma_a \epsilon^l$,
$\bar{\epsilon}_3 = \bar{\epsilon}^k$ and $\epsilon_4 =
\gamma_b \epsilon^j$.

\chapter{Equations of motion}
\label{app:motion}
The equations of motion of the lagrangian (\ref{genn2d5gau})in the presence of hypermultiplets and vector multiplets are
\eqn
-\RR_{\mu\nu} +a_{IJ} F^I_{\mu a} F_\nu^{Ja} +g_{XY} D_\mu q^X D_\nu q^Y 
+g_{x y} D_\mu \phi^x D_\nu \phi^y -\frac 16 |F|^2 g_{\mu \nu}
+\frac 23 \VV g_{\mu \nu} =0 \cr
&&\label{einstein_v}
\feqn
from which it follows in particular
\eqn
-\frac 32 R +\frac 14 |F|^2 +\frac 32 |q^\prime |^2 +\frac 32
|\phi^\prime |^2 - \frac 94 |a^I K_I|^2 + 5 \VV =0  \label{particular_v}
\feqn
Variation with respect to the gauge fields gives
\eqn
\DD_a (a_{IK} F^{Kae}) +\frac 1{2 \sqrt 6} C_{IJK} \epsilon^{abcde}
F^J_{ab} F^K_{cd} -g K^X_I D^e q^Y g_{XY}=0 \label{maxwell_v}
\feqn
Finally the equations for the scalars are
\eqn
& & \hat{D}_\mu D^\mu q^W +gA^{\mu I} D_\mu K_I^W = g^{WX} \partial_X 
\VV \label{qq_v} \\
& & \hat{D}_\mu D^\mu \phi^x +gA^{\mu I} D_\mu K_I^x = g^{x y} 
\partial_y \VV +\frac 14 g^{x y} \partial_y a_{IJ} F^I_{\mu \nu}
F^{J\mu \nu} \label{fifi_v} 
\feqn
Here $\DD$ is the covariant derivative with respect to the spin connection and 
$\hat{D}$ is a totally covariant derivative, ie with respect
to all the indices. So for example
\eqn
\hat{D}_\mu D^\mu \varphi^\Lambda =\DD_\mu D^\mu \varphi^\Lambda 
+\Gamma^\Lambda_{\Sigma\Theta} D_\mu \varphi^\Sigma D^\mu \varphi^\Theta \nonumber
\feqn
and in general
\eqn
D_\mu f^* (q ,\phi )= D_\mu q^X \partial_X f^* + D_\mu \phi^x
\partial_x f^* 
\feqn
Now we specialize the above relations to the problem studied in the chapter \ref{vet}.
Due to symmetry of the class of solutions considered only the Einstein equations for the components $(tt)$, $(rr)$ and $(\theta \theta)$ are independent:
\eqn
& & -e^{v-w} \partial_r (v^\prime e^{v-w}) -3\frac {v^\prime}r  e^{2(v-w)}
+e^{2(v-w)} (v^\prime \Lambda + \Lambda^\prime)^2 + 4 g^2 e^{2v} W^2~~~~~~~~~~~~~~~~~ \cr
& & ~~~~~~~- 3 g^2 e^{2v} (1- 3 \Lambda^2) [\frac 1{1-\Lambda^2} g^{xy} \partial_x W \partial_y W + g^{XY} \partial_X W \partial_Y W] = 0
\label{e-tt_v} 
\feqn
\newpage
\eqn 
& & e^{w-v} \partial_r (v^\prime e^{v-w}) -\frac 3r w^\prime
- (v^\prime \Lambda + \Lambda^\prime)^2 - 4 g^2 e^{2w} W^2 \cr
& & ~~~~~~~~~~~~~~~~~~~~~~~+ 3 g^2 e^{2w} (1 + 3(1-\Lambda^2))[
\frac 1{1-\Lambda^2} g^{xy} \partial_x W \partial_y W + g^{XY}
\partial_X W \partial_Y W] = 0
\cr 
& &\label{e-rr_v} \\
& & re^{-2w} (v^\prime -w^\prime ) -2(1-e^{-2w}) 
+ \frac 12 r^2 e^{-2w} (v^\prime \Lambda + \Lambda^\prime)^2 - 4 g^2 r^2 W^2 \cr & & ~~~~~~~~~~~~~~~~~~~~~~~+ 3 g^2 r^2 [\frac 1{1-\Lambda^2} g^{xy}
\partial_x W \partial_y W + g^{XY} \partial_X W \partial_Y W] = 0
\label{e-thetatheta_v} \\
& & \sqrt {\frac 32 }ge^v a^I K_{IX} q^{\prime X} =0 \label{e-rt_v}
\feqn
where we use the BPS relations for $\phi^{\prime x}$, $q^{\prime X}$ and
$|K|^2=6 g^{XY} \partial_X W \partial_Y W$. Following the manipolations of
\cite{noialtri} we consider the sum of (\ref{e-tt_v}) and (\ref{e-rr_v}) multiplied by $e^{2(v-w)}$
\eqn
\frac{(v^\prime + w^\prime)}r e^{-2w} & = & 3 g^2 [\frac 1{1-\Lambda^2} g^{xy} \partial_x W \partial_y W + g^{XY} \partial_X W \partial_Y W] \\
& = & \pm g \frac{e^{-w}}{\sqrt{1-\Lambda^2}} W^{\prime}
\label{tt+rr_v}
\feqn
The above expression is the direct generalization of the one in \cite{noialtri}
and is identically satisfied by (\ref{1fr_v}) and (\ref{b-c_v}). Now by the
substitution of (\ref{tt+rr_v}) in (\ref{e-tt_v}), (\ref{e-rr_v}) and
(\ref{e-thetatheta_v}) it is easy to check that also these expressions are
identically satisfied by the set of BPS equations.
Finally (\ref{e-rt_v}) is solved by (\ref{333_v}) and the equation
$q^{\prime Z} K_Z =0$ which follows for example from (\ref{erreeffe_v}) and
(\ref{3restrizione_v}).

Next consider the equatons for the gauge fields:
\eqn
& & K^I_X q^{\prime X}=0 \label{em-r_v} \\
& & \partial_r (a_{IJ} e^{-w} r^3 (v' a^J +a^{\prime J})) -e^w r^3 g^2
g_{XY} K_I^X K_J^Y a^J =0 \label{em-t_v}
\feqn
It is convenient to project these equations on the base $(h_I , h_J^x )$.
The contraction of (\ref{em-r_v}) with $h_I$ gives $K_X q^{\prime X} =0$ which
we already shown to be consequence of BPS equations.

The contraction with $h_{Ix}$ gives $q^{\prime X} K_X^I h_{Ix} =0$ which by
means of (\ref{erreeffe_v}) is equivalent to $\partial^X W K_X^I h_{Ix} =0$. But
from (\ref{6restrizione_v}) and (\ref{chiq_v}) we have
\eqn
\partial_x K^Z =\frac {\partial_x W}W K^Z  \label{simply_v}
\feqn
so that
\eqn
\partial^X W K_X^I h_{Ix} =\sqrt {\frac 32} \partial^X W \partial_x K_X
=\sqrt {\frac 32} \frac {\partial_x W}W \partial^X W K_X =0
\feqn
After an integration by parts and using (\ref{key_v}) the contraction of
(\ref{em-t_v}) with $h^I$ gives
\newpage
\eqn
\partial_r [r^3 e^{-w} (v' \Lambda +\Lambda')] +\sqrt {\frac 23} r^3 e^{-w}
\phi^{\prime x} h_{Ix} (v' a^I +a^{\prime I}) -e^w r^3 g^2 g_{XY} K^X K^Y_J
a^J =0\cr
&&
\feqn
and using (\ref{bene_v})
\eqn
\partial_r [r^3 e^{-w} (v' \Lambda +\Lambda')] \mp \frac {2gr^3 \Lambda
W'}{\sqrt {1-\Lambda^2}} \pm \frac {2gr^3 \Lambda q^{\prime X}
\partial_X W}{\sqrt {1-\Lambda^2}} -e^w r^3 g^2 g_{XY} K^X K^Y_J a^J =0 \cr
&&
\feqn
which after use of (\ref{333_v}) can be easily related to the computations
in \cite{noialtri} with $\Lambda$ in place of $a$.

The contraction with $h_y^I$ gives
\eqn
& & r^3 e^{-w} (\Lambda' +v' \Lambda) \sqrt {\frac 23} \phi^{\prime x} h_y^I
h_{Ix} \mp h_y^I \partial_r h_I^x
\frac {\partial_x W \sqrt 6 g\Lambda r^3}{\sqrt {1-\Lambda^2}}~~~~~~~~~~~~~~~~~~~~~~~~~~~~~~ \cr
& & ~~~~~~~~~~~~~~~~~~~~~~~\mp \partial_r
\left( \frac {\partial_y W \sqrt 6 g\Lambda r^3}{\sqrt {1-\Lambda^2}} \right)
=-g^2 r^3 e^w \sqrt {\frac 32} \partial_y K^Z K_Z
\feqn
From (\ref{1fr_v}) and (\ref{strana_v}) we find
\eqn
-rW \phi^{\prime x} =3\partial^x W
\feqn
and from (\ref{corol_v}), (\ref{1fr_v})
\eqn
g^2 e^{2w} rW =-2(W' -\phi^{\prime x} \partial_x W)
\feqn
Using these last equations together with (\ref{simply_v}) and (\ref{b-c_v})
we have
\eqn
h_y^I \partial_r h_I^x \frac {\partial_x W \Lambda r^2}W +\Lambda \partial_r
\left( \frac {\partial_y W r^2}W \right) =-r^2 \Lambda \frac {\partial_y W}W
\phi^{\prime x} \frac {\partial_x W}W  \label{chefatica_v}
\feqn
By means of (\ref{chiq_v}) and some integration by parts
\eqn
h_y^I \partial_r h^x_I \partial_x W &=& \sqrt {\frac 23} h_y^I \partial_r
h_I^x \partial_x h^J P_J^s Q_s \cr
&=& -h_y^I \partial_r (h_I^x h_x^J )\frac 23 P^s_J Q_s -
\sqrt {\frac 23} \partial_r \partial_y h^J P^s_J Q_s \cr
&=& h_y^I \partial_r (h_I h^J) \frac 23 P_J^s Q_s -\sqrt {\frac 23} \partial_r
[\partial_y (P^s Q_s)] +\sqrt {\frac 23} q^{\prime X} \partial_X
[\partial_y (P^s Q_s)] \cr
&=& \frac 23 g_{yx} \phi^{\prime x} W -\partial_r (\partial_y W) +q^{\prime X}
\partial_X (\partial_y W) \cr
&=& -2r\Lambda \frac {\partial_x W}W-\partial_r (\partial_y W) +q^{\prime X}
\partial_X (\partial_y W) \label{fondam_v}
\feqn
where in the last step we have used (\ref{strana_v}) and (\ref{1fr_v}).
This together with (\ref{erreeffe_v}) and (\ref{7restrizione_v}) shows that
(\ref{chefatica_v}) is identically satisfied.

The equations of motion for the hyperini are
\eqn
& &e^{-(v+w)} r^{-3} \partial_r (r^3 e^{v-w} g_{ZY} q^{\prime Y})
-\frac 12 q^{\prime X} \partial_X g_{ZY} q^{\prime Y} e^{-2w} +\frac 34 g^2
\partial_Z g_{XY} a^I K_I^X a^J K_J^Y \cr
& & +\frac 32 g^2 g_{XY} a^I a^J \partial_Z K_I^X K_J^Y =
g^2 \partial_Z \left( -6W^2 +\frac 34 K^2 +\frac 92 g^{xy} \partial_x W
\partial_y W \right)
\feqn
Using (\ref{333_v}), (\ref{erreeffe_v}), (\ref{7restrizione_v}) and (\ref{strana_v})
it becomes
\eqn
\pm e^{-w} 9 \frac gr \sqrt {1-\Lambda^2} \partial_Z W &\pm& 3(v'-w') ge^{-w}
\sqrt {1-\Lambda^2} \partial_Z W \cr
& \pm& e^{-2w} 3g \partial_Z W \partial_r (e^w \sqrt {1-\Lambda^2}) =-12g^2 W\partial_Z W
\cr
& & \feqn
which follows from (\ref{1fr_v}) and the considerations in \cite{noialtri}.

The equations of motion for the gaugini are
\eqn
& &\frac 34 \partial_x a_{IJ} e^{-2w} (a^{\prime I} +v' a^I) (a^{\prime J}
+v' a^J) -\frac 12 \partial_x g_{yz} \phi^{\prime y} \phi^{\prime z} e^{-2w}
+r^{-3} e^{-(v+w)} \partial_r (r^3 e^{v-w} \phi^{\prime y} g_{xy}) \cr
& & -g^2 \partial_x \left( -6W^2 +\frac 34 K^2 +\frac 92 g^{zy} \partial_z W
\partial_y W \right)=0
\feqn
From (\ref{strana_v}), (\ref{erreeffe_v}) and (\ref{7restrizione_v}) we find
\eqn
& & r^{-3} e^{-(v+w)} \partial_r (r^3 e^{v-w} \phi^{\prime y} g_{xy}) =
\pm 9\frac gr \frac {e^{-w} \partial_x W}{\sqrt {1-\Lambda^2}}
\pm 3gv' \frac {e^{-w} \partial_x W}{\sqrt {1-\Lambda^2}} \pm 3g
e^{-w} \partial_x W \partial_r \frac 1{\sqrt {1-\Lambda^2}} \cr
& &~~~~~~~~~~~~~~~~~~~~~~~~~~~~~~~~~~~~ +3g \frac {e^{-w}}{\sqrt {1-\Lambda^2}} \left[
\frac {3ge^{w} \partial_y\partial_x W \partial^y W}{\sqrt {1-\Lambda^2}}
+3ge^w \partial_x \partial_X W \partial^X W \sqrt {1-\Lambda^2} \right] \cr
&&
\\
&& -g^2 \partial_x \left( -6W^2 +\frac 34 K^2 +\frac 92 g^{zy} \partial_z W
\partial_y W \right) = 12g^2 W\partial_x W -9g^2 \partial_x \partial_Y W
\partial^Y W \cr
&&~~~~~~~~~~~~~~~~~~~~~~~~~~~~~~~~~~~~~~~~~~~~~~~~~~~~~~ -  \frac 92 g^2 \left(\partial_x g^{yz} \partial_y W \partial_z W + 2 \partial_x \partial_y W \partial^y W\right)\\
& & -\frac 12 \partial_x g_{yz} \phi^{\prime y} \phi^{\prime z} e^{-2w} =
-\frac 92 g^2 \frac {\partial_y W \partial^y W}{\sqrt {1-\Lambda^2}}
\label{elaborate_v}
\feqn
In a similar way as  (\ref{fondam_v}) one finds
\eqn
\partial_x h_I^y \partial_y W =\frac 23 W h_{Ix} +h_I \sqrt {\frac 23}
\partial_x W -h_I^y \partial_x \partial_y W
\feqn
From this and (\ref{key_v}), (\ref{bene_v}) and (\ref{a_v}) we find
\eqn
& & \frac 34 \partial_x a_{IJ} e^{-2w} (a^{\prime I} +v' a^I) (a^{\prime J}+v' a^J)
=\mp \frac {6ge^{-w} \partial_x W}{\sqrt {1-\Lambda^2}}
\left[ v'-\frac {1-\Lambda^2}r \right] +\frac {6\Lambda^2 g^2
W\partial_x W}{1-\Lambda^2} \cr
& & ~~~~~~~~~~~~~~~~~~~~~~~~~~~~~~~~~~~~~~~~~~~~~~~-\frac 92 g^2 \frac{\Lambda^2}{1-\Lambda^2} \left(\partial_x \partial_y W \partial^y W -2 \partial_x g_{yz} \partial^y W \partial^z
W\right)\cr &&
\feqn
Summing up all the terms
\eqn
0 &=& \left( \pm 9\frac gr e^{-w} \mp 3ge^{-w} v' \pm 3ge^{-w}
\frac {\Lambda \Lambda'}{1-\Lambda^2} \right)
\frac {\partial_x W}{\sqrt {1-\Lambda^2}}
+12 g^2 W\partial_x W \cr
&+& 6\Lambda^2 g^2 \frac {W\partial_x W}{1-\Lambda^2} \pm 6ge^{-w} \partial_x W
\sqrt {1-\Lambda^2}
\feqn
Using (\ref{1fr_v}) to eliminate $e^{-w}$ from the first and the last term and
next (\ref{a_v}) to eliminate $\Lambda' +\Lambda v'$ we finally see that the
gaugini equation also are satisfied.

\chapter{Some computations} \label{computation}
\section{Derivation of Eq.(\ref{5_g})}
At first we consider the commutator
\eqn
& & \left[ \partial_\mu +\frac 14 \omega_\mu^{cd} \gamma_{cd}
\ ,\  \frac i{4\sqrt 6} 
\gamma_{nab} F^{ab} e_\nu^n \right] \epsilon_i
= \frac i{4 \sqrt 6} \gamma_{nab} \partial_\mu (F^{ab} e^n_\nu) +
\frac i{16\sqrt 6} \omega_\mu^{cd} F^{ab} e^n_\nu [\gamma_{cd},
\gamma_{nab}] \cr
& & =\frac i{4 \sqrt 6} \gamma_{nab} \partial_\mu (F^{ab} e^n_\nu) 
-\frac i{4 \sqrt 6} \omega_\mu^{cd} e_{c\nu} F^{ab} \gamma_{dab}
+\frac i{2\sqrt 6} \omega_{\mu a}^{\quad d} F^{ab} e^n_\nu \gamma_{dnb}
\label{5I}
\feqn
where we have used
\eqn
[\gamma_{cd}\ ,\ \gamma_{nab}] =-2\eta_{cn} \gamma_{dab} +2\eta_{ca} \gamma_{dnb}
-2\eta_{cb} \gamma_{dna} +2\eta_{dn} \gamma_{cab} -2\eta_{da} \gamma_{cnb} 
+2\eta_{db} \gamma_{cna} \cr
&&\label{2per3}
\feqn
If we note that
\eqn
de^n =-\omega^n_{\ a} \wedge e^a \ \Rightarrow \ 
\partial_\mu e^n_\nu -\partial_\nu e^n_\mu =
-\omega_{\mu \ c}^{\ n} e^c_\nu +\omega_{\nu \ c}^{\ n} e^c_\mu
\feqn
we finally obtain
\eqn
\left[ \partial_\mu +\frac 14 \omega_\mu^{cd} \gamma_{cd}
\ ,\  \frac i{4\sqrt 6} 
\gamma_{nab} F^{ab} e_\nu^n \right] \epsilon_i
-(\mu \leftrightarrow \nu) =
\frac i{4\sqrt 6} \left[ \gamma_{\nu ab} \DD_\mu (F^{ab}) -
\gamma_{\mu ab} \DD_\nu (F^{ab}) \right] \epsilon_i \cr
&&\label{I5}
\feqn
Next consider the commutator
\eqn
& & \left[ \partial_\mu +\frac 14 \omega_\mu^{cd} \gamma_{cd} \ ,
\ -\frac i{\sqrt 6} e_{\nu a} \gamma_b F^{ab} \right] \epsilon_j =
-\frac i{\sqrt 6} \partial_\mu \left[ e_{\nu a} \gamma_b F^{ab} \right]
-\frac i{\sqrt 6} \gamma_c \omega_{\mu \ b}^{\ c} e_{\nu a} F^{ab}\cr
&&
\feqn
where we have used
\eqn
[\gamma_{cd} \ , \ \gamma_b] =2\gamma_c \eta_{db} -2\gamma_d \eta_{cb}
\feqn
As before the total contribution is then
\eqn
\left[ \partial_\mu +\frac 14 \omega_\mu^{cd} \gamma_{cd} \ ,
\ -\frac i{\sqrt 6} e_{\nu a} \gamma_b F^{ab} \right] \epsilon_j 
-(\mu \leftrightarrow \nu)=
-\frac i{\sqrt 6} \left[ e_{\nu a} \gamma_b \DD_\mu (F^{ab}) -
e_{\mu a} \gamma_b \DD_\nu (F^{ab}) \right] \epsilon_i \cr
&&\label{II5}
\feqn
(\ref{I5}) and (\ref{II5}) together give (\ref{5_g}).

\section{Derivation of Eq.(\ref{19})}
First consider
\eqn
-\frac 1{96} F^{ab} F^{cd} e^m_\mu e^n_\nu [\gamma_{abm} \ , \ \gamma_{cdn}]
\label{comincia}
\feqn
In five dimensions one has $\gamma_0 \gamma_1 \gamma_2 \gamma_3 \gamma_4
=i^k I$ for some integer $k$. Squaring we see that it must be $i^k =\pm i$.
 From these considerations we easily find
\eqn
\gamma_{abc} =\pm \frac i2 \epsilon_{abc}^{\quad de} \gamma_{de}
\feqn
which implies
\eqn
[\gamma_{abm} \ , \ \gamma_{cdn}] & = & -\frac 14 \epsilon_{abm}^{\quad \ xy}
                       \epsilon_{cdn}^{\quad \ wz} [\gamma_{xy} \ , \gamma_{wz}] \cr & = &
-\frac 12 \epsilon_{abm}^{\quad \ xy} \epsilon_{cdn}^{\quad \ wz}
[\gamma_{xz} \eta_{\mu \nu} -\gamma_{xw} \eta_{yz} +\gamma_{yw} \eta_{xz}
-\eta_{xw} \gamma_{yz}] \cr
& =& -2 \cdot 4! \eta_{a a'} \eta_{bb'} \eta_{mm'} \gamma_x^{\ w} \delta_{c\ d\ n\ w}^{a'b'm'x}
\feqn
(\ref{comincia}) becomes then equal to
\eqn
& &\frac 12 F^{ab} F^{cd} e_{m\mu} e^n_\nu \delta_{cdnw}^{abmx} \gamma_x^{\ w} =
\cr
& &~~~~~~~~~~~\frac 1{48} F^{ab} F^{cd} e_{m\mu} e^n_\nu \gamma_x^{\ w} \left[ \delta^a_c ( -\delta^b_d \delta^m_w \delta^x_n +\delta^b_n \delta^m_w 
\delta^x_d +\delta^b_w \delta^m_d \delta^x_n -\delta^b_w \delta^x_d \delta^m_n ) \right.\cr
& &~~~~~~~~~~~- \delta^a_d ( -\delta^b_c \delta^m_w \delta^x_n +\delta^b_n \delta^m_w 
\delta^x_c
+\delta^b_w \delta^m_c \delta^x_n -\delta^b_w \delta^x_c \delta^m_n ) \cr
&& ~~~~~~~~~~~- \delta^a_n ( -\delta^b_d \delta^m_w \delta^x_c +\delta^b_c \delta^m_w  \delta^x_d
+\delta^b_w \delta^m_d \delta^x_c -\delta^b_w \delta^x_d \delta^m_c ) 
\cr 
& & ~~~~~~~~~~~\left.-  \delta^a_w ( -\delta^b_d \delta^m_c \delta^x_n +\delta^b_n \delta^m_c  \delta^x_d
+\delta^b_c \delta^m_d \delta^x_n -\delta^b_c \delta^x_d \delta^m_n 
+\delta^b_d \delta^m_n \delta^x_c -\delta^b_n \delta^x_c \delta^m_d ) \right] \cr
&&~~~~~~~~~~~= \frac 1{24} F^2 \gamma_{\mu \nu} +\frac 1{12} [F^{2d}_\nu \gamma_{d\mu}
-F^{2d}_\mu \gamma_{d\nu} +F_\mu^{\ a} F_\nu^{\ b} \gamma_{ab}] \label{I19}
\feqn
For the remaining terms we easily have
\eqn
-\frac 16 e_{\mu a} e_{\nu c} F^{ab} F^{cd} [\gamma_b ,\gamma_d ]=-\frac 13 F_\mu^{\ b}
F_\nu^{\ d} \gamma_{bd} \label{II19}
\feqn
and
\eqn
\frac 1{24} e_{\mu a} F^{ab} F^{cd} [\gamma_b,\gamma_{\nu cd}] -(\mu \leftrightarrow \nu)
=-\frac 16 F_{[\mu}^{\quad b} F^{cd} \gamma_{\nu]bcd} \label{III19}
\feqn
(\ref{I19}), (\ref{II19}) and (\ref{III19}) together give (\ref{19}).

\section{Collecting the terms}
To obtain rapidly Eq.(28) it is convenient to collect the terms in the following
way:
\eqn
& & {\mbox{(\ref{1_g})$+$(\ref{2_g})$+$(\ref{7_g})}} \Longrightarrow \qquad -\frac 14 
\Omega_{\mu \nu}^{\quad ab} \gamma_{ab} \delta_i^{\ j} +\partial_\mu q^Y
\partial_\nu q^X R_{YXi}^{\quad j} ~~~~~~~\\
& & {\mbox{(\ref{4_g})$+$(\ref{9})}} \Longrightarrow \qquad
-g (\partial_\mu
A_\nu -\partial_\nu A_\mu ) P_{i}^{\quad j} -2g A_{[\nu} D_{\mu]}
P_{i}^{\quad j} ~~~~~~~ 
\feqn
\eqn
& & {\mbox{(\ref{5_g})}} \Longrightarrow ~~~
\frac i{4\sqrt 6} \left[ \gamma_{\nu ab} \DD_\mu F^{ab} 
-\gamma_{\mu ab} \DD_\nu F^{ab} \right]\delta_i^{\ j} 
+\frac i{\sqrt 6}
\left[ e_{\mu a} \gamma_b \DD_\nu F^{ab} -e_{\nu a} \gamma_b \DD_\mu 
F^{ab} \right] \delta_i^{\ j}\cr
&&
\\& & {\mbox{(\ref{6_g})$+$(\ref{11})}} \Longrightarrow \qquad
\frac {ig}{\sqrt 6} 
(\gamma_\mu D_\nu P_i^{\ j} -\gamma_\nu D_\mu P_i^{\ j})\\
& & {\mbox{(\ref{19})}} \Longrightarrow \qquad
\frac 1{24} F^2 \gamma_{\mu \nu} \delta_i^{\ j}
+\frac 1{12} \left[ F_{\ \nu}^{2 \ d} \gamma_{d\mu} -F_{\ \mu}^{2 \ d} 
\gamma_{d\nu}  \right] \delta_i^{\ j} 
-\frac 14 F_{\mu}^{\ a} F_{\nu}^{\ b} \gamma_{ab} \delta_i^{\ j}\\
& & {\mbox{(\ref{19})$+$(\ref{20})}} \Longrightarrow \qquad 
-\frac 16 F_{[\mu}^{\quad b} F^{cd} \gamma_{\nu]bcd} \delta_i^{\ j}
+\frac 16 g \gamma_{\mu \nu ab} F^{ab} P_i^{\ j}\\
& & {\mbox{(\ref{20})$+$(\ref{21})}} \Longrightarrow \qquad
\frac 23 g F_{[\mu}^{\quad b} \gamma_{\nu]b} P_i^{\ j}
+\frac {g^2}3 \gamma_{\mu \nu} |P|^2 \delta_i^{\ j} 
\feqn
where with $\DD$ we indicate the covariant derivative with respect to the spin connection. Put into the collected terms this gives Eq.(\ref{integral})

\chapter{Domain walls and the Randall-Sundrum mechanism}\label{domain wall}
\section{The Domain Wall solution}

\setcounter{equation}{0}

The conventional lore is that a vacuum of gravity or supergravity is
a configuration with maximal symmetry, namely with Lorentz invariance
$\mathrm{SO(1,d-1)}$ in $d$--dimensions. Adding translation
invariance one ends up with either Poincar\'e or de Sitter or anti de
Sitter symmetry which forces the vacuum expectation values of all
scalar fields to be constant. The new insight provided by the role of
the domain wall solutions  suggests that we might also consider vacua where
there is Poincar\'e invariance in one dimension less $\mathrm{ISO(1,d-2)}$
and where  the vacuum value of the scalar fields depends on the last $dth$
coordinate. These are precisely the domain wall vacua which are
expected to be a distinguished property of gauged supergravities.
Yet, as I have already mentioned, these wall geometries are
like solitons or kinks that interpolate between conventional vacua that we have discussed in sect.\ref{gaugd5theo}.

We start by the general assumption that the action that describes a $(d-2)$-brane can be consistently truncated to \cite{stellebrane},\cite{mbrastelle,mbratownsend}
\begin{equation}
  A_{D-Wall}^{[d]}=\int \, d^dx \, \sqrt{-g} \,\left[ 2 \, R[g]
  +\frac{1}{2}\partial^\mu \, \phi \partial_\mu \phi -
  2 \, \Lambda \, e^{-a \, \phi}\right]
\label{Dwaction}
\end{equation}
This admits a distinguished class of solutions that are called \textbf{domain walls} since at each instant of time
a brane of this type
separates the space manifold  into two adjacent non overlapping
regions:
\begin{eqnarray}
ds^2_{DW} & = & H(y)^{2\alpha} \left( dx^\mu \otimes dx^\nu \eta_{\mu \nu } \right)
+ H(y)^{2\beta} \, dy^2 \label{dwmet1}\\
e^\phi & = & H(y)^{-\ft {2a}{\Delta}}\label{dwdila1}\\
H(y) &=&c  \, \pm \, Q \, y \label{harmdw}
\end{eqnarray}
where $y$ is the single coordinate transverse to the wall, $c$ is an arbitrary integration
constant and the other parameters appearing in the above formulae have the following values:
\begin{equation}
  \alpha = \frac{2}{\Delta (d-2)} \quad ; \quad \beta =2\,
  \frac{d-1}{\Delta(d-2)}\quad ; \quad Q= \sqrt{\Lambda \, \Delta}
\label{DWconstant}
\end{equation}
in terms of $\Delta$ which is:
\begin{equation}
  \Delta= a^2 -2 \frac{d-1}{d-2}
\label{Dwdelta}
\end{equation}
The form (\ref{harmdw}) of the function $H$ is easy to
understand because in one--dimension a harmonic function is just a
linear function. 
Since $a^2$ is a positive quantity, $\Delta$ is bounded from below by
the special value $\Delta_{AdS}$ that corresponds to the very simple
case of pure gravity with a negative cosmological constant (case
$a=0$ in eq.(\ref{Dwaction})):
\begin{equation}
  \Delta \geq \Delta_{AdS} \equiv -2 \frac{d-1}{d-2}
\label{Dads}
\end{equation}
The name given to $\Delta_{AdS}$ has an obvious explanation. As it was
originally shown by L\"u, Pope and Townsend in \cite{popino1}, for
$a=0$ the domain wall solution (\ref{dwmet1}) describes a region of
the anti de Sitter space AdS$_d$. To verify this statement it
suffices to insert the value (\ref{Dads}) into (\ref{DWconstant})
and (\ref{dwmet1}) to obtain:
\begin{equation}
  ds^2_{DW} = H^{-2/(d-1)}(y)\left( dx^\mu \otimes dx^\nu \eta_{\mu \nu } \right)
+ H(y)^{-2} \, dy^2
\label{steppo1}
\end{equation}
Performing the coordinate transformation:
\begin{equation}
  r = \frac{1}{Q} \, \ln \, (c\pm Q\, y)  
\label{trasfo}
\end{equation}
the metric becomes:
\begin{equation}ds^2_{DW} = e^{-2\lambda r}\, \eta_{\mu\nu}\, dx^\mu \, dx^\nu + dr^2
\label{adsmetra}
\end{equation}
where
\begin{equation}\lambda= \sqrt{\ft{2\Lambda}{(d-1)(d-2)}}= (d-1) Q \label{lampara}
\end{equation}
In the same coordinates the solution for the dilaton field is:
\begin{equation}
  e^{\phi} = \exp \left[ -\frac{2\, a \,\lambda}{\Delta \, (d-1)} \,r\right]
\label{dilatoinr}
\end{equation}
Eq.(\ref{adsmetra}) is the metric of
$\mathrm{AdS}$ spacetime, in horospherical coordinates.  Following \cite{popino1}
we can verify this statement
by introducing the $(d+1)$ coordinates $(X,Y,Z^\mu)$
defined by
\begin{eqnarray}
X&=& \frac1{\lambda}\, \cosh\lambda r +\ft12\lambda\, \eta_{\mu\nu}\,
x^\mu x^\nu  \,
e^{-\lambda r} \nonumber\\
Y&=& -\frac1{\lambda}\, \sinh\lambda r -\ft12\lambda\, \eta_{\mu\nu}\,
x^\mu x^\nu \,
e^{-\lambda r} \label{embed}\\
Z^\mu&=&x^\mu\, e^{-\lambda r} \nonumber
\end{eqnarray}
They satisfy
\begin{eqnarray}
\eta_{\mu\nu}\, Z^\mu Z^\nu +Y^2 -X^2 &=&-1/\lambda^2 \label{emcon}\\
\eta_{\mu\nu}\, dZ^\mu dZ^\nu +dY^2 -dX^2&=&
e^{-2\lambda r} \eta_{\mu\nu}\, dx^\mu\, dx^\nu + dr^2
\end{eqnarray}
which shows that (\ref{adsmetra}) is the induced metric on the algebraic
locus (\ref{emcon}) which is the standard hyperboloid
corresponding to the $AdS$ space--time manifold. The signature of
embedding flat space is $(-,+,+,\cdots, +,-)$ and therefore
the  metric (\ref{adsmetra}) has the right $SO(2,d-1)$
isometry of the  $\mathrm{AdS}_d$ metric.
\par
Still following the discussion in \cite{popino1} we note that in horospherical coordinates
$X+Y=\lambda^{-1}\, e^{-\lambda r}$ is non-negative if $r$ is real. Hence
the region $X+Y<0$ of the full $\mathrm{AdS}$ spacetime is not accessible
in horospherical coordinates. Indeed this  coordinate patch
covers one half of the complete $\mathrm{AdS}$ space , and the metric describes
$\mathrm{AdS}_d /\mathbb{Z}_2$ where $\mathbb{Z}_2$ is the antipodal
involution $(X,Y, Z^\mu)\rightarrow
(-X, -Y, -Z^\mu)$. If $d$ is even, we can extend the metric
(\ref{steppo1}) to cover the whole anti de Sitter spacetime by setting the integration constant
$c=0$ which implies $H= Q\,y$.  So doing
the region with $y<0$ corresponds to the previously inaccessible region
$X+Y<0$.  If odd dimensions, we must restrict $H$ in
(\ref{steppo1}) to be non-negative in order to have a real metric and thus
in this case we have to choose $H=c + Q |y|$, with $c \ge 0$.  If the constant
$c$ is zero, the metric describes $\mathrm{AdS}_d/\mathbb{Z}_2$, while if $c$ is positive, the
metric describes a smaller portion of the complete $\mathrm{AdS}$ spacetime.  In any
dimension, if we set:
\begin{equation}
  H=c +Q |y|
\label{sofchoic}
\end{equation}
the solution can be interpreted as a domain wall
at $y=0$ that separates two regions of the anti de Sitter spacetime, with a delta function
curvature singularity at $y=0$ if the constant $c$ is positive.

\section{The Randall Sundrum  mechanism}
The magic of this
solution is that, as shown by Randall and Sundrum in \cite{RS2},
it leads to the challenging phenomenon of \textbf{gravity trapping}.
These authors have found that because of the exponentially rapid
decrease of the factor
\begin{equation}
  \exp [-\lambda |r|] \quad \mbox{with} \quad \lambda >0
\label{decrease}
\end{equation}
away from the thin domain wall that separates the two asymptotic
anti de Sitter regions it happens that gravity in a certain sense is
localized near the brane wall. Instead of the $d$--dimensional
Newton's law that gives:
\begin{equation}
   \mbox{force} \sim \frac{1}{R^{d-2}}
\label{Dnewton}
\end{equation}
one finds the the $d-1$--dimensional Newton's law
\begin{equation}
   \mbox{force} \sim \frac{1}{R^{d-3}} + \mbox{small corrections $ \mathcal{O}\left(
   \frac{1}{R^{d-2}}\right) $}
\label{D-1newton}
\end{equation}
This can be seen by linearizing the Einstein equations for the metric fluctuations
around any domain wall background of the form:
\begin{equation}
  ds^2 = W(r)\, \eta_{\mu\nu}\, dx^\mu \, dx^\nu + dr^2
\label{warpfac}
\end{equation}
that includes in particular the $AdS$ case (\ref{adsmetra}). In a very sketchy
way if one sets:
\begin{equation}
  h_{\mu \nu }(x,y) = \exp \left[ \mbox{i} p\cdot x\right]  \,
  \psi_{\mu\nu}(y)
\label{factorization}
\end{equation}
one finds that the linearized Einstein equations translate into an
analog  Schroedinger equation  for the wave--function
$\psi(y)$. This problem has a potential that is determined by the warp
factor $W(y)$. If in the spectrum of this quantum mechanical problem
there is a normalizable zero mode then this is the wave function of
a $d-1$ dimensional graviton. This state is indeed a bound state and
falls off rapidly when leaving the brane. Since the extra dimension
is non compact the Kaluza Klein states form a continuous spectrum
without a gap. Yet $d-1$ dimensional physics is extremely well
approximated because the bound state mode reproduces conventional
gravity in $d-1$ dimensions while the massive states simply
contribute a small correction.
\par
It is clearly of utmost interest to establish which domain walls have
this magic trapping property besides the anti de Sitter one.
This has been done by Cvetic, L\"u and Pope in \cite{cveticdelta}.
For obvious phenomenological reasons the case $d=5$ is the most interesting one and there have been a lot of attempts to construct explicit models in ${\cal N}=2$ gauged supergravity \cite{Behrndt:2000kz},\cite{flat-domainwall},\cite{cardoso:2002},\cite{berhndt:2002},\cite{curved-domainwall}. 
Let me summarize which are the requirements for the embedding of the Randall--Sundrum scenario inside a supersymmetric five--dimensional field theory
\begin{itemize}
  \item The candidate theory admits at least two different  anti de
  Sitter vacua with the same vacuum energy, namely two extrema  of the scalar potential
  $\varphi_0^{[1]}$ and $\varphi_0^{[2]}$ with $0>\mathcal{V}\left(\varphi_0^{[1]} \right)=
  \mathcal{V}\left(\varphi_0^{[2]} \right) $.
  \item The two $\mathrm{AdS}$ vacua  $\varphi_0^{[1]}$ and $\varphi_0^{[2]}$
  are stable, that is the spectrum of small fluctuations around these
  points satisfies the Breitenl\"ohner Friedman bound.\footnote{ For a
  review of this bound see for instance \cite{castdauriafre} Volume
  I.}
  \item There exists a smooth domain wall solution interpolating
  between these two vacua.
\end{itemize}
It is clear that the task is quite difficult due to the intrinsic complexity that the theory must have to satisfy these conditions and to escape the various no--go theorems \cite{renandrei}, \cite{Behrndt:2000tr}. Although some partial results have been obtained some general questions still remain like the uplifting of the five dimensional solutions to M--theory solutions.

\addcontentsline{toc}{chapter}{Bibliography}
%
%

\end{document}